\documentclass[format=acmsmall, review=false, screen=true]{acmart}

\usepackage{booktabs} % For formal tables

\usepackage[ruled]{algorithm2e} % For algorithms

\SetAlFnt{\small}
\SetAlCapFnt{\small}
\SetAlCapNameFnt{\small}
\SetAlCapHSkip{0pt}
\IncMargin{-\parindent}

\acmJournal{TOSEM}
\acmVolume{45}
\acmNumber{X}
\acmArticle{XX}
\acmYear{2023}
\acmMonth{06}
\copyrightyear{2023}
\setcopyright{acmlicensed}
\acmDOI{0000001.0000001}

\usepackage{alltt}
\usepackage{multirow}
\usepackage{booktabs}
\usepackage{listings}
\usepackage{graphicx}
\usepackage{float}
\usepackage{verbatim}
\usepackage{mathtools}
\usepackage{amsmath}
\usepackage{url}
\usepackage{standalone}
\usepackage{pgfplots}

\usepackage{natbib}     
\usepackage{syntax}
\usepackage{enumitem}
\usepackage{threeparttable}
\usepackage{framed}
\usepackage{hhline}
\usepackage{algpseudocode}
\usepackage{balance}
\usepackage{color, colortbl}
\usepackage{appendix}

\usepackage{longtable}
\usepackage{pifont}
\usepackage{array}

\usepackage{tikz}
\usetikzlibrary{calc}
\usepackage{verbatim}
\usetikzlibrary{arrows,shapes,backgrounds}

\tikzstyle{vertex}=[ellipse,fill=black!25,minimum size=20pt, inner sep=0pt]
\tikzstyle{edge} = [draw,thin,-]
\tikzstyle{glabel} = [text width=1cm,text centered,font=\bf]
\pgfdeclarelayer{bg}    % declare background layer
\pgfsetlayers{bg,main}  % set the order of the layers (main is the standard layer)

\usepackage{expl3}
\ExplSyntaxOn
\newcommand\latinabbrev[1]{
	\peek_meaning:NTF . {% Same as \@ifnextchar
		#1\@}%
	{ \peek_catcode:NTF a {% Check whether next char has same catcode as \'a, i.e., is a letter
			#1., \@ }%
		{#1., \@}}}
\ExplSyntaxOff

\newcommand{\pie}[1]{%
	\begin{tikzpicture}
	\draw (0,0) circle (1ex);\fill (1ex,0) arc (0:#1:1ex) -- (0,0) -- cycle;
	\end{tikzpicture}%
}

\newcommand{\CASE}[1]{\STATE \textbf{case} #1\textbf{:} \begin{ALC@g}}
	\newcommand{\ENDCASE}{\end{ALC@g}}

\newcommand{\DEFAULT}{\STATE \textbf{default:} \begin{ALC@g}}
	\newcommand{\ENDDEFAULT}{\end{ALC@g}}
\newcommand{\DEFAULTLINE}[1]{\STATE \textbf{default:} }
\colorlet{shadecolor}{gray!10}
\colorlet{framecolor}{black}
\newenvironment{frshaded}{%
	\MakeFramed {\FrameRestore}}%
{\endMakeFramed}

\newsavebox{\supbox}% Superscript box
\newcommand{\bsup}{\begin{lrbox}{\supbox}$\tt\scriptstyle}% Superscript begin
	\newcommand{\esup}{$\end{lrbox}{}^{\usebox{\supbox}}}% Superscript end
\def\eg{\latinabbrev{e.g}}
\def\ie{\latinabbrev{i.e}}

\algnewcommand{\LineComment}[1]{\State \(\triangleright\) #1}
\algdef{SE}[DOWHILE]{Do}{doWhile}{\algorithmicdo}[1]{\algorithmicwhile\ #1}%

\definecolor{lightpurple}{rgb}{0.8,0.8,1}
\definecolor{codebg}{RGB}{255,255,255}
\definecolor{commentcolor}{RGB}{11,140,11}
\begin{document}
% Title portion. Note the short title for running heads
\title[A Systematic Review of Automated Query Reformulations in Source Code Search]{A Systematic Review of Automated Query Reformulations in Source Code Search}

\author{Mohammad Masudur Rahman}
\orcid{0000-0003-3821-5990}
\authornote{This work was in-part completed as a PhD candidate at the University of Saskatchewan, Canada}
\affiliation{%
  \institution{Dalhousie University}
  %\streetaddress{110 Science Place}
  \city{Halifax}
  \state{NS}
  \postcode{B3H 1W5}
  \country{Canada}}
\email{masud.rahman@dal.ca}

\author{Chanchal K. Roy}
\orcid{0000-0003-3821-1234}
\affiliation{%
	\institution{University of Saskatchewan}
	%\streetaddress{110 Science Place}
	\city{Saskatoon}
	\state{SK}
	\postcode{S7N 5C9}
	\country{Canada}}
\email{chanchal.roy@usask.ca}

\begin{abstract}

Fixing software bugs and adding new features are two of the major maintenance tasks. Software bugs and features are reported as change requests. Developers consult these requests and often choose a few keywords from them as an ad hoc query. Then they execute the query with a search engine to find the exact locations within software code that need to be changed. Unfortunately, even experienced developers often fail to choose appropriate queries, which leads to costly trials and errors during a code search. Over the years, many studies attempt to reformulate the ad hoc queries from developers to support them. In this systematic literature review, we carefully select 70 primary studies on query reformulations from 2,970 candidate studies, perform an in-depth qualitative analysis (e.g., Grounded Theory), and then answer seven research questions with major findings. First, to date, eight major methodologies (e.g., term weighting, term co-occurrence analysis, thesaurus lookup) have been adopted to reformulate queries. Second, the existing studies suffer from several major limitations (e.g., lack of generalizability, vocabulary mismatch problem, subjective bias) that might prevent their wide adoption. Finally, we discuss the best practices and future opportunities to advance the state of research in search query reformulations.

\end{abstract}

%
% The code below should be generated by the tool at
% http://dl.acm.org/ccs.cfm
% Please copy and paste the code instead of the example below.
%


%
% End generated code
%


\maketitle

% The default list of authors is too long for headers.
\renewcommand{\shortauthors}{Rahman and Roy}

\section{Introduction}
Software maintenance costs up to 80\% of the total budget in modern software development \cite{maintenance-cost,modernizing}. 
One of the major aspects of maintenance is to deal with software bugs and failures, which cost the global economy billions of dollars every year \cite{17T}. A few of these bugs could even lead to 
massive fatalities (e.g., Boeing-737 MAX crashes \cite{boeing-etho,boeing-bug,boeing-bug2,boeing-indonesia}, Therac-25 accidents\footnote{https://bit.ly/2KU9IR2}). Software developers thus deal with hundreds of bugs and failures to ensure the high quality of 
their software products \cite{mozilla-bug}. They might also need to frequently enhance the features of their software 
to stay competitive in the market. We commonly call these maintenance tasks as \emph{change tasks} for the sake of brevity. Each of these change tasks is triggered by a \emph{software change request} either in the form of a bug report \cite{versionhistoryjsep} or simply a feature request containing plain texts \cite{kevic,saner2015masud, hillicse09}. Developers consult these change requests to implement the required changes in the software code. 

As a part of change implementation, the location of code that needs a change must be identified. Traditionally, various static and dynamic analyses have been used for this task. Since the last decade, various code search techniques, especially powered by Information Retrieval methods, have drawn the attention of software engineering community. 
To find the location of code requiring a change,
%These bugs and requested features are reported as 
%as a part of continuous evolution. In short, fixing software bugs and enhancing the software features are two of the major challenges of %software maintenance and evolution.
%Software bugs and requested features are reported to the developers in the form of bug reports and feature requests respectively, which are also commonly known as \emph{change requests} in the literature \cite{kevic,saner2015masud}.
developers often choose a few important keywords from a change request as an ad hoc query. Then they execute the query with a code search engine (\eg\ Lucene \cite{lucene}) and attempt to find out the exact locations within the software code that need to be changed as a part of fixing bugs or enhancement of features. Unfortunately, this has been challenging and even the experienced developers often fail to choose the right search queries from a change request \cite{kevic}. 
%For example, they
%According to \citet{kevic}, the developers 
%might fail up to 88\% of the time in choosing the right search queries \cite{kevic}.
As a result, they experience difficulties in detecting the appropriate locations within the code and spend the majority of their time in numerous trials and errors \cite{kevic,sitir,ko-88-percent}. One might think of using all the keywords of a change request as a query. 
However, the change requests are not originally written to be used as search queries. Thus, they 
%as a whole 
%which makes the keyword selection even more difficult. While choosing the right search keywords is a major challenge, the use of a 
%whole change request as a query is also not a viable choice either. 
%The whole texts from change requests 
often make verbose queries, which are noisy and ineffective \cite{verbose}. 
The developers might also attempt to improve their ad hoc queries by adding suitable keywords from the software code.   
However, \citet{vocaprob} suggest that there is a little chance (\eg\ 10\%--15\%) that they might guess the right keywords from the software code. Besides, the software code has an unrestricted vocabulary that evolves over time due to frequent code-level changes (e.g., addition of new features)
%, which could be larger 
%than that of regular texts (\eg\ change requests, bug reports) 
\cite{devanbu-fse}.
In short, selecting the right keywords either from the change requests or from the software code  is a major challenge during code search.
Thus, the developers are badly in need of automated supports for
constructing appropriate search queries and detecting relevant software code during the maintenance.  

Developers extensively search for code within a local codebase when dealing with bugs and features. However, they also frequently search for relevant code snippets on the Internet (\eg\ GitHub \cite{github-code-search}) 
to accomplish various programming tasks including code reuse, fixing bugs, and feature enhancement \cite{lemos-type-qr}. According to an earlier study \cite{twostudy}, the developers spend about 19\% of their programming time searching for relevant code on the Internet. During this code search, they also choose ad hoc search queries using a few keywords. However, 76\% of their queries need one or more reformulations to succeed \cite{koderlog, koderlog-msr}.
Thus, regardless of the problem context, choosing the right queries and locating the desired code are challenging, which warrants  appropriate automated supports for the developers.

%appropriate query construction targeting any code search is a major challenge for the developers. 
%Source code has a relatively smaller vocabulary than that of regular texts (\eg\ change requests, news article) \cite{devanbu-fse}
%Thus, preparing appropriate queries for the source code 
%Possible ways to help the developers overcome the above challenges are to (1) automatically suggest high quality queries from a change request for the \emph{local code searches} (\eg\ concept location, bug localization), and (2) automatically reformulate a query that is poorly chosen by the developers. Such supports involve (1) identification of appropriate keywords from relevant sources (\eg\ change requests, code under development), (2) removal of poor keywords and retaining of salient keywords, and (3) addition of more appropriate and complementary keywords to a given poor query. 
%The studies on Internet-scale code search reformulate a free-form NL query with complementary information such as relevant API classes and methods.
%Such reformulated query generally contains
%relevant API classes, methods or expressions that could be helpful for code search.
%Like the studies above, they also make use of relevance feedback \cite{tsc2016,active-cs,iman-tse,cexchange}, well established thesauri (\eg\ WordNet) \cite{lemos-type-qr,thesaurus-qr-idcs,pwordnet} and software repository mining \cite{portfolio,conngraph,rack,stolee2015,tsc2016} for reformulating a free-form NL query.  

Automated support for constructing queries and then for searching the code of interest (e.g., software bugs, features) has been an active area of research for decades. A number of studies \cite{gayg,shepherd,hillicse09,refoqus,sisman,qeffect,observed,verbose,ccmapping,infer,kevic,saner2015masud,saner2017masud,ase2016masud,ase2017masud,fse2018masud} attempt to support developers  
either (a) by constructing search queries for them from a change request
or (b) by reformulating their chosen ad hoc queries. These studies adopt various methodologies and algorithms as follows.
First, there have been several studies that construct queries from a change request by employing term weighting algorithms \cite{kevic,saner2017masud,fse2018masud,time-aware-term-weighting,twkraft-jsep},
natural language discourse analysis \cite{observed} and meaningful heuristics \cite{kevic}. Second, there have been other studies that reformulate (or reconstruct) a given search query by employing relevance feedback mechanism \cite{gayg,qeffect,refoqus,sisman,traceability-sliver-bullet,ase2017masud}, spatial code proximity analysis \cite{sisman,proximity-prf,sisman-jsep}, query difficulty estimation \cite{refoqus,qquality,ase2017masud,qperf}, term-query co-occurrence analysis \cite{ase2016masud,wembedding,hillicse09}, thesaurus lookup \cite{shepherd,hillicse09} and software repository mining \cite{ccmapping,infer,ase2016masud,wembedding}. Third, there have been another group of studies that reformulate queries for Internet-scale code search by capturing complementary keywords from crowdsourced contents \cite{tsc2016,iman-tse}, well established thesauri (\eg\ WordNet) \cite{lemos-type-qr,thesaurus-qr-idcs,pwordnet} and software repositories \cite{portfolio,conngraph,rack,stolee2015,tsc2016,cexchange}.       
Thus, there have been a significant number of studies that \emph{reformulate} queries to support source code search. Unfortunately, to the best of our knowledge,
there exists no systematic literature review that analyzes, categorizes or critically examines these studies, which is essential to advance the state of  research on this topic-- \emph{automated query reformulations for source code search}.

\textbf{Contributions:} In this systematic literature review, we investigate the existing researches on automated query reformulations supporting code search that were conducted during the last 15+ years. We carefully select 70 primary studies out of 2,970 candidates that were collected from 11 widely used publication databases (e.g., ACM Digital Library, IEEE Xplore, Fig. \ref{fig:study-selection}). Then we critically examine each of these primary studies by going through their full texts and by identifying their methodologies for reformulating queries, evaluations, and limitations. We apply the \emph{Grounded Theory approach} in our analysis using three levels of coding (open, axial, and selective) \cite{grounded-theory}. The method has been widely used for literature reviews since it can derive important theories that are firmly grounded within the qualitative data \cite{observed,slr,gt-survey}. Using this method, we categorize our selected studies into several categories in terms of their adopted methodologies and reported limitations. We also provide statistical evidence of conducted researches during the last two decades. We also compare and contrast between query reformulations in local code searches and that of Internet-scale code search. Finally, we discuss several recommended practices, open questions in automated query reformulations supporting code search, and suggest the future research directions in this domain. We thus make the following contributions in this article.

\begin{enumerate} [itemsep=3pt, topsep=3pt]

	 %We also provide a replication package \cite{slr-replication-package} to reproduce our work.
	
	\item \textbf{(RQ$_1$) Taxonomy based on methodology and algorithm.} We identify the methodologies or algorithms used by primary studies to reformulate their queries and classify the studies based on these dimensions. We found that, to date, eight major methodologies and algorithms have been used to reformulate queries for code search (Table \ref{table:methodology}). 
	%They are: term weighting, term-query co-occurrence analysis, thesaurus lookup, ontology lookup, search log mining, machine learning and use of ad hoc heuristics. 	
	About 40\% of these studies use term weighting and relevance feedback mechanism, 46\% make use of semantic relations, word co-occurrences, and thesaurus lookup, whereas 44\% of the studies rely on data mining and API recommendations to reformulate their search queries.

	%whereas 30\% look into the thesauri to select their query keywords.
	
	\item \textbf{(RQ$_2$) Evaluation and validation.} We analyze the evaluation methods, performance metrics, and validation targets used by the primary studies and report our findings. About 64\% of our primary studies use re-enactment-based evaluation, 26\% perform developer studies, and 9\% of the studies use a combination of both methods. They use a total of 30 performance metrics
	where a few metrics (e.g., Hit@K, MAP, Recall) (Fig. \ref{fig:popular-performance-metrics}) were frequently used. However, human participation is generally low (Fig. \ref{fig:participant-count-stats}), and the number of queries used to evaluate queries in the Internet-scale code search is often small (Table \ref{table:context-query-count}). About 30\% of our primary studies were also selected for comparison by the later studies.
	
	%They employ two types of evaluation methods: re-enactment based and developer-oriented. About 50\% of the primary studies use re-enactment based evaluation whereas 43\% of studies involve the developers in their experiments.
	
	\item \textbf{(RQ$_3$) Limitations and challenges of primary studies.} We analyze the limitations and challenges of the primary studies that were indicated either explicitly or implicitly in their papers. We find that their approaches suffer from eight
	major limitations or challenges (Table \ref{table:challenge}) including noisy keywords in their queries, vocabulary mismatch problem, lack of generalizability, human-induced biases, weak evaluation, and other prevailing issues that might prevent them from adoption by the software practitioners.
	
	 %types of limitations. About 30\% of these studies suffer from the vocabulary mismatch problem whereas another 30\% introduce extra cognitive burden on the developers during query reformulations. Most of the primary studies might also suffer from the lack of generalizability.
	
	\item \textbf{(RQ$_5$) Similarities and differences between local and Internet-scale code search.} About 58\% of our 70 primary studies perform query reformulations to support local code searches (e.g., concept location, bug localization, feature location) whereas the remaining 42\% focus on Internet-scale code search. 
	%In local code search, keywords are often selected using 
	%The studies supporting local code search rely on 
	%various term weighting methods (e.g., TF-IDF) whereas data mining and API recommendations are adopted
	%in the context of Internet-scale code search. 
	The primary studies from these two searches differ in their adopted methodologies (Fig. \ref{fig:compare-methodology}), reformulation types (Fig. \ref{fig:compare-qtype}), and evaluation methods (Fig. \ref{fig:compare-eval-type}). However, surprisingly, both groups of studies suffer from a common set of issues such as a lack of generalizability, weak evaluation, and noisy queries (Fig. \ref{fig:compare-challenge}). Future work on software code search can benefit from these empirical insights.
	
	%We offer a set of guidelines to develop effective tools and techniques for query reformulations and code searches.
	%. by carefully analysing the strengths and limitations of the primary studies. 
	%We also compare between the query reformulations targeting local code search (59\% studies) and the query reformulations targeting Internet-scale code search (41\% studies), and discuss several actionable insights.  
	
	\item \textbf{(RQ$_6$) Recommended practices in query reformulations for code search.} We identify several recommended practices in query reformulations based on experimental evidence and the general understanding of software engineering community. For example, collecting queries from professional developers \cite{refoqus} and selecting both open and closed-source systems \cite{flt-model-qr,open-close,oss-css} are recommended for designing an experiment. Reformulated queries should be executed against multiple search engines \cite{trconfig} to ensure a sound evaluation. Similarly, each proposed technique should be compared against state-of-the-art techniques through replication \cite{replication2,compare-sota}. Furthermore, the replication package should be shared publicly to advance the current state of knowledge \cite{replication}.
	 
	\item \textbf{(RQ$_7$) Open challenges and research opportunities.} We identify several open challenges in automated query reformulations supporting code search and  
	%targeting code search and 
	determine the scope of future work (Section \ref{sec:future-work}). We found that the state-of-the-art keyword selection algorithms might not be enough to identify appropriate keywords either from change requests or from software code.
	Future work can focus on adding more contexts (e.g., time-awareness) to the keyword selection algorithms, designing an appropriate fitness function for GA-based solutions \cite{cmills-icsme2018}, leveraging the structures from source code \cite{ase2017masud,concept-ase}, or using neural language modeling (e.g., word embeddings \cite{wembedding, word2vec,fasttext}) to deliver better quality queries for the code search.
	
	\item \textbf{(RQ$_1$) Replication package.} We provide a replication package\footnote{https://bit.ly/3QpZkSx} containing experimental metadata to reproduce our findings. It contains a comprehensive collection of 70 primary studies on automated query reformulations supporting code search that were conducted over the last 15+ years (Table \ref{table:year-venue-stat}). They were carefully selected from a total of 2,970 candidate
	studies (and 11 publication databases) through six levels of noise filtration \cite{slr,regression-testing-survey} (Fig. \ref{fig:study-selection}, Section \ref{sec:methodology}). We also present the statistical evidence of conducted researches using various analyses (e.g., Tables \ref{table:year-venue-stat}, \ref{table:venue-year-stat}, Fig. \ref{fig:year-cs-qr}) and discuss the recent trends of this topic (RQ$_4$). 
	
	 %They could be complemented with recent researches on distributional word semantics (e.g., word embeddings) , deep learning, and evolutionary computing (e.g., Genetic algorithms) .		
\end{enumerate}

%First, the term weighting approaches and query reformulation techniques that were adopted from Information Retrieval (IR) domain might not be directly applicable to the structured source code. These approaches were originally designed for unstructured natural language texts (\eg\ news articles) rather than source code.
%Besides, source code and regular texts often have very different semantics for the same words \cite{semantictool}. One way to possibly address this issue is to adopt context-aware term weighting methods (\eg\ TextRank \cite{rada}, Word embeddings \cite{wembedding}) for source code documents. Second, determination of the quality/goodness of a query without its execution is highly challenging, and available tools or methodologies from the literature fall short according to our investigation. In fact, the notion of query quality is highly complex, and the quality is perturbed by even single word change \cite{verbose}. Thus, query quality especially in the context of Software Engineering warrants further research. Third, although change requests sometimes contain software artefacts other than regular texts such as code snippets, stack traces and test cases \cite{goodbugreport}, most of the existing concept/bug localization techniques fail to exploit them. Significant amount of work are warranted on how such non-text artefacts could be leveraged in the query reformulation for the change tasks.  

\textbf{Structure of the article:} The rest of this article is organized as follows. Section \ref{sec:background} presents the background concepts and terminologies of automated query reformulations. Section \ref{sec:methodology} discusses the methodology of our systematic literature review. Section \ref{sec:results} presents our review results where Section \ref{sec:qr-techniques} focuses on the methods, algorithms and data sources used in the primary studies. Section \ref{sec:validation} discusses the evaluation methods of the primary studies whereas Section \ref{sec:challenge} points out their limitations and challenges.   
Section \ref{sec:qualitative} compares and contrasts between local code searches and Internet-scale code search in terms of their query reformulations. Section \ref{sec:recommended-practice} focuses on the recommended practices.
%presents our guidelines for developing 
%effective tools and techniques targeting query reformulation in the context of code search.  
Section \ref{sec:future-work} discusses the open challenges and scopes for future research in this domain. 
Section \ref{sec:threats} identifies the threats to validity of our findings,
Section \ref{sec:related} discusses the related parallel surveys, and finally, Section \ref{sec:conclusion} concludes our article with the major findings.

\section{Background}\label{sec:background}

\subsection{Automated Query Reformulation}
Searching for relevant code is a frequent activity in modern software development \cite{twostudy}. One major part of this search operation is to choose an appropriate query that reflects the information need. Unfortunately, during code search, software developers fail even 88\% of the time to choose the right search queries \cite{kevic,ko-88-percent}. When their search query fails, they attempt to reformulate the query (a) by adding new keywords, (b) by removing noisy keywords, or (c) by replacing the existing keywords with more appropriate ones. When these modifications are performed using automated means, they are called \emph{automated query reformulations} \cite{refoqus, shepherd,traceability-sliver-bullet}. Thus, the query reformulation tasks are three types as follows:
\begin{enumerate}[itemsep=3pt, topsep=3pt]
	\item \textbf{Query expansion:} A given search query is expanded 
	with synonyms, semantically similar, or complementary keywords.
	Developers often need to expand their initial queries since they are generally short (\eg\ 2--3 keywords on average) \cite{collective-experience, stolee2015}. According to existing studies \cite{stolee2015,koderlog}, up to 76\% of developer
	queries need to be expanded during code search.

	\item \textbf{Query reduction:} A given search query is reduced by removing the noisy, ambiguous, or less discriminating keywords. The keywords that occur in more than 25\% of the documents in a corpus are less discriminating \cite{refoqus,ase2016masud}. They are often removed from a query during code search.
	%One of the widely used heuristics for query reduction is \emph{document coverage}. That is, keywords that are found in more than 25\% of the source code documents within a corpus are removed from the query since they are not specific enough \cite{refoqus,ase2016masud}. Earlier studies also apply best sub-query selection \cite{query-reduction}, and search keyword selection using term weighting methods \cite{saner2017masud,saner2015masud} as a form of query reduction. 
	
	\item \textbf{Query replacement:} The keywords of a given query are replaced with more appropriate keywords. Examples of query replacement include spelling corrections \cite{multi-rec}, query generalization, and query specialization \cite{pwordnet,wordnet}. During code search, developers often learn new information from their unsuccessful searches and then calibrate their search by replacing the old query keywords with more appropriate keywords.
	
	 %replace their old search query with more appropriate query. 
	 %Such replacement could be intended for spelling corrections \cite{multi-rec}, query generalization or even for query specialization \cite{pwordnet,wordnet}. Developers often learn new information from browsing the search results, redefine their information needs, and then replace the initial search query altogether with more appropriate ones \cite{hard-trace-ase}. 
	 
\end{enumerate}

\begin{figure}[!t]
	\centering
	%\resizebox{3.5in}{!}{%-
	%\includegraphics{local-code-search.pdf}
	\begin{tikzpicture}[scale=.90, auto,swap]
		
		\begin{pgfonlayer}{bg}    % select the background layer
			
			%\node at (-1.2,1.5)[ellipse,draw,fill=yellow!20] (qe) {QE};
			
			\node[inner sep=0pt] (user0) at (-8.8,0)
			{\includegraphics[width=.45in]{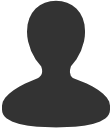}};
			
			\node[inner sep=0pt] (cr) at (-6.5,0)
			{\includegraphics[width=.5in]{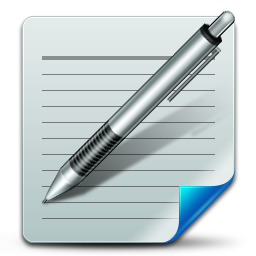}};
			
			\node[inner sep=0pt] (wheel0) at (-6.5,-3)
			{\includegraphics[width=.40in]{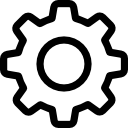}};
			
			\node[inner sep=0pt] (user) at (-4,0)
			{\includegraphics[width=.5in]{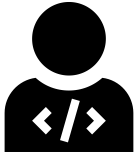}};
			
			\node[inner sep=0pt] (initq) at (-2,0)
			{\includegraphics[width=.40in]{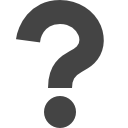}};
			
			\node[inner sep=0pt] (se) at (0,0)
			{\includegraphics[width=.50in]{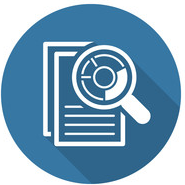}};
			
			\node[inner sep=0pt] (refq) at (0,-3)
			{\includegraphics[width=.40in]{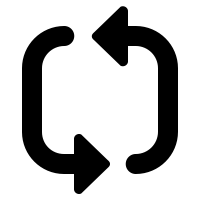}};
			\node[inner sep=0pt] (wheel) at (-1.1,-3)
			{\includegraphics[width=.40in]{./query-reform/setting3.png}};

			\node[inner sep=0pt] (repo) at (2.5,0)
			{\includegraphics[width=.5in]{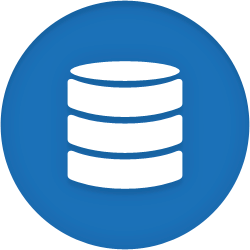}};
			
			%\node[inner sep=0pt] (repos) at (2.5,-3)
			%{\includegraphics[width=.50in]{./query-reform/repos.png}};
			
			%\node[inner sep=0pt] (summary) at (4.5,0)
			%{\includegraphics[width=.35in]{./query-reform/summary.png}};

			\draw[->,thick] (user0) -- (cr);
			\draw[->,thick] (cr) -- (user);
			\draw[->,thick] (user) -- (initq);
			\draw[->,thick] (initq) -- (se);
			\draw[<->,thick] (se) -- (repo);
			%\draw[<->,thick] (se) -- (repos);	
			\draw[->,thick] (se) -- (refq);	
			\draw[->,thick] (cr) -- (wheel0);
			\draw[->,thick] (wheel0) -| (user);	
			\draw[->,thick] (wheel) -| (-3.8,-.8);	
			
			\node at (-8.8,-1) (b) {\small User};
			\node at (-6.5,1) (b) {\small Change request};
			\node at (-3.9,1) (b) {\small Developer};
			\node at (-2,-1) (b) {\small Search query};	
			\node at (0,1) (b) {\small Code search};	
			\node at (2.5,-1) (b) {\small Local codebase};	
			\node at (-.5,-3.8) (b) {\small Query reformulation};	
			\node at (-6.5,-3.8) (b) {\small Query suggestion};	
			
			%%adding the markers%%
			\node at (-7.5,.3) [inner sep=2pt,circle,draw] (1) {\small 1};
			\node at (-4.8,.3) [inner sep=2pt,circle,draw] (2) {\small 2};
			\node at (-7.5,-3) [inner sep=2pt,circle,draw] (3) {\small 3};
			\node at (-2.7,.3) [inner sep=2pt,circle,draw] (4) {\small 4};
			\node at (-1,.3) [inner sep=2pt,circle,draw] (5) {\small 5};
			\node at (-.5,-2.2) [inner sep=2pt,circle,draw] (7) {\small 7};
			%\node at (-2.4,-.3) [inner sep=2pt,circle,draw] (8) {\small 8};	
			\node at (1.5,.3) [inner sep=2pt,circle,draw] (6) {\small 6};
			
		\end{pgfonlayer}
	\end{tikzpicture}
	\vspace{-.2cm}
	\caption{Automatic query reformulations in local code search}
	\label{fig:qr-ctask}
	\vspace{-.3cm}
\end{figure}

\subsection{Context of Query Reformulation}
\begin{figure}[!t]
	\centering
	%\resizebox{3.5in}{!}{%
	%\includegraphics{internet-scale-code-search.pdf}
	\begin{tikzpicture}[scale=.90, auto,swap]
		
		\begin{pgfonlayer}{bg}    % select the background layer
			
			%\node at (-1.2,1.5)[ellipse,draw,fill=yellow!20] (qe) {QE};
			
			\node[inner sep=0pt] (user) at (-4,0)
			{\includegraphics[width=.5in]{./query-reform/developer.png}};
			
			\node[inner sep=0pt] (initq) at (-2,0)
			{\includegraphics[width=.40in]{./query-reform/initquery.png}};
			
			\node[inner sep=0pt] (se) at (0,0)
			{\includegraphics[width=.50in]{./query-reform/search.png}};
			
			\node[inner sep=0pt] (refq) at (0,-3)
			{\includegraphics[width=.40in]{./query-reform/reformulate.png}};
			\node[inner sep=0pt] (wheel) at (-1.1,-3)
			{\includegraphics[width=.40in]{./query-reform/setting3.png}};

			\node[inner sep=0pt] (repos) at (2.5,0)
			{\includegraphics[width=.5in]{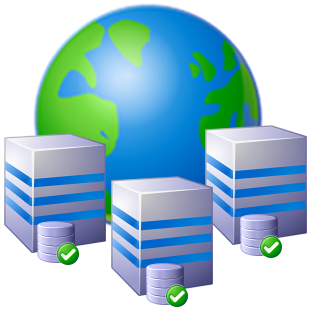}};

			\draw[->,thick] (user) -- (initq);
			\draw[->,thick] (initq) -- (se);
			\draw[<->,thick] (se) -- (repos);	
			\draw[->,thick] (se) -- (refq);	
			\draw[->,thick] (wheel) -| (user);	
			
			\node at (-4,1) (b) {\small Developer};
			\node at (-2,-1) (b) {\small Search query};	
			\node at (0,1) (b) {\small Code search};	
			\node at (2.5,-1) (b) {\small Internet-scale};
			\node at (2.5,-1.4) (b) {\small code repositories};
			\node at (-.5,-3.8) (b) {\small Query reformulation};	
			
			\node at (-4.8,.4) [inner sep=2pt,circle,draw] (1) {\small 1};
			\node at (-2.7,.4) [inner sep=2pt,circle,draw] (2) {\small 2};
			\node at (-1,.4) [inner sep=2pt,circle,draw] (3) {\small 3};
			\node at (-.5,-2.3) [inner sep=2pt,circle,draw] (5) {\small 5};
			\node at (1.7,.4) [inner sep=2pt,circle,draw] (4) {\small 4};
			
		\end{pgfonlayer}
	\end{tikzpicture}
	\vspace{-.2cm}
	\caption{Automatic query reformulations in Internet-scale code search}
	\label{fig:qr-cs}
	\vspace{-.3cm}
\end{figure}

Context refers to the environment where a task is performed.
It has been found to be useful for
%It plays a key role in 
many software engineering tasks (e.g., code completion, exception handling) \cite{context,cscc,holmes,surfexample,ghafari2017}. 
Similarly, the context plays a key role in query reformulation for the code search.
%is also important for constructing the right queries during the code search. 
In particular, the queries and their reformulation practices often depend on the type of code search they are intended for. Based on the locality and type of a corpus, code searches 
can be divided into two major categories as follows.   
\begin{enumerate}[itemsep=3pt, topsep=3pt]
	\item \textbf{Local code search}: The code is searched within a local codebase that contains only one software project. Fig. \ref{fig:qr-ctask} shows how queries are constructed from a change request in the context of local code search. The developer selects either the whole texts or a few important keywords from the change request as an ad hoc query (Steps 1-4, Fig. \ref{fig:qr-ctask}). Then the query is executed by a code search engine that returns the source code that needs to be modified (Steps 5-6, Fig. \ref{fig:qr-ctask}). If the query fails, it is reformulated with automated techniques by carefully analyzing the change request texts and the retrieved source code (Step 7, Fig. \ref{fig:qr-ctask}). 
	There also have been several attempts \cite{kevic,saner2017masud,observed} to support the developer with an initial query from the change request to avoid frequent trials and errors during the code search (Step 3, Fig. \ref{fig:qr-ctask}). Software bugs and features are located within source code using the local code search. 
	
	\item \textbf{Internet-scale code search:} The code is searched within an Internet-scale codebase that contains thousands if not millions of software projects (e.g., GitHub\footnote{https://github.com}, SourceForge\footnote{https://sourceforge.net/}). Fig. \ref{fig:qr-cs} shows 
	how search queries are reformulated in the context of Internet-scale code search.  The developer selects a few keywords as an ad hoc query and then attempts to find out the relevant code from the corpus using a search engine (Steps 1-4, Fig. \ref{fig:qr-cs}). If the query fails, it is reformulated with automated techniques using thesaurus lookup \cite{iman-tse,pwordnet,rack} or intent refinement \cite{intent-prediction}. 
	%Since the corpus contains a large number of projects,  
	Software developers spend about 19\% of their programming time searching for relevant, reusable code snippets from the Internet \cite{twostudy}.

	%the schematic diagram of query reformulations in the context of Internet-scale code search. The search is initiated by a developer by submitting a free-form natural language query to a code search engine (\eg\ GitHub code search). Recent studies \cite{google-for-code-search,stolee2015} also suggest that developers frequently use general-purpose web search engines (\eg\ Google) for their code searches. Unlike the local code search, this search is performed over thousands if not millions of code repositories across various application domains. 
	%Furthermore, construction of an initial query is a bigger challenge in this case since supporting materials such as change requests might not be available. Thus, several studies attempt to discover the intent behind a given search query \cite{rack,iman-tse,intent-prediction}, its linguistic or semantic issues \cite{pwordnet,stolee2016}, and then reformulate the query. This search targets the retrieval of one or more relevant code segments that meet the developer's need reflected in the query. 

\end{enumerate}

\subsection{Steps of Automated Query Reformulation}\label{sec:step-qr}
During code search, queries are reformulated in multiple steps. These steps might vary based on either type of reformulation or type of code search. However, most of the existing approaches \cite{gayg,refoqus,sisman,qperf,saner2017masud,ase2017masud} share a common set of steps (e.g., Fig. \ref{fig:steps-qr}) as follows.

\begin{enumerate} [itemsep=3pt, topsep=3pt]
\item \textbf{Query feedback collection:}
The first step of query reformulation is to collect feedback on a query (Steps 2--4, Fig. \ref{fig:steps-qr}). One way to collect the feedback is \emph{relevance feedback method} \cite{prf}, which has been widely used in the Information Retrieval domain. In this method, each of the documents retrieved by a given query is examined and annotated as either \emph{relevant} or \emph{irrelevant} to the query \cite{traceability-sliver-bullet,gayg}. 
The relevance feedback can be of three types as follows.

\begin{figure}[!t]
	\centering
	%\resizebox{3.5in}{!}{%
	%\includegraphics{steps-qr.pdf}
	\begin{tikzpicture}[scale=.90, auto,swap]
		
		\begin{pgfonlayer}{bg}    % select the background layer
			
			%\node at (-1.2,1.5)[ellipse,draw,fill=yellow!20] (qe) {QE};
			
			\node[inner sep=0pt] (user) at (-4,0)
			{\includegraphics[width=.45in]{./query-reform/developer.png}};
			
			\node[inner sep=0pt] (initq) at (-2,0)
			{\includegraphics[width=.40in]{./query-reform/initquery.png}};
			
			\node[inner sep=0pt] (se) at (0,0)
			{\includegraphics[width=.45in]{./query-reform/search.png}};
			\node[inner sep=0pt] (repo) at (1.1,0)
			{\includegraphics[width=.45in]{./query-reform/repository.png}};
			
			\node[inner sep=0pt] (prf) at (3,0)
			{\includegraphics[width=.45in]{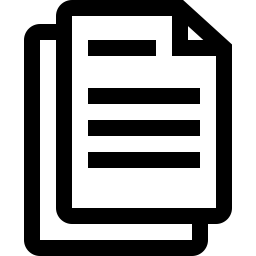}};
			
			\node[inner sep=0pt] (mine) at (5,0)
			{\includegraphics[width=.45in]{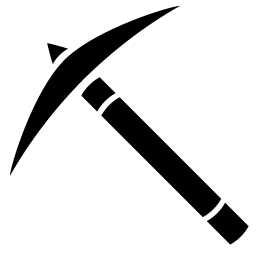}};
			
			\node[inner sep=0pt] (tw) at (5,-3)
			{\includegraphics[width=.45in]{./query-reform/setting3.png}};
			
			\node[inner sep=0pt] (ranking) at (2.8,-3)
			{\includegraphics[width=.45in]{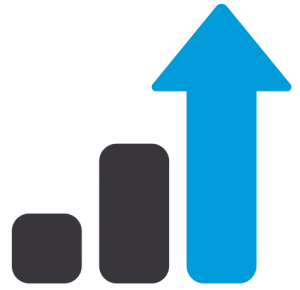}};
			
			\node[inner sep=0pt] (bestq6) at (0.3,-3)
			{\includegraphics[width=.45in]{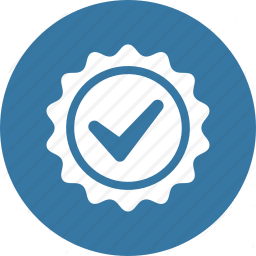}};

			\node[inner sep=0pt] (add) at (-2,-3)
			{\includegraphics[width=.45in]{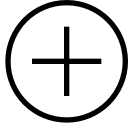}};
			
			\node[inner sep=0pt] (refq) at (-4,-3)
			{\includegraphics[width=.45in]{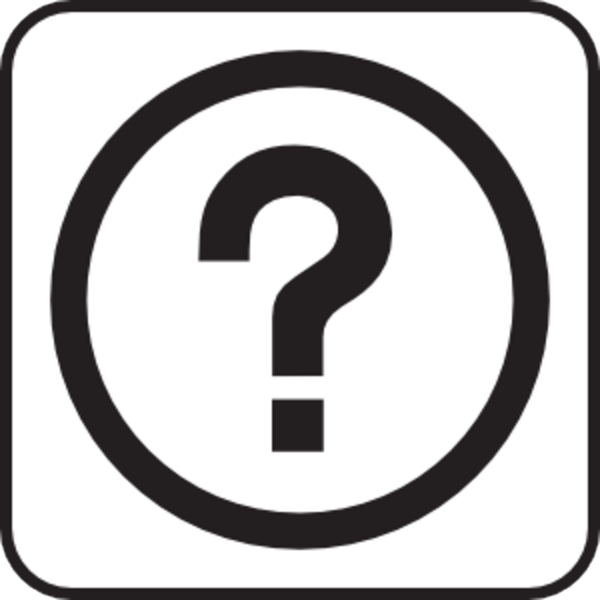}};

			\draw[->,thick] (user) -- (initq);
			\draw[->,thick] (initq) -- (se);
			\draw[->,thick] (repo) -- (prf);
			\draw[->,thick] (prf) -- (mine);
			\draw[->,thick] (5,0.005) -- (tw);		
			\draw[->,thick] (tw) -- (ranking);
			\draw[->,thick] (ranking) -- (bestq6);	
			\draw[->,thick] (bestq6) -- (add);
			\draw[->,thick] (initq) -- (add);
			\draw[->,thick] (add) -- (refq);			
			\draw[->,thick] (refq) -- (user);	
			
			\node at (-4,1) (b) {\small Developer};
			\node at (-2,1) (b) {\small Search query};	
			\node at (0.55,1) (b) {\small Code search};	
			\node at (3,-1) (b) {\small Pseudo-relevance};
			\node at (3,-1.4) (b) {\small feedback};
			\node at (5,1) (b) {\small Text mining};
			\node at (5,-4) (b) {\small Candidate};
			\node at (5,-4.4) (b) {\small term weighting};
			\node at (2.8,-4) (b) {\small Term ranking};
			\node at (0.3,-4) (b) {\small Best candidate};
			\node at (0.3,-4.4) (b) {\small selection};
			\node at (-4,-4) (b) {\small Reformulated};
			\node at (-4,-4.4) (b) {\small query};
			
			\node at (-4.7,.4) [inner sep=2pt,circle,draw] (1) {\small 1};
			\node at (-2.7,.4) [inner sep=2pt,circle,draw] (2) {\small 2};
			\node at (-1,.4) [inner sep=2pt,circle,draw] (3) {\small 3};
			\node at (2.2,.4) [inner sep=2pt,circle,draw] (4) {\small 4};
			\node at (5.5,.4) [inner sep=2pt,circle,draw] (5) {\small 5};
			\node at (5.8,-2.5) [inner sep=2pt,circle,draw] (6) {\small 6};
			\node at (5.8,-2.5) [inner sep=2pt,circle,draw] (6) {\small 6};
			\node at (2.4,-2.5) [inner sep=2pt,circle,draw] (7) {\small 7};
			\node at (-.5,-2.5) [inner sep=2pt,circle,draw] (8) {\small 8};
			\node at (-1.6,-2.1) [inner sep=2pt,circle,draw] (9) {\small 9};
			\node at (-4.4,-2.0) [inner sep=2pt,circle,draw] (10) {\small 10};
			
		\end{pgfonlayer}
		
	\end{tikzpicture}
	%}
	\vspace{-.2cm}
	\caption{Steps of automated query reformulation using pseudo-relevance feedback}
	\label{fig:steps-qr}
	\vspace{-.2cm}
\end{figure}

\begin{itemize}[itemsep=3pt, topsep=3pt]
	\item \textbf{Explicit relevance feedback:} Developers provide \emph{explicit} feedback and annotate each of the retrieved documents (by a search query) as either relevant or irrelevant \cite{gayg}. Although the explicit feedback could be accurate and meaningful, capturing it regularly from the developers is time-consuming and sometimes even impossible.
	
	\item \textbf{Implicit relevance feedback:} Since explicit feedback is costly, 
	several studies capture low-cost feedback that is \emph{implicitly} provided by the developers \cite{implicit-feedback}. This feedback is derived from a developer's reactions towards the retrieved documents such as eye movements, document examination patterns, and keyword deletion or retention patterns.
	
	\item \textbf{Pseudo-relevance feedback:} Unlike the above two types, this feedback does not warrant any developer intervention. That is, the Top-K (e.g., K=5) source documents retrieved by a given query are naively considered as \emph{relevant} to the query \cite{prf,refoqus,hillicse09}. It is also known as \emph{blind relevance feedback}. There has been significant evidence
	\cite{refoqus,ase2016masud,ase2017masud,fse2018masud} that suggests the effectiveness of this relevance feedback mechanism in query reformulation.
\end{itemize}

\item \textbf{Candidate keyword selection:} Once relevance feedback is collected, the next step is to select appropriate keywords 
to reformulate the given query using the feedback. 
Many studies \cite{refoqus,sisman,saner2017masud} analyze the annotated documents collected as a part of feedback and attempt to find out important keywords from them using various term weighting algorithms (Steps 5--6, Fig. \ref{fig:steps-qr}). Term weight is a numerical proxy to the relative importance of a keyword within a document \cite{tfidf}. Many of the term weighting algorithms (e.g., Section \ref{sec:termweight}) are borrowed from the Information Retrieval domain. 

%attempt to find out important terms from various information sources including the relevance feedback documents. In order to do that, they often make use of various term weighting methods borrowed from Information Retrieval (IR) domain (\ie\ Steps 5--6, Fig. \ref{fig:steps-qr}). According to our literature review, existing term weighting methods adopted in the query reformulation process can be classified into three broad categories as follows: based on their weights, and only Top-K candidates are suggested as a reformulation 
%\cite{saner2017masud,sisman,rocchio,qsurvey}. These important keywords are then used to automatically expand or replace the original poor query. 

\item \textbf{Reformulation of the search query:} Once the candidate keywords are collected, they are ranked and the top few keywords are used to expand a given query (Step 7, Fig. \ref{fig:steps-qr}). Less important keywords and the keywords non-existent in the corpus are discarded from the given query \cite{rocchio,refoqus}.
A few studies \cite{refoqus,ase2017masud,qperf-mills-tosem} argue that the same term weighting algorithm might not always work. Thus, they construct multiple reformulated versions of a given query using multiple term weighting algorithms and then deliver the best-reformulated query using supervised machine learning (Steps 8-10, Fig. \ref{fig:steps-qr}).
 
 %\citet{refoqus} argue that a single reformulation strategy might not be appropriate for all queries. That is, different queries might need different types of reformulations (\eg\ expansion, reduction, replacement). Towards this goal, several studies \cite{refoqus,trconfig,ase2017masud,qperf} adopt query difficulty analysis and machine learning to suggest the best reformulated query. In particular, they (1) construct multiple reformulation candidates for a given query, (2) determine their quality using 21--28 query quality metrics from Information Retrieval domain \cite{qsurvey}, and then (3) return the best candidate using machine learning as the reformulated query  (Steps 8-10, Fig. \ref{fig:steps-qr}).
	
\end{enumerate}

\subsection{Term Weighting} \label{sec:termweight}
 Term weighting is a popular method to determine the relative importance of individual terms within 
 %for selecting important keywords from 
 any body of texts (e.g., change request, source code) \cite{tfidf}. It is often used to select candidate keywords during
 %an important step of 
 automated query reformulations (Step 6, Fig. \ref{fig:steps-qr}). Several term weighting algorithms had been proposed and widely used over the last few decades. They can be classified into three broad categories as follows.
 
 \begin{enumerate}[itemsep=3pt, topsep=3pt]
	\item \textbf{Frequency-based term weighting:} The importance of a keyword is determined based on its occurrence frequencies. 
	The arguably most popular frequency-based method is
	%One of the widely used frequency-based methods is 
	TF-IDF \cite{tfidf}.  	
	TF-IDF stands for Term Frequency (TF) times Inverse Document Frequency (IDF). While TF counts the occurrences of a term (or keyword) within a document, the DF counts the number of documents (within a corpus) containing the term. On the other hand, IDF, the inverse of document frequency, determines the informativeness of a term by penalizing the highly frequent terms (e.g., stop words).
	Thus, TF-IDF determines a term's importance without considering the dependencies that it might have on other terms.

	\item \textbf{Graph-based term weighting:} 
	Unlike the frequency-based methods, graph-based methods \cite{blanco,saner2017masud,fse2018masud} capture the dependencies among keywords during their weight calculation. 
	Existing studies capture co-occurrences \cite{rada}, syntactic dependencies \cite{blanco}, and hierarchical dependencies \cite{fse2018masud} among the keywords, and transform a text document into a graph structure. In this graph, the keywords are represented as nodes and their dependencies as edges. Then the important nodes (or keywords) are identified
	%Then these studies identify the most important nodes (i.e., keywords) 
	using a graph-based, recursive algorithm namely PageRank \cite{pagerank}.
Source code has an unrestricted, evolving vocabulary that could be larger than that of regular texts \cite{devanbu-fse}. Furthermore, the code is rich in terms of structures and dependencies (e.g., static relations). Thus, graph-based algorithms leveraging dependencies could be a suitable choice to select important keywords from the source code during query reformulations \cite{ase2017masud,fse2018masud}. 
	
	%While the above two approaches rely on statistical properties of texts such as term occurrences and term co-occurrences, another group of studies 
	    	
\item \textbf{Probabilistic term weighting:} Unlike the above two types, probabilistic term weighting methods make use of Information Theory to calculate the term weights \cite{aqe-divergence,info-theory-carpineto,carmelbook}.
	% In particular, 
	%adopt a probabilistic view towards the term weighting \cite{aqe-divergence,info-theory-carpineto,carmelbook}. They determine the weight of a candidate term based on (1) its co-occurrence probability with the query keywords \cite{ehill} and (2) its divergence from the probability of randomness \cite{aqe-divergence}. In particular, they make use of Information theory, and
	One of the popular probabilistic methods is 
	%they estimate the weight of a given keyword $t$ using 
	Kullback--Leibler Divergence (KLD) that determines the divergence of a keyword (within a document) from the random occurrence probability.
	%as follows:  
	%\begin{equation}\label{eq:termprob}
	%S(t) = P_R(t) \times log(\frac{P_R(t)}{P_C(t)}) 
	%\end{equation} 
	%Here $R$ denotes the relevant documents that contain the keyword and $C$ refers to a corpus containing these documents. $P_R(t)$ estimates the probability of a keyword $t$ occurring in the relevant documents $R$ whereas $P_C(t)$ is the random probability within the corpus $C$. Thus, if $P_R(t)$ diverges significantly from the randomness, $P_C(t)$, then the keyword $t$ is considered as highly weighted within $R$, i.e., an import keyword.  Several studies \cite{traceability-sliver-bullet,logEntropy} combine probability and frequency to calculate term weights.  
	%with classical TF-IDF for query construction in the context of traceability recovery problem. According to them, the weight of a candidate term $S_{ij}$ can be calculated as follows:
	%as follows, and reported promising results for query reformulation . 
	%\begin{equation}
	%S_{ij}=log(tf_{ij}+1)\times \floor{1-\sum_{j=1}^{n} \frac{P_{ij}\times log(P_{ij})}{log(n)}}~~\text{where}~ P_{ij}=\frac{tf_{ij}}{\sum_{k=1}^{n}tf_{ik}}   
	%\end{equation}
	%Here $tf_{ij}$ refers to term frequency of the $i_{th}$ term in the $j_{th}$ document, $P_{ij}$ is the probabilistic weight of the same term across the corpus, and $n$ is the total number of documents in the corpus. 
	%According to them, 
	That is,  if a keyword is frequent across numerous documents within a corpus (a.k.a., generic keyword), it does not contain unique information, which makes it less important. On the contrary, a keyword that is specific to only a few documents often contains meaningful information, which makes it important.
	
	%a term's importance $S_{ij}$ is low if it is widely spread across multiple documents of the corpus, \ie\ with higher entropy. According to Information theory \cite{information-entropy}, such spread indicates that the term does not contain any unique information which could be of interest. 	
\end{enumerate}

\subsection{Implications of Automated Query Reformulation}
Choosing the right queries during code search is a challenging problem. Hence, automated supports for constructing queries are highly warranted for the developers.
%According to \citet{kevic}, developers often fail to select appropriate queries from a change request during code search within a local codebase. As several recent studies \cite{google-for-code-search,stolee2015} report, they also need to reformulate their queries frequently during code search on the web.
%code related queries more than those intended for regular web search when using Google. 
%In short, constructing an appropriate query for the code search is challenging regardless of the search context.
%A recent empirical finding \cite{google-for-code-search} suggests that queries intended for code search undergo more reformulations than those for regular web search.
%\citet{kevic} also suggest that developers generally fail to issue appropriate queries for code search at their first attempts (\ie\ 88\% of the times). 
%Thus, automated supports are highly warranted in the query construction for low cost code searches and software maintenance. 
However, automatic reformulations of queries have both positive and negative implications as follows: 
\begin{enumerate}
	\item \textbf{Benefits of query reformulation:} Developers often use short queries 
	to search for relevant code, which might not always 
	reflect their information need \cite{collective-experience}. 
	Thus, expanding these short queries with similar or complementary keywords 
	often improves their performance. According to existing literature \cite{emse2018masud}, $\approx$ 20\% performance gain could be achieved 
	in code search using automatically reformulated queries. Carefully reformulated queries might also help the developers locate their desired code with reduced cognitive or manual efforts.
	
	%often improves a poorly designed query. Existing studies have reported $\approx$ 20\% performance improvement as a result of query reformulations \cite{expand-reduce}. Furthermore, automatically reformulated queries and the documents retrieved by them help the developers (1) redefine their information needs and (2) retrieve the source code documents of interest more quickly. 
	
	\item \textbf{Costs of query reformulation:} Automatic reformulations might hurt the search queries that are already good (or appropriate) \cite{refoqus,ase2017masud,carmel}. Adding extra keywords makes them noisy \cite{query-drift}. Thus, no reformulation is better for them than inappropriate reformulations \cite{query-drift,semantictool, flt-model-qr}.
	%Hence, they might drift away from their original topics. Existing studies \cite{query-drift,semantictool, flt-model-qr} also argue that inappropriate reformulations of a search query are more harmful than no reformulation at all.
	There also exist a few difficult queries that cannot be improved using the traditional reformulation approaches \cite{carmel,hard-trace-ase}.
\end{enumerate}

Given these two-fold implications of query reformulations, contemporary approaches attempt to maximize the benefits and minimize the costs of reformulation. For example, several studies \cite{refoqus,ase2017masud,qperf} adopt supervised machine learning and query quality analysis to improve the poor queries and to preserve the good queries during the code search.

\section{Methodology}\label{sec:methodology}
We conduct a systematic literature review (SLR) \cite{slr,slr-org,regression-testing-survey} on the topic -- query reformulations for code search. 
%as was used by several earlier studies \cite{regression-testing-survey,rahman2018gaps,gt-survey}. 
We start our investigation by asking seven research questions, and then provide a comprehensive analysis of the topic -- \emph{automated query reformulations} in code search. We provide empirical evidence of the relevant research, and establish the state-of-the-art practices. Our goals are to summarize the current state of research in this domain and to outline the future research directions. We refer to the guidelines of \citeauthor{slr-org} for our systematic review as follows:  

\subsection{Research Questions}\label{sec:rq}
Specifying the research questions is one of the most important steps of the systematic literature review \cite{slr-org,slr}. We ask seven research questions in our survey. Our questions are classified into three categories -- \emph{general questions}, \emph{statistical questions} and \emph{focused questions}. Table \ref{table:rquestions} shows our seven research questions.

Our general questions target the \emph{generic} aspects of query reformulation approaches such as their algorithms, methodologies, or used corpora (\textbf{RQ$_1$}), their evaluation and validation methods (\textbf{RQ$_2$}) and the challenges or limitations they experience (\textbf{RQ$_3$}). 
Our statistical questions attempt to gather the \emph{statistical} evidence of the conducted researches over the years (\textbf{RQ$_4$}).
Finally, our focused questions concentrate on the \emph{specific} aspects of automated query reformulations targeting code search. They compare between local code search and Internet-scale code search in terms of their query reformulation practices (\textbf{RQ$_5$}), identify the best practices in query reformulations for code search (\textbf{RQ$_6$}), and also identify the future research directions (\textbf{RQ$_7$}).

\subsection{Search Strategy}
The next step of the systematic literature review is to collect a complete set of primary studies that can help us answer the research questions. 
This step involves the construction of appropriate search keywords on a research topic and then the accumulation of a non-biased set of studies from appropriate publication databases (\eg\ IEEE Xplore, ACM Digital Library).

\textbf{Construction of search keywords:} Choosing appropriate keywords and optimizing them for search are crucial to the discovery of available studies on a research topic. Research questions determine the scopes and motivations of a systematic review. \citeauthor{slr-org} suggest that these questions should be broken down into individual facets such as search keywords. Then they should be expanded with their synonyms, abbreviations, and alternative spelling using boolean operators.
Earlier studies often use a \textbf{PIO} (\textbf{P}opulation + \textbf{I}ntervention + \textbf{O}utcome) criterion \cite{poi,slr} 
to identify their search keywords. The PIO approach determines three types of keywords as follows:

\emph{Population} terms encompass all the aspects of a research topic including its application scopes, technologies, and standards. In the context of automated query reformulation targeting code search, we thus consider the following terms as the population of search keywords.
%\vspace{-.2cm}
\FrameSep.3em
\begin{frshaded}
	\noindent
	 Information retrieval, IR, text retrieval, TR, bug localization, concept location, feature location, FLT, concern location, Internet-scale code search, code search engine, search engine, local code search, code search, source code search, and code search query.
\end{frshaded}
%\vspace{-.2cm}
\emph{Intervention} terms focus on the specific aspect of a topic under study. In the context of automated query reformulation, we thus consider the following terms as our intervention keywords.
\begin{frshaded}
	\noindent
	Query reformulation, query expansion, query reduction, query formulation, query refinement, automated query expansion, AQE, query suggestion, query recommendation, term selection, query replacement, query difficulty, query quality, keyword selection, keyword extraction, search term identification, search query, search term, and search keyword.
\end{frshaded}
%\vspace{-.2cm}

\emph{Outcome} terms are related to the factors that are of importance to the stakeholders (\eg\ developers). In the context of query reformulation targeting code search, these outcomes could be the improvement in result ranks, reduced effort in code search, and reduced effort in localizing the bugs. However, we did not want to restrict our search too much.  Thus, we do not include the \emph{outcome} terms in our search keywords like the earlier study \cite{slr}.
Finally, we construct our keyword set for the systematic literature review as: \textbf{Population} AND \textbf{Intervention}. 

\textbf{Source of information:} Our goal was to collect as many as studies possible on the topic of interest from the literature. We choose 11 electronic publication databases for our study. These databases store most of the researches conducted in Computer Science and Software Engineering \cite{slr}. In particular, we collect all the journal articles and conference papers published between 1998
and 2021. To the best of our knowledge, 2004 was the year when the very first primary study (\citet{irmarcus}) was published.
However, 1998 was the inception year of Google's PageRank algorithm \cite{pagerank}, which had been frequently used since then both by Information Retrieval \cite{rada,blanco} and code search communities \cite{portfolio,portfolio-tosem}.
We thus choose 1998 as the starting year of our investigation to avoid missing any relevant studies accidentally.
%the algorithm that has revolutionized web search.
Our investigation finishes with 2021 as the ending year. First, 2,871 candidate studies, conducted between 1998 and 2018,   
were collected as a part of the PhD comprehensive exam of the first author. Recently, we include 99 more studies conducted between 2018 and 2021. We thus collect a total of \textbf{2,970} potentially relevant articles and papers on query reformulations targeting code search for our systematic review. 

\begin{table}[!t]
	\centering
	\caption{Research Questions}\label{table:rquestions}
	\vspace{-.35cm}
	\resizebox{4.8in}{!}{%
		\begin{threeparttable}
			\begin{tabular}{l|p{4.8in}}
				\hline
				\textbf{Ref\#} & \textbf{Question} \\
				\hline
				\hline
				\multicolumn{2}{c}{\textbf{General questions}}\\
				\hline
				\textbf{RQ$_1$} & Which methodologies (e.g., methods, algorithms) have been used for automated query reformulations targeting code search in the literature?\\
				\hline
				\textbf{RQ$_2$} & Which methods, metrics or subject systems have been used to evaluate and validate the researches on automated query reformulations?\\
				\hline
				\textbf{RQ$_3$} & What are the major challenges of automated query reformulations intended for code search? How many of them have been solved to date by the literature?\\
				\hline
				\multicolumn{2}{c}{\textbf{Statistical questions}}\\
				\hline
				\textbf{RQ$_{4}$} & 
				How many research articles have been published to date on automated query reformulations?
				%How much activities of research on automated query reformulations have been performed to date? 
				What are the venues that these researches got published at? \\
				\hline
				
				\multicolumn{2}{c}{\textbf{Focused questions}}\\
				\hline
				%\textbf{RQ$_4$} & What are the differences or similarities between query reformulation for code search and query reformulation for regular text search?\\
				%\hline
				\textbf{RQ$_5$} & What are the differences and similarities between query reformulations for local code search and query reformulations for Internet-scale code search? \\
				%\hline
				%\textbf{RQ$_6$} & Which one is more appropriate among term weighting, query-term co-occurrence and thesaurus-based approaches for query keyword selection?\\
				\hline
				\textbf{RQ$_6$} & What are the recommended practices in query reformulation for source code search?\\
				\hline
				\textbf{RQ$_7$} & What are the scopes for future work in the area of automated query reformulation targeting the code search?\\
				\hline

				%\textbf{RQ$_{9}$} & What are the venues that these researches got published at?\\ 
				%\hline
				%\textbf{RQ$_{10}$} & How much impacts do these researches have in terms of industry adoption?  \\   
				%\hline
			\end{tabular}
		\end{threeparttable}
		\vspace{-.3cm}
	}
\end{table}

\small
\begin{figure}[!t]
	\centering
	%\resizebox{3.5in}{!}{%
	%\includegraphics{selection-of-primary-studies.pdf}
	\begin{tikzpicture}[scale=.90, auto,swap,
		block/.style={
			draw,
			fill=white,
			align=center,
			rectangle, 
			text width={2cm},
			inner sep=3pt,
			fill=lightgray,
			font=\small},
		mylabel/.style={
			font=\small }]
		
		\begin{pgfonlayer}{bg}    
			
			%\node[block](acc) at (4,11.5) {InSpec};
			
			\node[block](acm) at (4,10) {\begin{tabular}{c}
					\small ACM Digital\\
					\small Library\\
			\end{tabular}};
			\node[block](crossref) at (4,9) {CrossRef};
			\node[block](dblp) at (4,8.2) {DBLP};
			\node[block](mendeley) at (4,7.4) {Mendeley};
			
			\node[block](google) at (4,6.4) {\begin{tabular}{cc}
					Google\\
					Scholar\\
			\end{tabular}};
			\node[block](ieee) at (4,5.4) {IEEE Explore};
			\node[block](proquest) at (4,4.7) {ProQuest};
			\node[block](science) at (4,4) {ScienceDirect};
			\node[block](springer) at (4,3.3) {SpringerLink};
			\node[block](wos) at (4,2.3) {\begin{tabular}{cc}
					Web of\\
					Science (ISI)\\
			\end{tabular}};
			\node[block](wiley) at (4,1.1) {\begin{tabular}{cc}
					Wiley Onlie\\
					Library\\
			\end{tabular}};
			\node[mylabel] (A) at (5.8,11.3){\begin{tabular}{cc}
					Initial \\
					search \\
			\end{tabular}};
			\node[mylabel](B) at (7.3,11.5){\begin{tabular}{cc}
					Irrelevant\\
					topics\\
					removal\\
			\end{tabular}};
			\node[mylabel] (C) at (8.8,11.5){\begin{tabular}{cc}
					Filtered \\
					by title\\
					(C1\&C2)\\
			\end{tabular}};
			\node[mylabel] (D) at (10.3,11.5){\begin{tabular}{cc}
					Filtered \\
					by abstract\\
					(C1\&C2)\\
			\end{tabular}};
			\node[mylabel](E) at (12,11.3){\begin{tabular}{cc}
					Combined \\
			\end{tabular}};
			\node[mylabel](F) at (13.7,11.3){\begin{tabular}{cc}
					Duplicate \\
					removal\\
			\end{tabular}};
			\node[mylabel](F) at (15.4,11.5){\begin{tabular}{cc}
					Filter by \\
					full texts\\
					(C3)\\
			\end{tabular}};
			\node[mylabel](G) at (17.2,11.3){\begin{tabular}{cc}
					Final \\
					selection\\
			\end{tabular}};
			
			% add the labels
			
			%adding the year
			\node (top-year) at (12,10.4){\textbf{(2004--2018)}};		
			
			\node (combinedl) at (12,5.4)[circle,fill=lightgray,inner sep=3pt]  {\large +};
			\draw[->,thick] (acm) -| (combinedl);
			\draw[->,thick] (crossref) -| (combinedl);
			\draw[->,thick] (mendeley) -| (combinedl);
			\draw[->,thick] (dblp) -| (combinedl);
			\draw[->,thick] (google) -| (combinedl);
			\draw[->,thick] (ieee) -- (combinedl);
			\draw[->,thick] (proquest) -| (combinedl);
			\draw[->,thick] (science) -| (combinedl);
			\draw[->,thick] (springer) -| (combinedl);
			\draw[->,thick] (wos) -| (combinedl);
			\draw[->,thick] (wiley) -| (combinedl);
			
			\draw[dashed] (5.8,10.8) -- (5.8,-2);
			\draw[dashed] (7.3,10.8) -- (7.3,-2);
			\draw[dashed] (8.8,10.8) -- (8.8,-2);
			\draw[dashed] (10.3,10.8) - -(10.3,-2);
			\draw[dashed] (13.7,10.8) -- (13.7,-2);
			\draw[dashed] (15.4,10.8) -- (15.4,-2);
			\draw[dashed] (17.2,10.8) -- (17.2,-2);
			
			\node (total) at (4,0.3) {\textbf{Total:}};
			\node (finish) at (17.8,0.3) {};
			\draw[thick] (total) -- (finish);
			\node (end) at (17.8,5.4){};
			\draw [->,thick] (combinedl) -- (end); 
			
			% ACM DL line
			\node (acm-init) at (6.3,9.8) {176};
			\node (crossref-init) at (6.3,8.8) {152};
			\node (dblp-init) at (6.3,8) {216};
			\node (mendeley-init) at (6.3,7.2) {203};
			\node (google-init) at (6.3,6.2) {200};
			\node (ieee-init) at (6.3,5.2) {432};
			\node (proquest-init) at (6.3,4.5) {137};		
			\node (science-init) at (6.3,3.8) {22};
			\node (springer-init) at (6.3,3.1) {313};
			\node (wos-init) at (6.3,2.1){999};
			\node (wiley-init) at (6.3,0.9){21};
			\node (total-init) at (6.4,0){\textbf{2,871}};	
			
			%impurity removal
			\node (acm-impure) at (7.8,9.8) {114};
			\node (crossref-impure) at (7.8,8.8) {101};
			\node (crossref-impure) at (7.8,8) {179};
			\node (crossref-impure) at (7.8,7.2) {123};
			\node (google-impure) at (7.8,6.2) {173};	
			\node (ieee-impure) at (7.8,5.2) {299};																											
			\node (prq-impure) at (7.8,4.5) {127};							
			\node (sd-impure) at (7.8,3.8) {14};
			\node (sd-impure) at (7.8,3.1) {265};
			\node (sd-impure) at (7.8,2.1) {912};
			\node (sd-impure) at (7.8,0.9) {10};
			\node (total-init) at (7.9,0){\textbf{2,317}};	
			
			%filter by title
			\node (acm-title) at (9.3,9.8) {41};
			\node (acm-abs) at (10.8,9.8) {16};
			\node (crf-title) at (9.3,8.8) {27};
			\node (crf-abs) at (10.8,8.8) {7};	
			\node (dblp-title) at (9.3,7.8) {75};
			\node (dblp-abs) at (10.8,7.8) {10};
			\node (mendeley-title) at (9.3,7.2) {59};
			\node (mendeley-title) at (10.8,7.2) {30};
			\node (google-title) at (9.3,6.2) {81};
			\node (google-abstract) at (10.8,6.2) {18};	
			\node (ieee-title) at (9.3,5.2) {65};
			\node (ieee-abstract) at (10.8,5.2) {41};
			\node (proquest-title) at (9.3,4.5) {69};	
			\node (proquest-abstract) at (10.8,4.5) {38};			
			\node (sd-title) at (9.3,3.8) {04};	
			\node (sd-abstract) at (10.8,3.8) {04};
			\node (sl-title) at (9.3,3.1) {46};	
			\node (sl-abstract) at (10.8,3.1) {03};
			\node (wos-title) at (9.3,2.1) {91};	
			\node (wos-abstract) at (10.8,2.1) {26};
			\node (wol-title) at (9.3,0.9) {04};	
			\node (wol-abstract) at (10.8,0.9) {02};
			\node (total-init) at (9.3,0){\textbf{562}};
			\node (total-init) at (10.8,0){\textbf{195}};
			\node (total-init) at (12.3,0){\textbf{109}};
			\node (total-init) at (14,0){\textbf{93}};
			\node (total-init) at (15.7,0){\textbf{62}};
			\node (total-init) at (17.5,0){\textbf{56}};
			\node (total-init) at (17.6,5.1){\textbf{\Large 56}};
			
			\node(extension) at (12,-1){\textbf{(2018--2021)}};
			\node[block](google2) at (4,-1.5) {\begin{tabular}{cc}
					Google\\
					Scholar\\
			\end{tabular}};
			\node (google-init2) at (6.3,-1.8) {99};
			\node (google-impure2) at (7.8,-1.8) {99};
			\node (google-title2) at (9.3,-1.8) {85};
			\node (google-abstract2) at (10.8,-1.8) {30};
			\node (total-init) at (12.3,-1.8){30};
			\node (dup-removal) at (14,-1.8){27};
			\node (google-c3) at (15.7,-1.8){14};
			\node (google-total) at (17.5,-1.8){\textbf{14}};
			
			\node (google-end) at (17.6,-1.5) {};
			\draw [->,thick] (google2) -- (google-end); 
			
		\end{pgfonlayer}
		
	\end{tikzpicture}
	
	%}
	\vspace{-.3cm}
	\caption{Selection of primary studies}
	\label{fig:study-selection}
	\vspace{-.5cm}
\end{figure}
\normalsize

\subsection{Study Selection}\label{sec:study-selection}
We exclude irrelevant studies from the 2,970 initial results, and select only the relevant ones for our systematic review. We apply three different filters, and choose the ones as our primary studies that meet certain conditions and quality standards. In the context of our systematic literature review, we carefully exclude such studies that:
\begin{itemize}
	\item \textbf{C$_1$}: do not address \emph{code search}
	\item \textbf{C$_2$}: do not address \emph{query reformulation}
	\item \textbf{C$_3$}: do not address \emph{query reformulation for code search}
\end{itemize} 
Fig. \ref{fig:study-selection} shows different steps of our study selection process. We apply multiple filters to the initial search results across several steps, and choose the final set as follows:

\textbf{(a) Removal of irrelevant topics:} Initial search results often contain entries that are not related to software engineering. This might happen due to an accidental keyword matching between the query and the full texts from the papers.
We use a semi-automatic approach to remove these irrelevant studies and to ensure the quality of our results. 
%we use a semi-automated approach for this removal. 
First, we analyze the keywords from the \emph{title} of each result and carefully identify the words that might be \emph{irrelevant} to
source code search and query reformulation. For example, we found that the results containing ``library" or ``libraries" in their title mostly deal with document search from digital libraries. On the other hand, the results containing ``mobile" in their title mostly deal with mobile-based location search or recommendation, which is not related to our topic of interest -- \emph{query reformulation in code search}. Thus, we discard the studies that contain any of these keywords in their \emph{title}.
	\begin{frshaded}
\noindent
Image, multimedia, multilingual,
bilingual, video, library, libraries, digital,
database, databases, db, chinese, japanese, trec,
mobile, medical, geo, music, sql, audio, and speech
	\end{frshaded}
   \noindent
To ensure that we are not losing any relevant studies, we also analyze the discarded results by these keywords: \{``library", ``libraries", ``mobile"\}. We found that only 2 out of their 62 results were related to code search but did not deal with any query reformulations, which shows the viability of our filtration step. 
 %query reformulations have been adopted in the application domains above, which are clearly not related to source code search. 
Thus, we removed 554 (18.65\%)  irrelevant results, and our initial studies got reduced to 2,416 (81.35\%) after this step.  
 
 \textbf{(b) Noise filtration using title:} After the removal of irrelevant topics, we are still left with thousands of results. Manually analyzing them for relevance is still impractical. We thus adopt another heuristic to reduce the noise from our results. Since we are interested about the studies related to query reformulation and code search, we discard any studies that \textbf{do not} contain 
 at least one of the following important keywords in their \emph{title}.
\begin{frshaded}
	query, code, developer, change, bug, concept, feature, concern, and software
\end{frshaded} 
\noindent
We believe that the studies that miss these keywords in their titles might not be relevant to our systematic review. It should be noted that we do not include ``search" into the above list. We found that the keyword “search” might be very generic and could lead to various irrelevant studies related to document search or location search rather than code search. This semi-automated filtration step discarded 1,769 (59.56\%) irrelevant studies, and our collection got reduced to 647 (21.78\%) potentially relevant studies.  
We also repeat the filtration step including “search” keyword in the above list, analyze the results from three publication databases – ACM Digital Library, IEEE Xplore, and Google Scholar – and found no change in the total number of primary studies. 

\textbf{(c) Noise filtration using abstract:} Noise filtration using title might not be sufficient enough, and the collection still could contain irrelevant studies. We thus manually analyze the titles from 647 studies and their corresponding abstracts in order to understand the underlying research. We noticed that despite having the desired keywords, many of these studies do not deal with query reformulations and code search. We thus manually discarded 422 (14.21\%) studies, and our collection got reduced to 225 (7.58\%) relevant studies. 

\textbf{(d) Result merging \& duplicate removal:} In this step, we merge the relevant studies from each of the 11 publication databases, and discard the duplicate studies. Since the same study could be indexed by multiple databases, the duplication might occur during retrieval. Even multiple versions of the same study (\eg\ peer-reviewed, pre-print) could be retrieved from multiple databases. We thus manually analyzed 225 studies and discarded a total of 105 (3.54\%) duplicate studies. This step left us with a total of 120 (93 + 27) relevant studies.    

\textbf{(e) Noise filtration using full texts:} We were unable to determine the subject matter of several studies even after reading their titles and abstracts. We thus read the full texts of 120 research papers, and  
discard several irrelevant studies. Many of these studies  discuss either code search algorithms or query reformulations in general but not the techniques for query reformulations. Since our topic was \emph{query reformulation for code search}, we considered them irrelevant for our survey and discarded them from the collection. This filtration left us with 76 studies. We also discarded short papers (\eg\ poster, doctoral symposium) and non peer-reviewed items (\eg\ technical reports, dissertation). However, a few short papers (e.g., new idea, tool) were retained due to their strong relevance to our topic of interest. Since our focus was only on the \emph{techniques} for query reformulations, we also discarded such studies that conducted developer surveys, interviews, or case studies on concept location, feature location, and Internet-scale code search. This filtration step led us to a total of 64 primary studies. It should be noted that despite careful, time-consuming steps above, relevant studies might be still omitted due to sub-optimal queries or the limitations of search engines provided by the publication databases. We thus added a few studies that were relevant (according to our prior work experience) but missed by the search process accidentally. Finally, we chose a collection of \textbf{70 (2.36\%)} papers as the primary studies for our systematic review.

\subsection{Quality Assessment}\label{sec:assessment}
Table \ref{table:selected-studies} (Appendix \ref{appendix:B})  provides an overview of our selected primary studies, their adopted methodologies, and offered features. As a standard practice, we investigate whether these studies meet certain quality criteria or not \cite{slr,slr-org}. In particular, we carefully go through each of these 70 studies, capture their subject matters, and then answer 10 standard quality questions as follows:
\begin{itemize}
	\item Q$_1$: Is there a clear statement about the aim of the research?
	\item Q$_2$: Is there an adequate description about the context of the research?
	\item Q$_3$: Is there a review about the related work from the literature? 
	\item Q$_4$: Is there an adequate description of a query reformulation technique targeting code search (\eg\ bug localization, concept location, Internet-scale code search)?
	\item Q$_5$: Has the approach been validated?
	\item Q$_6$: Does the conclusion reflect the aim or purpose of the conducted study?
	\item Q$_7$: Is there a clear statement of findings?
	\item Q$_8$: Does the study discuss its limitations or threats to validity?
	\item Q$_9$: Does the study recommend further research?
	\item Q$_{10}$: Does the study provide any supports for the replication (\eg\ replication package)?   
\end{itemize}
Table \ref{table:quality-ass} (Appendix \ref{appendix:A})  shows responses to each of the questions above. We record the response as either ``Yes" ({\small\pie{360}}), ``No" ({\small\pie{0}}) or ``Somewhat Yes" ({\small\pie{180}, \small\pie{270}}). We record ``Yes" when the supporting evidence is clearly visible, ``No" when the supporting materials are not found at all, and ``Somewhat Yes" when the relevant evidence is weak or ambiguous. From Table \ref{table:quality-ass}, we see that majority of our selected studies meet the quality criteria. 
While a few studies do not provide an exhaustive description of their query reformulation approach, they are still retained in this survey due to their overall relevance. Many of these studies also do not provide a publicly available replication package. However, it does not hurt the overall goal of our literature review. Thus, they were also retained. 

\subsection{Grounded Theory Based Analysis}\label{sec:gtheory}
We use Grounded Theory approach \cite{grounded-theory} to analyze our primary studies. It is widely used in the social science researches to derive meaningful theories that are firmly grounded on the data (\eg\ interview scripts). Recently, this approach  has also found applications in the Software Engineering researches \cite{bug-fix-thesis-arshad,observed, icsme2020masud}. We systematically analyze the full texts from each of the primary studies, derive important theories using this approach, and then answer several of our research questions. Grounded theory approach involves three stages of coding as follows:

\textbf{(a) Open coding} involves breaking down the gathered data into identifiable, interesting chunks and annotating them with appropriate key phrases (a.k.a., open codes) \cite{icsme2020masud}. We analyze the full text of each primary study, attempt to understand a concept of interest (e.g.,  methodology, validation, shortcomings), and then record our observations using suitable key phrases. The key idea was to keep an open mind and choose as many codes as needed to represent each study. 
For example, we used 209 open codes \cite{slr-replication-package} to represent the \emph{methodologies} adopted by our 70 primary studies to reformulate their queries. We spent $\approx$100 man-hours in this open coding task.

\textbf{(b) Axial coding} establishes relationships among the open codes. In this stage, we place our open codes into a spreadsheet, and establish connections among them using various colours. We annotate the key phrases with the same colour if they
are somehow related. We consider not only lexical overlap but also the semantic relatedness to connect the key phrases \cite{icsme2020masud}. Our goal was to convert the open codes into low-level categories. 
This stage provided a set of 21 tentative categories \cite{slr-replication-package} that explain the methodologies adopted by our 70 primary studies.

\textbf{(c) Selective coding} identifies the core variables behind a given phenomenon \cite{grounded-theory,bug-fix-thesis-arshad}. In our case, automated reformulation of a given query is the target phenomenon. While the axial coding provides a set of low-level categories, we carefully analyze them and merge them into high-level categories based on their themes and semantic relatedness.
This stage provided eight key categories of methodologies (Table \ref{table:methodology}) that were used to reformulate search queries by our primary studies. Each of these categories is represented using a set of semantically related key phrases \cite{slr-replication-package}. 

\begin{table}[!t]
	\centering
	\caption{Methodologies used for query reformulation}\label{table:methodology}
	\vspace{-.3cm}
	\resizebox{3.8in}{!}{%
		\begin{threeparttable}
			\begin{tabular}{l|l}
				\hline
				%\rowcolor{LightGray}
				\textbf{ID} & \textbf{Methodology}\\
				\hline
				\hline
				M1 & Use of term weights and relevance feedback \\
				\hline
				M2 & Mining of dependency graphs \\
				\hline
				M3 & Extraction of semantic relations from co-occurrences and thesauri \\
				\hline
				M4 & Use of conceptual and syntactic relations\\
				\hline
				M5 & Machine learning\\
				\hline
				M6 & API recommendation using data mining \\
				\hline
				M7 & Genetic algorithms \\
				\hline
				M8 & Miscellaneous \\
				\hline
			\end{tabular}	
		\end{threeparttable}
		\vspace{-.2cm}
	}
\end{table}

\section{Results}\label{sec:results}
We present our detailed analyses and findings in this section. We carefully go through each of the 70 primary studies, consult their 
major sections (e.g., problem descriptions,  methodologies, algorithms), 
and then summarize our non-trivial observations using Grounded Theory approach \cite{grounded-theory}.
We divide our findings and discussions into multiple logical sections, and attempt to answer each of our research questions (Table \ref{table:rquestions}) as follows:

\subsection{Answering RQ$\mathbf{_1}$: Methodologies for automated query reformulations}\label{sec:qr-techniques}
We identify eight key methodologies used by our primary studies for query reformulation. 
We use the Grounded Theory, a popular qualitative analysis approach, to come up with them.
Tables \ref{table:methodology}, \ref{table:study-methodology} show the methodologies and the primary studies that adopt them.
%These categories differ from one another in terms of their underlying algorithms/methodologies as well as the corpora (\eg\ thesaurus, codebase) that they use for suggesting the reformulated queries. Table \ref{table:study-category} shows how each of the primary studies falls into one or more categories. 
We discuss each of these methodologies, their reported strengths and weaknesses as follows.

\textbf{(M1) Use of term weights and relevance feedback:} Term weighting determines the \emph{relative importance} of a term (or word) within a body of texts (\eg\ document). It has been an integral part of Information Retrieval domain \cite{salton,tfidf}. \citet{tfidf} first introduced a term weighting method namely TF-IDF for Vector Space Model (VSM)-based document retrieval. According to Vector Space Model, a text document can be modeled as a vector of distinct words that are found in the document. On the other hand, TF-IDF stands for Term Frequency (TF) $\times$ Inverse Document Frequency (IDF) (check Section \ref{sec:termweight} for details). That is, the terms that are frequently used within a target document but rarely used across the whole corpus, are considered to be \emph{important} within the target document. Existing studies extract these terms as important keywords from a document \cite{manning-ir}. Besides TF-IDF, its variants (e.g., TF, IDF) are also used for term weighting on an ad hoc basis.  

\begin{table}[!t]
	\centering
	\caption{Primary studies and their adopted methodologies}\label{table:study-methodology}
	\vspace{-.3cm}
	\resizebox{4.4in}{!}{%
		\begin{threeparttable}
			\begin{tabular}{l|p{4in}| c}
				\hline
				%\rowcolor{LightGray}
				\textbf{ID} & \textbf{Primary studies} & \textbf{Total}\\
				\hline
				\hline
				M1 & S1, S2, S4, S7, S9, S10, S11, S12, S14, S19, S21, S22, S24, S25, S26, S30, S34, S38, S40, S43, S45, S50, S51, S54, S59, S61, S64, S70 &  28 (\textbf{40}\%)\\
				\hline
				M2 & S4, S19, S22, S24, S25, S31, S33, S40, S46, S56, S57, S61 & 12 (17\%)\\
				\hline
				M3 & S2, S3, S4, S6, S13, S15, S16, S17, S18, S19, S20, S21, S23, S24, S27, S28, S35, S37, S38, S39, S40, S41, S42, S43, S45, S49, S50, S57, S58, S60, S65, S70 & 32 (\textbf{46}\%) \\
				\hline
				M4 & S1, S5, S12, S16, S17, S18, S36, S37, S42, S43, S47, S48, S49, S51, S53, S62 & 16 (23\%)\\
				\hline
				M5 & S8, S10, S11, S22, S29, S32, S35, S58, S62, S63, S66, S67 & 12 (17\%) \\
				\hline
				M6 & S2, S7, S12, S15, S16, S17, S18, S19, S20, S26, S28, S34, S35, S37, S38, S39, S41, S43, S46, S49, S51, S52, S54, S55, S57, S58, S61, S63, S65, S66, S70 & 31 (\textbf{44}\%) \\
				\hline
				M7 & S9, S62, S67, S68, S69 & 5 (7\%)\\
				\hline
				M8 & S1, S3, S4, S5, S6, S7, S11, S12, S13, S14, S15, S17, S21, S24, S26, S29, S37, S39, S41, S44, S46, S53, S57, S60, S62, S67 & 26 (\textbf{37}\%) \\
				\hline
			\end{tabular}	
		\end{threeparttable}
		%\vspace{-.4cm}
	}
\end{table}

A number of primary studies \cite{gayg,time-aware-term-weighting,kevic,cocabu,active-cs,flt-model-qr,qperf,qperf-mills-tosem,icsme2018masud,icpc2019rodrigo, concept-ase} use term weighting methods (e.g., TF-IDF) to reformulate their search queries. In particular, they use these methods in association with \emph{relevance feedback mechanism} \cite{prf}. Relevance feedback is a way to capture complementary information on a query that can be used to reformulate the query
%evaluate a search query before it can  
 (check Section \ref{sec:step-qr} for details). 
First, a set of tentatively relevant documents are collected using the feedback mechanism. 
Second, candidate terms are extracted from these documents using one or more term weighting methods. 
%by determining their \emph{relative importance} through term weighting. 
Third, the candidate terms are then ranked based on their relative importance (a.k.a., term weights) and only Top-K terms are used to reformulate a query. 
\citet{candidate-link} first introduce relevance feedback and term weighting (e.g., TF-IDF) in the context of software engineering problems.
They reformulate search queries to recover traceability links and  
%recovery problem, and 
to reduce false-positive trace links. \citet{traceability-sliver-bullet} critically examine and extend their work by 
conducting multiple case studies and using various software artifacts (e.g., use cases, requirements, test cases). 
However,  \citet{gayg} were the first to use the term weighting and relevance feedback  
to reformulate search queries in the context of code search (\eg\ concept location).
%was first introduced by . 
Since then many studies adopt the term weighting and relevance feedback in their query reformulation approaches \cite{active-cs,sisman,cocabu,refoqus,ase2017masud}. \citet{refoqus} employ Rocchio's expansion \cite{rocchio}, Dice similarity \cite{qsurvey}, and Robertson Selection Value (RSV) \cite{rsv} to reformulate a query 
%for concept location 
where they use TF-IDF as a proxy of term importance.
Their query difficulty models also make use of several variants of TF-IDF (\eg\ avgIDF, maxIDF) \cite{qperf,qperf-mills-tosem}. \citet{flt-model-qr} adopt similar query reformulation techniques and term weighting methods to find relevant features from software model corpus. They later combine expert feedback with Rocchio's method to expand their search queries \cite{low-cost-crowd-qr, collab-feature-location}. \citet{twkraft-jsep} assign varying weights to the terms from various locations of a source document to identify the target software features. 
%source code terms to improve feature location task.
\citet{active-cs} use explicit feedback from developers to expand their queries (i.e., Rocchio's expansion) in the context of Internet-scale code search. However, they use their expanded query to re-rank the results retrieved by a given query rather than the code search.

%\begin{table}[!t]
%\centering
%\caption{Overlap of primary studies across the methodologies} \label{table:methodology-study-overlap}
%%\vspace{-.3cm}
%\resizebox{3.8in}{!}{%
%	\begin{threeparttable}
%		\begin{tabular}{c||c|c|c|c|c|c|c|c}
%			\hline
%			%\rowcolor{LightGray}
%			 & \textbf{M1} & \textbf{M2} & \textbf{M3} & \textbf{M4} & \textbf{M5} & \textbf{M6} & \textbf{M7} & \textbf{M8}\\
%			\hline
%			\hline
%			M1 & - & 8 &13 &6 &4 &12 &1 &13 \\
%			\hline
%			M2 & 8 &- &6 &0 &2 &6 &0 &4  \\
%			\hline
%			M3 & 13 &6 & - &8 &2 &19 &0 &13 \\
%			\hline
%			M4 & 6 &0 &8 & - &1 &8 &1 &8\\
%			\hline
%			M5 & 4 &2 &2 &1 & - &4 &2 &4 \\
%			\hline
%			M6 &  12 &6 &19 &8 &4 & - &0 &10  \\
%			\hline
%			M7 & 1 &0 &0 &1 &2 &0 & - &2  \\
%			\hline
%			M8 & 13 &4 &13 &8 &4 &10 &2 & - \\
%			\hline
%		\end{tabular}	
%	\end{threeparttable}
%	%\vspace{-.4cm}
%}
%\end{table}

To reformulate a search query, several studies capture relevant API classes, tags, and software-specific terminologies from a programming Q\&A site, Stack Overflow, where they employ various keyword selection methods including term weighting
\cite{icsme2018masud,tsc2016,li2016,cexchange,ase2016masud}. \citet{icsme2018masud} collect relevant Q\&A  threads from Stack Overflow for a given query using pseudo-relevance feedback, identify the important API classes from their embedded code segments using TF-IDF, and then use the top-ranked API classes to expand the query.
%suggest a ranked list of candidate APIs for query expansion. 
 \citet{tsc2016} employ Rocchio's expansion method, and identify software-specific keywords from the Q \& A threads of Stack Overflow using TF-IDF. \citet{cexchange} select frequent API classes from the relevant source documents retrieved by a query to expand their query.
 %as the candidates for query expansion. 
 \citet{kevic} use TF-IDF and three other lightweight heuristics (e.g., part of speech, position) to identify search keywords from a change request to support the concept location task. 
  %as a feature in their Logistic Regression model and automatically identify suitable search terms from a change request for concept location. 
 \citet{time-aware-term-weighting} extend the classical term frequency metric (i.e., TF) with time-awareness and improve the feature location task. They suggest that important terms within a source document are not only more frequent but also more recent than the noisy terms.
 %That is, according to their term weighting approach, 
 %such terms are more important within a source code document that are not only more frequent but also more recently introduced than the others. 
 They make use of version control history of a software system to gather the term frequency and timing related meta data. \citet{hard-trace-ase} make use of relevant web pages from three search engines (e.g., \emph{Google}, \emph{Bing} and \emph{Yahoo!}) to expand the \emph{stubborn} trace queries. In particular, they extract complementary key phrases from these web pages based on their occurrences (e.g., domain term frequency) and expand the trace queries.  
 
% using . Then they extract the key phrases from them by calculating their \emph{domain term frequency}, \emph{domain specificity} and \emph{concept generality} for query replacement. 

Although TF-IDF has been a popular method of term weighting, it suffers from a major limitation. TF-IDF and other frequency based methods fail to capture the dependencies among terms, which can be important to determine their relative importance \cite{rada,blanco}. These dependencies could be statistical, syntactic, semantic or hierarchical (check Section \ref{sec:termweight}).
%For example, the phrase, \emph{``query reformulation"}, conveys 
%a more comprehensive semantic about a concept than either \emph{``query"} or \emph{``reformulation"} does alone. In other words, both \emph{``query"} and \emph{``reformulation"} share each other's semantics temporarily in this phrase and thus establish a temporal semantic dependency due to their co-occurrence.
 Many existing studies from Information Retrieval domain \cite{rada,blanco} make use of these dependencies to determine term weights. However, \citet{saner2017masud} were the first to leverage both statistical and syntactic relationships among terms 
 to formulate queries for code search.
%in the context of query reformulation targeting code search.
They first transform a change request into a text graph using these dependencies and then apply PageRank algorithm to the graph to identify the top search keywords for concept location. 
%graph-based term weighting (\eg\ Equation \ref{eq:textrank}) on the graph to select search keywords for concept location. 
They later extend and adapt their approach to source code \cite{ase2017masud}, stack traces \cite{fse2018masud} and reformulate search queries for IR-based bug localization. They also report significant benefit of using graph-based term weighting over traditional alternatives (e.g., TF-IDF) in query reformulation.
%benefits of using graph-based term weights over frequency based term weights especially in automated query reformulation.
Several later studies \cite{kim-bl-qr, manq-qr,genqr-tse-perez} also adopt graph-based approaches to reformulate their code search queries.
%with syntactic dependencies among the terms using \emph{Jespersen Rank theory} \cite{saner2017masud}. \citeauthor{ase2017masud} also employ graph-based term weighting on source code \cite{ase2017masud} and stack traces \cite{fse2018masud} as a part of query reformulations that target IR-based bug localization. Several of their studies \cite{saner2017masud,ase2017masud,fse2018masud} report the benefits of graph-based term weights over frequency based term weights especially in automated query reformulation. However, such benefits are often achieved at a higher cost.
%\vspace{-.1cm} 
\begin{frshaded}
\noindent
About \textbf{40\%} (28/70) of the primary studies  use relevance feedback and term weighting to reformulate their queries during various code search operations (e.g., concept location, bug localization, Internet-scale code search). While relevance feedback captures complementary items on a query (e.g., relevant documents), the term weighting methods identify important keywords from them to reformulate the query. Existing studies leverage several proxies (e.g., frequency, dependencies) to determine the relative importance of search keywords. Thus, the effectiveness of their reformulated queries depend on the appropriateness of these adopted proxies for term importance.
%Term weighting determines the relative importance of keywords 
%within a document that is often captured using relevance feedback mechanism. About 40\% of our primary studies use term weighting and relevance feedback to reformulate their queries 
%based reformulation studies identify important keywords from a document (\eg\ bug report, source code) based on either term frequencies and/or semantic/syntactic dependencies among the terms. That is, they choose their reformulation candidates regardless of the keywords from a given search query.
%Thus, effectiveness of query reformulation with these studies partly depends on the relevance of the documents under analysis to the given query.     
\end{frshaded}
\vspace{-.1cm}

\textbf{(M2) Mining of dependency graphs:} Several primary studies \cite{auto-query,query-by-example, qsynthesis, active-cs,fse2018masud} leverage dependencies among items (e.g., words, API classes) to construct queries for code search. 
\citet{auto-query} mine an appropriate sub-graph from the program dependency graph (PDG) of a given code segment 
and construct a dependency query to find structurally similar code segments from a codebase.
Their goal was to support code-level changes by automatically detecting the similar code segments. 
They later incorporate explicit feedback from developers in their code search \cite{active-cs}.
\citet{precise-scalable-icpc} extract abstract syntax tree (AST) from a given code example and reformulate it into an XPath query to detect structurally similar code examples. 
\citet{query-by-example} also address the same problem by extracting appropriate sub-trees from the AST as a reformulated query.
Designing search query from a given code segment is often called \emph{query by example} in the literature \cite{query-by-example,precise-scalable-icpc}.
%during code search. 
 \citet{finelocator} expand the object-oriented methods (i.e., corpus documents)
 %an  (a.k.a., corpus document) 
 with neighbouring methods where the neighbours are detected based on their call dependencies, temporal proximity, and semantic similarity. 
\citet{multi-faceted-flt} detect syntactic and semantic facets from the results retrieved by a given query and use them to reformulate the query for concept location task. \citeauthor{fse2018masud} also mine three types of graphs -- 
trace graphs \cite{fse2018masud}, text graphs \cite{saner2017masud}, and API co-occurrence graphs \cite{icsme2018masud,ase2017masud} -- to reformulate queries for various code search operations (e.g., bug localization, concept location, Internet-scale code search). Several later studies \cite{kim-bl-qr, manq-qr, genqr-tse-perez} extend their work where they use graph-based mining to reformulate their queries.
\begin{frshaded}
	\noindent
	About \textbf{16\%} (11/70) of the primary studies mine dependency graphs to reformulate queries for their code search. They extract  dependencies from both structured (e.g., source code) and unstructured items (e.g., texts) to develop these graphs. Unlike texts, the graphs have the potential to capture more contextual information (e.g., call dependencies, AST, syntactic dependencies), which could significantly benefit the reformulated queries. 
\end{frshaded}

\textbf{(M3) Extraction of semantic relations from co-occurrences and thesauri:} Unlike the term weighting and graph-mining studies above, 
 a number of primary studies \cite{qr-log-mining,irmarcus,sisman,correlation-cs,multi-rec,swordnet,ccmapping,ase2016masud} leverage
 semantic relations between
 %co-occurrences between 
 query keywords and candidate terms to reformulate their search queries. 
 % between query keywords and candidate terms for query reformulation. 
 \citet{sisman} first use \emph{spatial code proximity} and suggest that terms co-occurring with query keywords in close proximity within source code are suitable candidates for query expansion. They were the first to expand search queries for IR-based bug localization.
 %for query reformulation in the context of IR-based bug localization.
 %They suggest that the terms that co-occur in close proximity with the keywords of a given query within the source code are suitable candidates for the query expansion. 
They extend their idea later with term ordering and language modeling \cite{sisman-jsep}.
\citet{swordnet} mine the mapping between code and corresponding comment,
%comment-code mapping from source code documents, 
and construct a software-specific thesaurus -- \emph{SWordNet}. Then they 
reformulate a given query with semantically similar, alternative words collected from their thesaurus to support the concept location task.
%using this thesaurus for local code search (\eg\ concept location). 
Similarly, \citet{ccmapping} analyze method signatures and their corresponding leading comments from source code, and extract semantically similar word pairs to support query reformulations. Both approaches above are subject to the availability of sufficient comments or documentations in the source code.

 \citet{shepherd} extract Verb-Direct Object (V-DO) pairs from method signatures and code comments, and recommend equivalent verbs and DO
 to support the developers in their query reformulation. They suggest that 
 % for a semi-automated query reformulation by the developers. 
%According to their approach, 
 if two V-DO pairs share the same direct object, then their verbs could be considered as equivalent.  
However, all free-form queries cannot be modeled as a V-DO pair \cite{hillicse09}. \citet{hillicse09} extend this work by relaxing the constraints of V-DO pairs and by suggesting phrasal representations of the method signatures. Developers thus could locate query keywords within these phrases, determine the relevance of corresponding methods, and then could manually choose the candidate terms for query expansion. \citet{irmarcus} construct a latent semantic space using \emph{Latent Semantic Indexing (LSI)} and suggest semantically similar words
to reformulate a given query. They also consider the co-occurrences between query keywords and candidate terms within source code to estimate the 
semantic relatedness.

Several other studies leverage the co-occurrences among items (e.g., terms, tags) within programming Q\&A threads \cite{ase2016masud,correlation-cs,li2016} and search logs \cite{qr-log-mining,qr-stackoverflow} for query reformulation.
Stack Overflow, a popular programming Q\&A site, discusses software-specific concepts and terminologies in millions of questions and answers. These questions and answers have been frequently used to support several 
%it is a suitable corpus for extracting semantically similar word pairs targeted for 
Software Engineering tasks \cite{ase2016masud,wordsim,icsme2018masud,li2016}.
\citet{correlation-cs} leverage explicit and implicit co-occurrences of tags from 
the same question and 
%\emph{implicit} co-occurrences from 
the duplicate questions respectively, and reformulate a given query with semantically similar tags.
%In particular, 
They calculate Mutual Information (MI) between any two tags with the query keyword being one of them, and then improve a given query using alternative tags.
\citet{ase2016masud} analyze the contextual words of both query keywords and candidate terms in the \emph{titles} of Stack Overflow questions, and suggest appropriate terms for query reformulation 
%reformulation terms for a query 
based on their contextual similarity. \citet{qr-log-mining} mine co-occurrences between query keywords and candidate terms in the past queries from code search logs, and suggest the frequently co-occurred terms for query expansion. 

Although co-occurrence between query keywords and candidate terms has been a popular idea for query reformulation, two similar words might not always co-occur in the same context \cite{highorder,highorder2}. According to \citet{highorder}, high order co-occurrences could be a possible solution to this issue. However, most of the primary studies above consider only the first order co-occurrences among terms to estimate their semantic relatedness, which might not be enough.

%in semantic relevance estimation which might not be sufficient enough.
%It should be noted that several graph-based approaches \cite{saner2015masud,saner2017masud} also capture term co-occurrences as a proxy to semantic dependencies among the terms. However, they do not warrant the presence of query keywords in such co-occurrences. Thus, they identify the candidate terms for query reformulation disregarding the keywords of a given query.
  %\textbf{(c) Thesaurus Based Studies:} Unlike the above two types of studies, thesaurus-based studies 
  
 Several primary studies \cite{shepherd,lemos-type-qr,correlation-cs,multi-rec,pwordnet,thesaurus-qr-idcs,anne} adopt well-established thesauri such as WordNet \cite{wordnet, wordnet-text-clustering} to reformulate their search queries.
  %reformulating a given natural language (NL) query. 
  %English thesaurus 
  %In fact, WordNet 
  %has been extensively used in various natural language processing tasks such as word sense disambiguation \cite{wordnet} and text clustering \cite{wordnet-text-clustering}. 
  %Several of our primary studies \cite{shepherd,lemos-type-qr,correlation-cs,multi-rec,swordnet,pwordnet,thesaurus-qr-idcs,kevicdict,anne} also make use of WordNet and software-specific thesauri (\eg\ SWordNet) in their query reformulations.
  \citet{shepherd} accept a Verb-Direct Object (V-DO) pair as a search query, and suggest synonyms from WordNet to expand their query.
  %for query expansion. 
  Developers are then responsible to choose the right keywords for their queries during concept location. \citet{multi-rec} later extend this work 
  %with pre-search and post-search recommendations where they  
   %of keywords for the query construction.
  %They provide auto-completion supports for identifiers and Verb-DO pairs, and suggest frequently co-occurred terms before search. If the search is not successful, they 
  with multi-level keyword suggestion and typo correction leveraging WordNet. \citet{pwordnet} replace each keyword of a given query with synonyms extracted from WordNet, and outperform the technique of \citet{hillicse09} in concept location task. 
  %another approach \cite{hillicse09} in locating concepts within the source code. 
  \citet{lemos-type-qr} also capture synonyms and antonyms from WordNet and 
 %reformulate a given query using synonyms and antonyms from WordNet and 
  predefined types from Java type thesaurus to reformulate queries 
  during Internet-scale code search. 
  
  Most of the studies above are inspired by parallel researches on automated query expansion from the Information Retrieval (IR) domain \cite{qsurvey}. However, existing findings \cite{semantictool} also suggest that the same word can hold two different semantics for regular texts and source code.
  %in regular texts and in source code. 
  For example, the term \texttt{new} represents an adjective in the regular texts whereas it means a verb (\eg\ object creation) in the object-oriented source code \cite{infer,swordnet}. In other words,  
  query expansion approaches solely based on English language thesaurus (e.g., WordNet) might not be effective for code search. 
  Given such a limitation, several primary studies \cite{infer,swordnet,ccmapping,correlation-cs} focus on developing software-specific thesauri to support query reformulations where they leverage the co-occurrences among keywords within the same context (e.g., method signatures, question tags).
  %for an effective reformulation of code search queries. 
  
  \citet{swordnet} first construct a software-specific thesaurus namely \emph{SWordNet} containing 8.5 million word pairs extracted from nine open source projects. This thesaurus was later used to collect semantically similar words and to reformulate queries for code search   
  \cite{swordnet,thesaurus-qr-idcs}. \citet{kevicdict} mine change requests, corresponding changed code, and developer's interactions with the IDE during code-level changes, and construct a dictionary that maps the problem domain vocabulary to solution domain vocabulary. 
  Then they translate a search query (a.k.a., change request)  into corresponding 
  source code elements (\eg\ methods, classes) using the dictionary to improve 
  %and then improve 
  the concept location task. 
  %\citet{correlation-cs} construct a software term database by using synonymous and related tags from Stack Overflow Q \& A site, and reformulate a given query with equivalent tags. 
  \citet{anne} mine programming concepts (\eg\ \emph{integer array}) and corresponding syntactic representations (\eg\ \texttt{int arr[]}) from the Q \& A threads of Stack Overflow, and construct a mapping database. Then they augment the source code statements with appropriate programming concepts from the database to improve code comprehension tasks.
  %so that the source code could be retrieved against natural language queries. 
  Although thesaurus-based approaches are popular and widely used, they might suffer from the lack of generalizability. That is, the thesauri should be constructed from appropriate corpora (\eg\ codebase, Stack Overflow) and 
  must be updated frequently to remain relevant.
  
  %so that they can deliver appropriate alternative keywords for a given NL query during code search.

  \vspace{-.1cm}
  \begin{frshaded}
  	\noindent
  	About \textbf{46\%} (32/70) of our primary studies make use of term co-occurrences and 
  	%co-occurrences between query keywords and candidate terms or 
  	well-established thesauri (e.g., WordNet, SWordNet) to reformulate their queries during code search.
  	However, semantically similar
  	%Direct co-occurrence between query keywords and candidate terms has been widely used by the primary studies in selecting candidates for query reformulation. However, semantically similar or relevant t
  	terms might not always co-occur in the same context, which warrants for higher order co-occurrences.
  	Thesauri-based approaches might suffer from the lack of generalizability and the thesauri might also require frequent updates to remain relevant.

  	 %within a certain context. High order co-occurrences (\eg\ second, third order) could be a potential solution to this issue.  
  \end{frshaded}
  \vspace{-.1cm}
  
  %\vspace{-.1cm}
  %\begin{frshaded}
  	%\noindent
  	%Thesaurus-based studies make use of English language thesauri (\eg\ WordNet) and/or software-specific thesauri (\eg\ SWordNet), collect synonyms, semantically similar or semantically relevant words against a given natural language query, and then reformulate the query using them. However, their thesauri might suffer from generalizability issues and might also require frequent updates.        
  %\end{frshaded}
  
\textbf{(M4) Use of conceptual and syntactic relations:} 
Several studies \cite{qr-service-discovery, nlsupport, qr-interactive-clarification, time-aware-term-weighting, api-book,hillicse09,pwordnet,anne,manq-qr,saner2017masud} leverage conceptual relations (e.g., domain ontology) and syntactic relations  among words (e.g., parts of speech tagging) to reformulate search queries. Ontology defines inter-relations among various concepts of an application domain (\eg\ Software Engineering) \cite{ontology}. Modern software systems are inherently complex and deal with numerous concepts ranging from high level problem descriptions to low level programming solutions. One way to preserve and manipulate this conceptual knowledge is to construct an ontology from these systems. A few of the primary studies \cite{qr-service-discovery, nlsupport, qr-interactive-clarification}
leverage the ontology to reformulate their queries during code search.

\citet{qr-service-discovery} reformulate a problem domain query into a solution service query using their ontology constructed from the source code. They capture the mapping between proven solution classes and requirement classes in their ontology.
%by using the ontology constructed from project source code. Their ontology encapsulates the knowledge on the mapping between proven service classes and corresponding problem domain classes. Thus, it can expand a given query from problem domain with relevant solution classes. 
\citet{nlsupport} construct another ontology using semantic web technology 
where they capture structural and data dependency relationships among classes, methods and attributes from a software system. 
Their approach accepts a quasi-natural language query from the developer, reformulates it into a structured SPARQL query, and then retrieves a list of relevant program elements. %(\eg\ identifiers) of developer's interest. 
Unlike \citeauthor{nlsupport}, \citet{qr-interactive-clarification} construct ontological models from multiple software repositories such as source code, bug-fixing history, and bug reports. Then they reformulate a free-form query into multiple candidate questions by capturing the query conditions and then mapping them to ontological concepts.
%. They first identify query conditions within an NL query, map these conditions to qualified concepts within the ontological models, and then choose a set of navigation paths based on their relevance to the query. Finally, they translate these navigation paths into human readable questions as the reformulated queries for the given NL query. 
Developers then could choose the most appropriate question from them and locate the relevant program elements.
Several other studies \cite{hard-trace-ase,concept-ase} also make use of domain-specific concepts to reformulate their queries and to improve their search.
%\{concept-ase} construct an API graph from source code and learn API embeddings from the graph using \emph{TransR} \cite{trans-r}.
Although ontology-based studies are reportedly useful to reformulate natural language queries into complex structured queries \cite{nlsupport}, they are still limited by their underlying technologies (\eg\ semantic web, SPARQL). Besides, maintaining an ontology is 
costly since the changes in source code are frequent and they need to be reflected in the ontology.

Besides conceptual relations (e.g., ontology), a number of primary studies \cite{time-aware-term-weighting, api-book,hillicse09,pwordnet,anne,manq-qr,saner2017masud} leverage the syntactic or linguistic properties of texts to reformulate their search queries. As a standard practice, they often perform parts of speech (POS) tagging and then select or emphasize the nouns and verbs in their query reformulation. The underlying idea is that these words convey more important semantics than the others. However, POS tagging is often noisy with technical texts (e.g., bug reports, change requests), which could hurt the query reformulation.

%challenging and time-consuming given that changes in the source code of a software system are \emph{frequent} and \emph{inevitable}. 
  \vspace{-.1cm}
\begin{frshaded}
	\noindent
	About \textbf{23\%} (16/70) of our primary studies make use of conceptual (e.g., ontology) and syntactic relations (e.g., POS tagging) to reformulate their queries during code search. Although they have been used by several studies, maintaining a software-specific ontology could be challenging and the POS tagging with technical texts could be noisy.
	%Ontology-based reformulation studies exploit the conceptual knowledge harnessed from one or more software repositories to reformulate a given query. They are reportedly found useful in locating the program elements and in answering conceptual questions during program comprehension.      
	%While these approaches support the developers in their code searches targeting especially program comprehension, 
	%However, they are also limited by their ontologies and associated technologies (\eg\ SPARQL). 
\end{frshaded}     
\vspace{-.1cm}
\textbf{(M5) Machine learning:} Several primary studies \cite{kevic, refoqus, qperf, qperf-mills-tosem,ase2017masud,neural-query-expansion,dl-qr-change-sequence,qr-stackoverflow, alteration-intent} adopt machine learning to reformulate their queries during code search.
%in reformulating a given query. 
\citet{kevic} capture \emph{part of speech}, \emph{term weight}, \emph{position} and \emph{notation} of each term from a change request, and train a \emph{Logistic Regression} model. Their model then accepts an incoming change request and returns Top-3 terms as a search query for concept location. \citet{neural-query-expansion} train an encoder-decoder model using deep neural networks and translate a natural language query into relevant API methods to support Internet-scale code search. 
\citet{alteration-intent} suggest that the changed version of code might hold the true intent of a search query that retrieves the original code.
Being motivated by this idea, \citet{dl-qr-change-sequence} train a deep belief network (DBN) with original code and their changed versions from thousands of GitHub projects, and then expand a given query with appropriate terms predicted from the changed code. 
\citet{lawrie-qr-bl} develop an automated summarizer using sequence-to-sequence model and reformulate a bug report into a summary to support the bug localization task. \citet{qr-stackoverflow} train a sequence-to-sequence model with an attention mechanism where they use original and reformulated queries from the search logs of Stack Overflow Q\&A site. Then they use their model to reformulate a given query during code example search at Stack Overflow. Although the sequence-to-sequence models have been popular for reformulating queries, they might require a large amount of training data.

Determining quality of a search query before its actual reformulation has been a popular idea in the Information Retrieval domain \cite{carmel}.
%Hence, query quality prediction has been a well studied problem . 
\citet{qperf} first analyze and predict the difficulty of queries in the context of code search. They train their model using 21 pre-retrieval metrics from the IR domain, and then predict the quality of a given query during concept location. \citet{qperf-mills-tosem} later extend this model with seven post-retrieval metrics (\ie\ 28 in total), and evaluate the model performance extensively.
%in the context of both concept location and traceability link recovery. 
Given a search query, \citet{refoqus} generate a list of four reformulation candidates
%candidate reformulations 
using Rocchio's expansion \cite{rocchio}, Robertson Selection Value (RSV) \cite{rsv}, Dice similarity \cite{qsurvey} and query reduction \cite{qsurvey}. Then they train a query difficulty model to suggest the best reformulated query.  
%employ their query difficulty model in delivering the best candidate from their candidate list as a reformulated query. 
\citet{ase2017masud} also extract four reformulation candidates by analyzing method signatures and field signatures from source code and 
then suggest the best candidate 
%of a source code document by employing a graph-based term weighting algorithm namely \emph{CodeRank}. Then they recommend the best reformulation candidate for a given query (targeting concept location) 
using the query difficulty analysis \cite{qperf}. Although the query difficulty analysis has been popular and cheap, its metrics might not be as accurate as the query performance metrics (e.g., mean average precision, query effectiveness). It should noted that query difficulty metrics (e.g., keyword frequency, keyword entropy) are often used to prevent poor queries from execution when the ground truth is not known. On the contrary, query performance metrics are computed using the ground truth information, which makes them more reliable.
%reformulated query to a given query for concept location task.

Over the last few years, word embedding technology has been leveraged to tackle the vocabulary mismatch problem in code search.
\citet{iman-tse} learn the embeddings of query keywords and API classes
from a corpus of $\approx$25,000 open source projects using a three-layer neural network and Continuous Bag of Words (CBOW) algorithm.
Then they translate a natural language query into relevant API classes to support code search. 
Similarly, \citet{icsme2018masud} learn the word embeddings from 1.40 million Q\&A threads of Stack Overflow using \emph{Skip-gram algorithm} \cite{fasttext}, and expand a natural language query with relevant API classes. 
Both studies above report significant performance improvement in their code search.
%\citet{qecc-intent} mine code change history and learn the mapping between changed code and original code. 
%Their model accepts a free-form natural language query, retrieves tentatively relevant source code using the query, and then predicts the potential code changes for the retrieved code.
%They call such code changes as the \emph{query intent}.
%Keywords from such code changes (\eg\ API classes) are then used to expand the given search query. 
%\vspace{-.2cm}

\begin{frshaded}
	\noindent
	About \textbf{17\%} (12/70) of our primary studies use machine learning, query difficulty analysis, and word embeddings to reformulate their queries during code search. While these technologies have high potential to improve query reformulations and code search,  
	%Machine learning models could improve the query reformulations significantly if they are equipped with high quality reformulation candidates and/or appropriate corpora.
	%used in association with other widely adopted approaches (\eg\ term weighting). 
	%However, 
	they are also restricted by their training datasets, learning algorithms, and model parameters.
	%They might also suffer from scalability and generalizability issues.
\end{frshaded} 

\textbf{(M6) API recommendation using data mining:} Many primary studies \cite{qr-log-mining,swim,qr-stackoverflow, codehow,cocabu,iman-tse,rack,icsme2018masud,concept-ase} mine code search logs and software repositories (e.g., API documentation, programming Q\&A website) to reformulate search queries through the recommendation of relevant API classes or methods.
%from  search engines to reformulate a given query. 
%In fact, there have been a number of studies \cite{boldi-qr-mining,term-retain,semantic-gap} in the Information Retrieval domain that mine web search logs, and deliver automated supports in the query construction (\eg\ Google's auto-complete suggestion). Thus, log-mining based primary studies in Software Engineering are mostly inspired by the parallel approaches from IR domain. 
\citet{koderlog} first capture the search logs from an Internet-scale code search engine-- \emph{Koders}, and 
analyze the patterns, topics, and user behaviours during code search from the logs. 
They report that the majority of natural language queries from these logs required query reformulations, which indicates the need for appropriate query reformulation techniques.
\citet{qr-log-mining} also analyze the logs from Koders \cite{koderlog-msr}, construct a term-term co-occurrence matrix from the past queries, and 
%based on the past queries. Their approach accepts a list of query keywords, and  
then suggest frequently co-occurred terms from the matrix to expand a given query.
%Manually performing these reformulations is both time-consuming and technically challenging. 
%Thus, appropriate tool supports are highly warranted for query reformulations in the context of Internet-scale code search. 
\citet{swim} analyze the \emph{click-through} data from Bing search engine that contains   
millions of (\texttt{query}, \texttt{URL}) pairs. They mine the logical mapping between each query and API classes from corresponding URL, construct a conditional probabilistic model, and then suggest relevant API classes to reformulate a given query.
\citet{qr-stackoverflow} analyze the search logs from Stack Overflow Q\&A website, train a sequence-to-sequence model using original and reformulated queries, and then generate a reformulated query for a given query. 
%establish a logical mapping between the query keywords and the API classes from corresponding webpage/URL. Then they develop a conditional probabilistic model using this mapping information. Their model accepts an NL query for code search, and then delivers the most likely API classes as a reformulated query. 
\citet{trace-query-transformation} also capture the query pairs from developers containing original and reformulated queries, 
%original and reformulated query pairs from the developers, 
mine query transformation rules, and use them to reformulate a query for traceability link recovery.

Several primary studies mine relevant API classes from API documentation \cite{codehow,sniff}, programming Q\&A websites \cite{rack,icsme2018masud,icpc2019rodrigo,cocabu,anne}, codebase, and version control history \cite{iman-tse,concept-ase, dl-qr-change-sequence,qecc-intent,qsynthesis} to reformulate a given query. \citet{codehow} identify relevant API classes to a query considering textual similarity between the query and candidate API documentation from MSDN, and use them to expand the query.
\citet{sniff} enrich the API invocations in source code with corresponding API documentation to tackle the \emph{vocabulary mismatch problem} \cite{vocaprob} in code search. 

\citet{rack} first capture the co-occurrences between query keywords and API classes in the Q\&A threads of Stack Overflow, and suggest relevant API classes to expand a given query. Several later studies \cite{icsme2020masud,icpc2019rodrigo} extend this work. Other primary studies reformulate a query with relevant keywords \cite{tsc2016,ase2016masud,cocabu} and tags from Stack Overflow \cite{correlation-cs,li2016}. \citet{tsc2016} capture software-specific keywords from the relevant Q\&A threads using pseudo-relevance feedback, question popularity, and term weighting method, and then expand a given query for code search. \citet{li2016} construct a software-specific tag database and reformulate a given query with synonymous and relevant tags from Stack Overflow questions.

\citet{iman-tse} mine relevant API classes from $\approx$25,000 open-source projects to expand a given query. \citet{concept-ase} use \emph{Recodoc} \cite{recodoc}, a TF-IDF based traceability recovery technique, to reformulate a given query into relevant API entities.  \citet{dl-qr-change-sequence} mine thousands of code commits from the version control history and 
suggest API elements from the changed code using Deep belief network (DBN) to reformulate a query. 
Similarly, \citet{qecc-intent} attempt to capture the intent of a query from 
the changes made to its initially retrieved code, and then expand the query with API elements extracted from these changes.
While expanding a given query with relevant API classes, keywords, and tags was found useful,  determining the true intent of a query still remains a challenge and unlocking that could lead to more accurate query reformulation.
  %\vspace{-.1cm}
\begin{frshaded}
	\noindent
	About \textbf{44\%} (31/70) of our primary studies mine relevant API classes, keywords, and tags from various sources such as search logs, API documentation, programming Q\&A site, open-source projects, and version control history. Many of these approaches are inspired by the parallel studies from Information Retrieval. While mining relevant items from these sources was found useful, determining the true intent of a search query still remains a challenge. Besides, the logs from widely used commercial search engines (\eg\ Google, Bing) might not be available for public use.
\end{frshaded} 

%\small
%\input{used-algorithm.tex}
%\vspace{-.2cm}
%\normalsize

\textbf{(M7) Genetic algorithms:} Several primary studies \cite{cmills-icsme2018, flt-model-qr, manq-qr,lawrie-qr-bl, genqr-tse-perez} use Genetic algorithms (GA) to reformulate their search queries. \citet{cmills-icsme2018} first use a Genetic algorithm to select the near-optimal query from a bug report to support bug localization. They demonstrate that bug reports often contain near-optimal queries although there might not be any explicit hints for bug localization (e.g., program elements, stack traces), which is interesting. However, their approach uses query effectiveness to determine the \emph{fitness} of query candidates. That is, the approach will need ground truth to find the near-optimal query, which makes it unsuitable for practical use. A few later studies \cite{lawrie-qr-bl,masud-emse2021} revisit their findings with different datasets and reach similar conclusions. According to \citet{masud-emse2021}, although the majority of bug reports contain near-optimal keywords, the GA-based approach might not be able to identify them from a few bug reports (e.g., $\approx$12\%) despite the availability of ground truth information.
\citet{manq-qr} perform multi-objective optimization using 15 fitness functions, and
reformulate a query using Genetic algorithm to support bug localization. 
On the other hand, \citet{genqr-tse-perez} use Genetic algorithms to reformulate corpus documents (e.g., model fragments) in the context of feature location task. They employ four query reformulation types-- expansion \cite{rocchio}, replacement \cite{hard-trace-ase}, reduction \cite{refoqus}, and selection \cite{saner2017masud} -- as the mutation operations and textual similarity between query and corpus document as the fitness function in their GA-based feature location. While they report benefit in reformulating the corpus documents (e.g., model fragments), the reformulation of queries was not found effective by one of their earlier work \cite{flt-model-qr}.

\begin{frshaded}
	\noindent
	About \textbf{7\%} (5/70) of our primary studies use Genetic algorithms (GA) to reformulate their search queries and corpus documents. While GA-based approaches have high potential for effective query reformulations, designing an appropriate, practical fitness function still remains a challenge. Besides, the GA-based approaches could also be costly due to their large search space constructed by all the keywords from either a bug report or a change request.
\end{frshaded} 

\textbf{(M8) Miscellaneous:} Many primary studies \cite{time-aware-term-weighting,testcase-cs-ist,api-book,cocabu,kevic,shepherd,hillicse09,cexchange,codehow,query-by-example,rack,querythecode,hard-trace-ase,sniff} use lightweight heuristics and ad hoc metrics to reformulate their queries during code search.
%and make ad hoc choices in the reformulation of a given query. 
 \citet{time-aware-term-weighting} select only nouns as query keywords from a feature request to 
 improve the feature location task. \citet{api-book} make use of nouns, verbs, and adjectives from the texts.
 \citet{shepherd} model search queries for concern location 
 %could be modelled as 
 as Verb-Direct Object (V-DO) pairs. 
%as a search query for concept location. 
\citet{hillicse09} expand this technique with other parts of speech.
%capture the uses of query keywords in the method/field signatures, and suggest the co-occurring terms for query reformulations that are frequently found within their contexts. 
\citet{kevic} consider four lightweight heuristics such as 
%in extracting the search query from a change request. They consider
\emph{part of speech}, \emph{term weight}, \emph{notation} and \emph{position} of each keyword 
within a change request to separate the query keywords from the rest using machine learning.
 %within the change request texts as the indicators of its suitability as a search keyword. 

\citet{rack} capture co-occurrences between query keywords and API classes in the Q\&A threads of Stack Overflow, 
apply three lightweight heuristics to detect the relevant API classes, and then use them to expand a query for Internet-scale code search.
%keywords (in the titles of Stack Overflow questions) and API classes (in the corresponding accepted answers), employ three advanced heuristics, and then translate a free-form NL query into relevant API classes for Internet-scale code search. 
Unlike \citeauthor{rack}, several studies \cite{sniff,codehow,concept-ase} rely on textual similarity between API documentations and a query to find the relevant API classes. Other studies simply extract the structural entities such as API classes or API methods from the queries \cite{api-book}, test cases \cite{testcase-cs-ist}, accepted answers of Stack Overflow \cite{cocabu}, and the relevance feedback documents \cite{cexchange}.

% for query reformulation. 

\citet{req-qr} replace a query with the most relevant requirement specification using textual similarity to improve feature location.
\citet{multi-faceted-flt} extract multiple facets (e.g., structural, intent, dependency) from the initial results of a query to help developers further refine the query interactively. \citet{verbose} adopt stepwise query reduction and suggest that removal of even only one noisy keyword 
can significantly improve the query.
%from the verbose query can provide significant benefit during bug localization.
%\citet{query-by-example} accept a generic code segment as query, transform it into an Abstract Syntax Tree (AST), and then retrieve the code segments that have similar ASTs. \citet{auto-query} accept a few code segments, compute program dependency graphs (PDG) from them, and then generate dependency queries for code search by employing a graph-mining algorithm.
\citet{observed} perform natural language discourse analysis, identify observed behaviour (OB), expected behaviour (EB), and steps to reproduce (S2R) from a bug report (a.k.a., query), and use the observed behaviour as a reduced version of the query for localizing bugs. 
\citet{finelocator} identify the neighbours of a given method in a codebase considering semantic similarity, temporal proximity, and call dependency, and expand the method's body with the neighbouring methods to improve feature location. 
\citet{sisman} also leverage spatial code proximity to reformulate their query for bug localization.
\citet{interactive-qr-word-relations} expand a search query leveraging several object-oriented relations (e.g., inheritance, implementation) from the source code. \citet{manq-qr} also apply several heuristics as fitness functions in their GA-based query reformulations to locate buggy code. 

%\citet{querythecode} pre-process a free-form NL query, extract the candidate keywords, and map them to the most likely API classes from the project source for the code search. \citet{precise-scalable-icpc} accept a code example as query, 
%transform the code example into AST and corresponding XPath query, and then assist the developers in code element search and program comprehension. As an alternative to query expansion, a few studies augment the code with API documentations \cite{sniff} and stereotype comments \cite{stereo-code} for improving code search. 
%\vspace{-.1cm}
\begin{frshaded}
	\noindent
	About \textbf{37\%} (26/70) of our primary studies use lightweight heuristics and ad-hoc metrics to reformulate their queries during code search. These heuristics and metrics often offer low cost alternatives to the established but costly approaches (e.g., thesaurus-based reformulation). However, they might not always be effective and thus need to be complemented with more sophisticated query reformulation techniques (e.g., term weighting).
	%the established methods which are more effective but costly. Thus, these heuristics are generally used in the preprocessing step to narrow down the overall search space and to identify the optimal query for code search with less efforts.
\end{frshaded}

\begin{figure}[!t]
	\centering
	\includegraphics[width=2.5in]{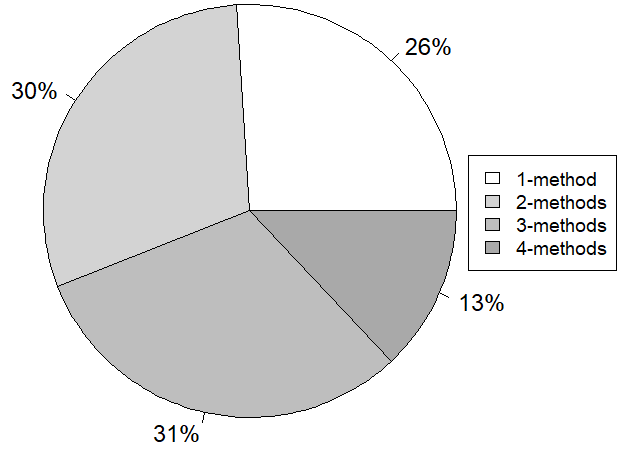}
	\vspace{-.2cm}
	\caption{Primary studies adopting multiple methodologies in the query reformulation}
	\vspace{-.3cm}
	\label{fig:multiple-method}
\end{figure}

\begin{figure}[!t]
	\centering
	\includegraphics[width=4.2in]{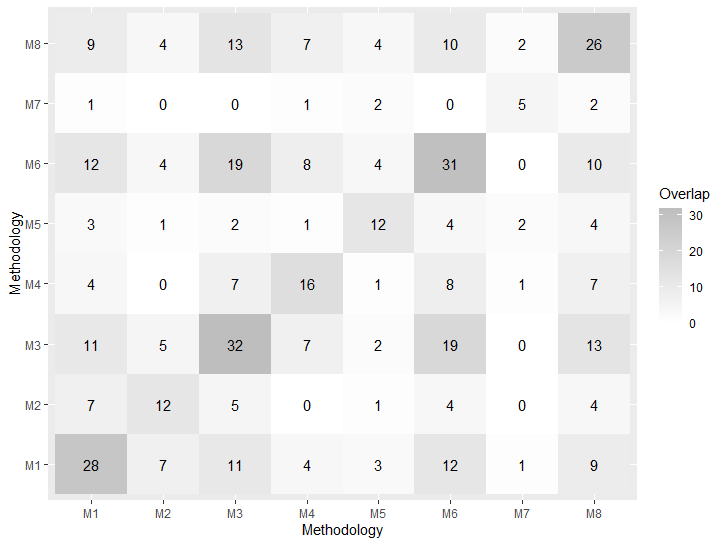}
	\vspace{-.1cm}
	\caption{Overlap of the methodologies used by the primary studies}
	\vspace{-.2cm}
	\label{fig:method-study-overlap}
\end{figure}

\begin{figure}[!t]
	\centering
	\includegraphics[width=4.2in]{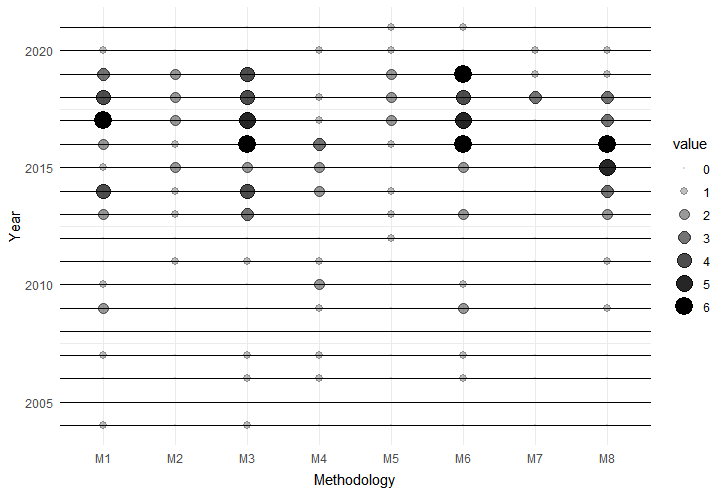}
	\vspace{-.1cm}
	\caption{Adoption of methodologies by the primary studies over the years}
	\vspace{-.2cm}
	\label{fig:year-method-overlap}
\end{figure}

\begin{figure}[!t]
	\centering
	\includegraphics[width=2.5in]{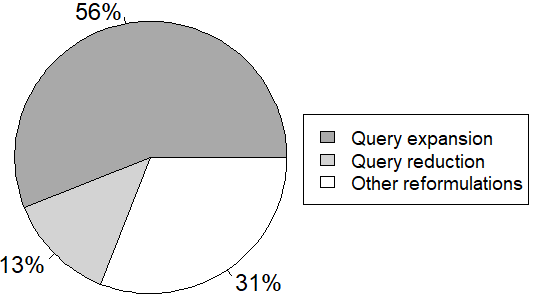}
	%\vspace{-.2cm}
	\caption{Primary studies performing different types of query reformulations}
	\vspace{-.2cm}
	\label{fig:study-qr-type}
\end{figure}

\begin{figure}[!t]
	\centering
	\includegraphics[width=2.8in]{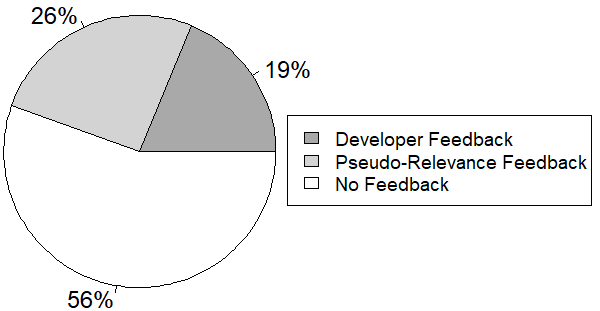}
	\vspace{-.2cm}
	\caption{Primary studies using different relevance feedback mechanisms in their query reformulations}
	\vspace{-.2cm}
	\label{fig:study-prf-type}
\end{figure}

We also provide the statistical summary of our study methodologies, their overlap, and several other items such as reformulation types, relevance feedback, and working contexts. Fig. \ref{fig:multiple-method} shows how one or more methodologies are used by the primary studies. We see that 26\% of the studies use only one of the eight methodologies (Table \ref{table:methodology}). The rest 74\% (52/70) of them use two or more methodologies to reformulate their queries during code search. That is, one methodology might not be sufficient for the majority of primary studies.
We also see that 13\% of the studies combine four different methodologies to reformulate their queries during code search.
Fig. \ref{fig:method-study-overlap} further demonstrates how the eight methodologies are combined together across the primary studies.
We see that three methodologies -- use of term weights and relevance feedback (M1),  extraction of semantic relations from co-occurrences and thesaurus (M3), andAPI recommendation using data mining (M6) -- are shared by 60\% (42/70) of the primary studies. Such a finding can be explained as follows. First, many term weighting methods consider such words as \emph{important} that frequently co-occur with query keywords in the relevant contexts such as search logs \cite{qr-log-mining}, change request \cite{saner2017masud,fse2018masud}, source code \cite{ase2017masud,sisman,sisman-jsep,shepherd,hillicse09}, stack traces \cite{fse2018masud}, and programming Q\&A websites \cite{ase2016masud,li2016}. Second, they might also capture such API classes to reformulate a query that frequently occur in the relevant contexts \cite{cexchange} and are semantically similar to the query \cite{iman-tse,icsme2018masud}. Third, the thesaurus-based approaches are connected to mining since
%frequently occurring API classes in the relevant contexts to reformulate a query \cite{icpc2019rodrigo,iman-tse} 
the construction of any software-specific thesaurus involves an extensive mining of software repositories  \cite{ccmapping,infer,swordnet,rack,icsme2018masud}. Fig. \ref{fig:year-method-overlap} further demonstrates the trends on adoption of various methodologies in query reformulations. We see that the use of thesauri (M3) and data mining (M6) was prevalent in search query reformulation, which might be inspired by the availability of well-established thesauri (e.g., WordNet) and open-source projects (e.g., GitHub). The use of term frequency and relevance feedback was also remarkable. The other methodologies were less frequent. It also should be noted that the research activities on search query reformulations has increased significantly since 2013.

Fig. \ref{fig:study-qr-type} shows how different types of query reformulations are used by our primary studies. We see that 56\% of studies expand their queries by adding appropriate keywords whereas 13\% reduce their queries by discarding the noisy keywords. On the other hand, 31\% of our primary studies adopt a combination of multiple techniques to reformulate their search queries.

Fig. \ref{fig:study-prf-type} shows how different mechanisms to collect relevance feedback are used by the primary studies. We see that 19\% of studies capture explicit developer feedback on the queries for their reformulations. Unfortunately, capturing feedback from the developers could be costly and sometimes infeasible. Thus, about 26\% of studies reformulate their queries using pseudo-relevance feedback without directly involving the developers. On the other hand, the remaining 56\% of studies do not make use of any relevance feedback mechanism to reformulate their queries.

\begin{figure}[!t]
	\centering
	\includegraphics[width=3in]{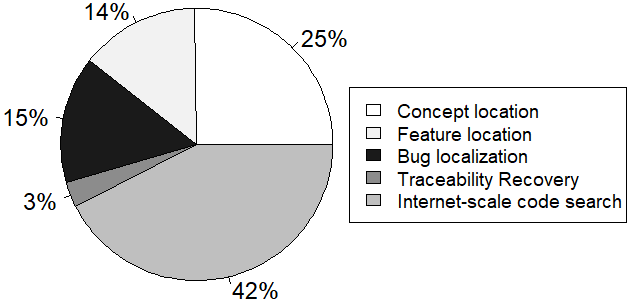}
	\vspace{-.2cm}
	\caption{Query reformulations used to search code in different working contexts}
	\vspace{-.4cm}
	\label{fig:study-wc-type}
\end{figure}

Fig. \ref{fig:study-wc-type} shows how the primary studies reformulate their queries to improve code search in various problem contexts.
For example, 25\% of studies reformulate their queries to search various software concepts implemented in the source code. Software features are a type of concepts that can be experienced and exercised by software users. About 14\% of studies reformulate their queries to locate features in the source code. Software bugs are the erroneous features that need to be removed from a software system. We see that 15\% of our primary studies reformulate their queries to find bugs in the source code. On the other hand, 42\% of the primary studies reformulate their queries to find relevant, reusable code segments from the Internet-scale code repositories such as GitHub, SourceForge, and programming Q\&A sites such as Stack Overflow. 

We also further investigate the above dimensions and report interesting patterns.
Table \ref{table:context-method-overlap} shows frequently used methodologies for query reformulations in various code search contexts.
We see that the local code searches such as concept location and feature location are more likely to 
%capture relevance feedback on the queries and  
use term weighting algorithms to expand their search queries. On the other hand, Internet-scale code search techniques are more likely to use various data mining approaches, semantic relations, and popular thesauri (e.g., WordNet) to expand their queries.
We also see that the studies focus more on query expansion rather than any other form of reformulation.

\begin{frshaded}
	\noindent
	\textbf{Summary for RQ$\mathbf{_1}$:} The primary studies on search query reformulations use eight major methodologies including term weighting and relevance feedback (40\%), semantic relations, term co-occurrences, and thesaurus (46\%), machine learning and query difficulty analysis (17\%), and data mining and API recommendation (44\%). The majority of these studies (56\%) expand their queries and almost 
	half of them capture relevance feedback on their queries. Furthermore, 57\% of studies help find relevant software entities (e.g., software bugs, features) whereas the remaining studies help find relevant, reusable code examples form the Internet-scale code repositories (e.g., GitHub).
\end{frshaded} 

\begin{figure}[!t]
	\centering
	\includegraphics[width=3.3in]{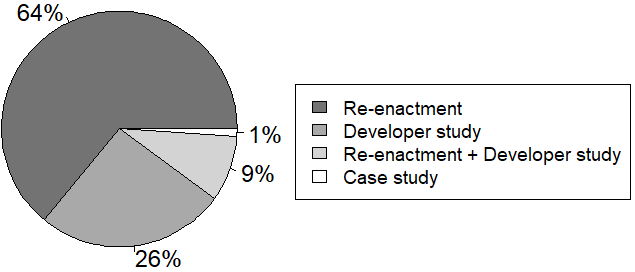}
	\vspace{-.2cm}
	\caption{Types of evaluation conducted by the primary studies}
	\vspace{-.4cm}
	\label{fig:empirical-developer-both}
\end{figure}

\begin{table}[!t]
	\centering
	\caption{Frequently used methodologies for query reformulation in various code searches}\label{table:context-method-overlap}
	\vspace{-.3cm}
	\resizebox{5.4in}{!}{%
		\begin{threeparttable}
			\begin{tabular}{l|p{3in}|c|l}
				\hline
				%\rowcolor{LightGray}
				\textbf{Code search} & \textbf{Frequently used methodologies} & \textbf{\#Study} & \textbf{Reformulation} \\
				\hline
				\hline
				\multirow{3}{*}{} & M1: Use of term weights and relevance feedback & 10  & \multirow{3}{*}{} \\
				Concept location &M4: Use of conceptual and syntactic relations & 7 & Query expansion \\
				&M3: Extraction of semantic relations from co-occurrences and thesauri & 6 & \\
				\hline
				\multirow{3}{*}{} &M1: Use of term weights and relevance feedback & 6 & \multirow{3}{*}{} \\
				Feature location &M3: Extraction of semantic relations from co-occurrences and thesauri  & 4 &  Expansion + reduction \\
				&M7: Genetic algorithms & 2 & \\
				\hline
				\multirow{3}{*}{} &M8: Miscellaneous  & 7  & \multirow{3}{*}{} \\
				Bug localization &M2: Mining of dependency graphs  & 4  & Query expansion \\
				&M5: Machine learning & 3 & \\
				\hline
			   \multirow{3}{*}{} & M6: API recommendation using data mining & 21  & \multirow{3}{*}{} \\
				Internet-scale code search &M3: Extraction of semantic relations from co-occurrences and thesauri & 19 & Query expansion \\
				& M8: Miscellaneous & 13 & \\
				\hline
			\end{tabular}	
		\end{threeparttable}
		\vspace{-.3cm}
	}
\end{table}

\subsection{Answering RQ$\mathbf{_2}$: Evaluation \& validation}\label{sec:validation}
%Although automated approaches for query reformulation are highly warranted for effective code search, they need to be evaluated and validated extensively before putting them into practice. Auto-generated entities (\eg\ queries, comments) often contain unexpected noise which not only hurts their overall quality but also prevents their adoption by the developer community. 
Our primary studies use various methods to evaluate and validate their performance and to place their techniques in the literature.
In particular, they use several evaluation methods, performance metrics, subject systems, query collections, ground truth, and validation targets. We discuss each of these evaluation dimensions with relevant statistics as follows.

\textbf{(a) Evaluation methods:} Search queries can be evaluated either through execution or without execution.
Based on the presence of query execution, evaluation methods can be divided into two categories--
%Two different methods are often used to evaluate auto-generated search queries--
(1) \emph{pre-retrieval} and (2) \emph{post-retrieval} \cite{refoqus}. 
Pre-retrieval methods evaluate a search query based on its linguistic and lexical aspects (e.g., coherence, specificity) without 
executing the query \cite{qsurvey}. Since these aspects of a query can be captured during 
the indexing of a corpus, the pre-retrieval evaluation methods are lightweight and cost-effective. 
Unfortunately, they are less reliable as they do not analyze the results retrieved by a query \cite{qeffect}.
On the contrary, post-retrieval evaluation methods are more reliable but costly. They execute a query and compare its results with the ground truth to evaluate the performance \cite{qsurvey,refoqus}.
%execute a given query and then analyse the retrieved results against the ground truth. Thus, these methods are costly but reliable for evaluation  
All of our selected primary studies adopt post-retrieval methods for their evaluation and validation. A few studies \cite{qperf,qperf-mills-tosem} also use linguistic and textual metrics to separate high-quality search queries from poorly designed queries.

Construction of ground truth can be another dimension to classify the evaluation methods. Based on this dimension, query evaluation methods can be divided into two categories-- (1) \emph{re-enactment based} and (2) \emph{developer study}. In case of re-enactment based evaluation, ground truth information is captured from historical data or existing benchmarks \cite{refoqus}. For example, to assess a search query from a bug report, the changed files from corresponding bug-fix commit are considered as the ground truth \cite{sisman,fse2018masud,observed}. 
%We call this evaluation as empirical evaluation throughout the rest of article. 
On the other hand, in the case of developer study, the participants are responsible for constructing the ground truth \cite{iman-tse} and for determining relevance of the retrieved results by a query \cite{codehow,active-cs,auto-query,lemos-type-qr,qr-log-mining}. From Fig. \ref{fig:empirical-developer-both}, we see that 64\% of our primary studies evaluate their queries through re-enactment whereas 26\% of them involve human participants for query evaluation. Interestingly, 9\% of the primary studies use both types of evaluation and 1\% of them employ case studies for their evaluation. Although the re-enactment based method uses solid ground truth, it might not always reflect a realistic evaluation scenario since the stakeholders (e.g., developers) are not directly involved.
On the contrary, involving human developers could also introduce subjective bias if appropriate mitigation strategy is not in place (e.g., inter-rater agreement analysis, polling \cite{iman-tse}). Besides, involving a large number of human participants could be costly. Thus, a combination of re-enactment and developer study is an ideal choice for effective evaluation. However, only 9\% of the primary studies adopt such a combination in their query evaluation.

\begin{figure}[!t]
	\centering
	\includegraphics[width=3.2in]{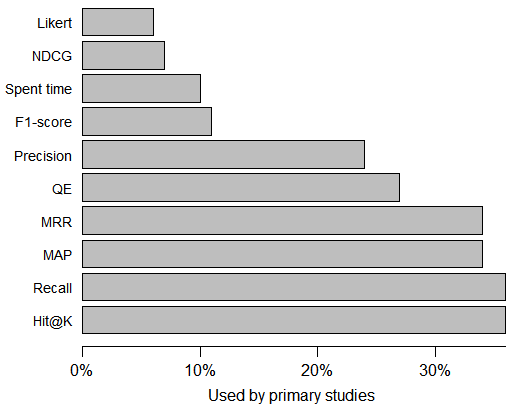}
	\vspace{-.2cm}
	\caption{Top-10 most frequently used performance metrics by the primary studies}
	\vspace{-.4cm}
	\label{fig:popular-performance-metrics}
\end{figure}

\begin{figure}[!t]
	\centering
	\includegraphics[width=4.2in]{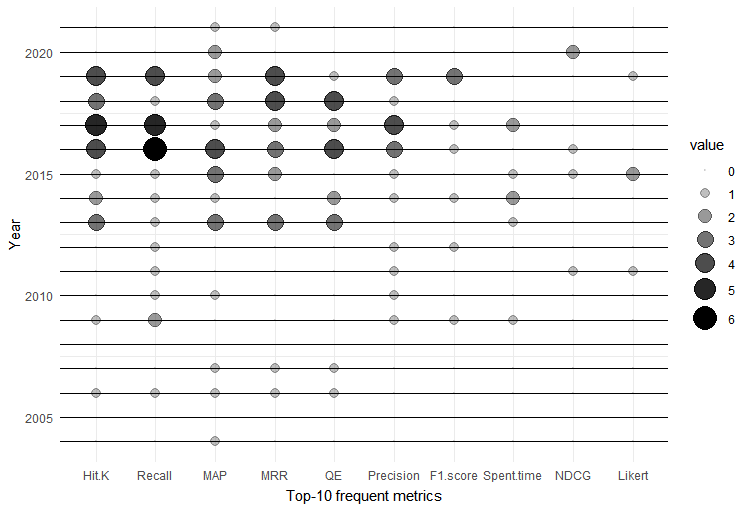}
	\vspace{-.1cm}
	\caption{Use of top-10 most frequently used performance metrics by primary studies over the years}
	\vspace{-.2cm}
	\label{fig:year-pop-metric-freq}
\end{figure}

\begin{figure}[!t]
	\centering
	\includegraphics[width=4in]{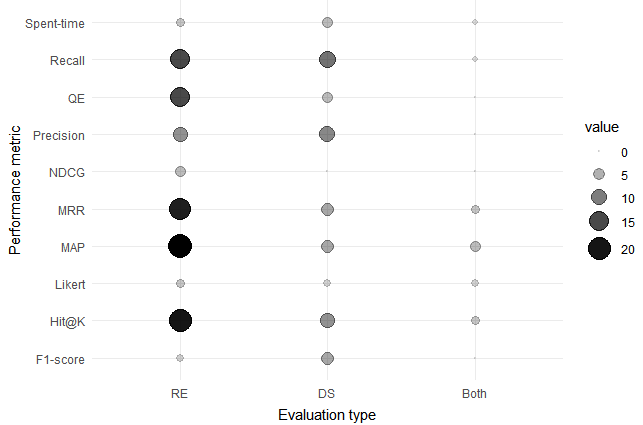}
	\vspace{-.1cm}
	\caption{Use of top-10 most frequently used performance metrics in different evaluation methods. (\textbf{RE}=Re-enactment, and \textbf{DS}=Developer Study)}
	\vspace{-.2cm}
	\label{fig:pop-metric-eval-type}
\end{figure}

\begin{figure}[!t]
	\centering
	\includegraphics[width=4.4in]{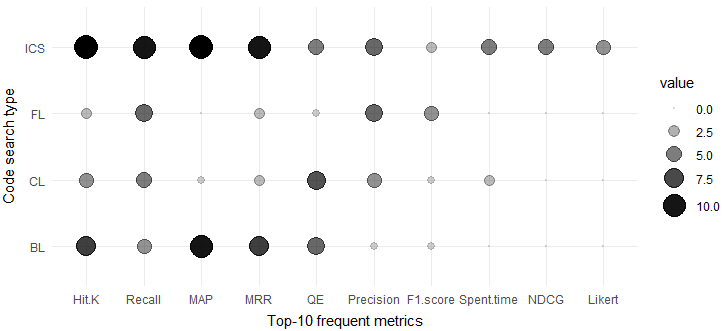}
	\vspace{-.1cm}
	\caption{Use of top-10 most frequently used performance metrics in different code searches over the years. (\textbf{CL}=Concept Location, \textbf{FL}=Feature Location, \textbf{BL}=Bug Localization, and \textbf{ICS}=Internet-scale Code Search)}
	\vspace{-.2cm}
	\label{fig:context-metric-overlap}
\end{figure}

\begin{figure}[!t]
	\centering
	\includegraphics[width=4.2in]{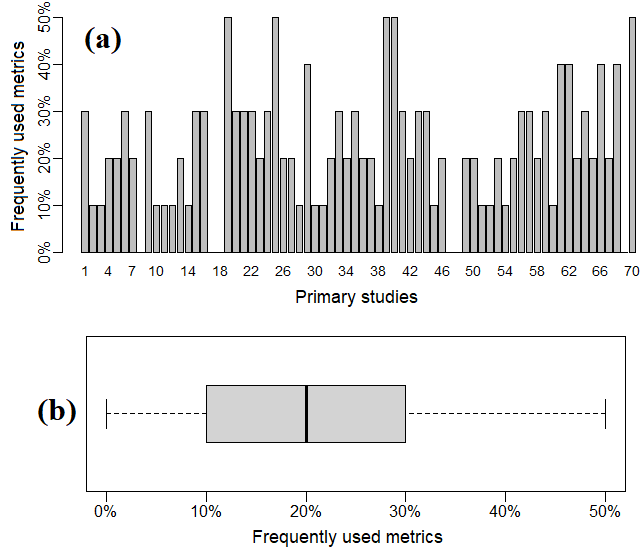}
	\vspace{-.2cm}
	\caption{(a) Percentage of top-10 most frequently used metrics by each primary study, and (b) Box plot of the percentage measures}
	\vspace{-.4cm}
	\label{fig:popular-metrics-used-by-studies}
\end{figure}

\textbf{(b) Performance metrics:} Each of our primary studies uses one or more performance metrics to evaluate 
%selected studies employs several performance metrics for the evaluation and validation of 
their reformulated queries. Since the majority of these studies borrow algorithms or techniques from several other research domains such as 
Information Retrieval, Machine Learning, and Recommendation Systems, the performance metrics are also taken from these domains. 
%An overview of the popular metrics can be found in Table \ref{table:performance-metrics}. 
%shows an overview of the popular, frequent metrics used to evaluate the query reformulation techniques. 
We discuss these metrics and their adoption by the primary studies as follows.

Fig. \ref{fig:popular-performance-metrics} shows the Top-10 performance metrics that are frequently adopted by our primary studies on automated query reformulation. An overview of these 10 metrics can be found in Appendix \ref{appendix:C}.
We see that Hit@K, recall, MAP, and MRR are used by more than 30\% of the studies. Hit@K determines whether there exists any ground truth within the top K results whereas recall determines the percentage of ground truth present in the top K results. 
On the other hand, both MAP and MRR emphasize on the position of ground truth within the retrieved results. Another performance metric that estimates a developer's effort to find the first ground truth within the retrieved results (a.k.a., QE), is used by 27\% of the primary studies.
There also exist other metrics such as NDCG and Likert that are used by less than 10\% of the studies.

We also investigate how the top 10 performance metrics have been used during the last 15+ years.
From Fig. \ref{fig:year-pop-metric-freq}, we see that most of these metrics such as Hit@K, MAP, MRR, and QE enjoyed increased adoption since 2013. Our investigation in Fig \ref{fig:pop-metric-eval-type} suggests that such an adoption might be connected to re-enactment based evaluation, which has been adopted by 64\% of the primary studies (Fig. \ref{fig:empirical-developer-both}). We also further investigate how the primary studies adopt these metrics in different working contexts. From Fig. \ref{fig:context-metric-overlap}, we note that Hit@K, Recall, MAP, and MRR were frequently used by the primary studies across different working contexts. While each of them was popular in Internet-scale code search, Hit@K, MAP, and MRR were found especially popular in bug localization and concept location. It also should be noted that QE was found more popular 
in bug localization and concept location than in Internet-scale code search. That is, detection of the first relevant element is very important in the local code searches (e.g., bug localization, concept location), which can help identify other relevant elements from a corpus through their static dependencies \cite{refoqus}.

We also further analyze how each study uses these top 10 metrics. Fig. \ref{fig:popular-metrics-used-by-studies}
 summarizes our analysis. We not only see the metric adoption ratio for each of the studies but also their distribution. According to Fig. \ref{fig:popular-metrics-used-by-studies}-(b), at least half of the studies select two metrics whereas 25\% of them use three or more metrics. On the other hand, at least 75\% of the studies use at least one of the 10 metrics discussed above. 
We also found that at least 5\% of the primary studies adopt five metrics to evaluate and validate their reformulated queries.
All these findings and statistics can serve as an important guideline for choosing appropriate performance metrics in future work.

%serve as an important guideline for the evaluation 

% The use of these appropriate and popular performance metrics is essential to mitigate the threats to construct validity of experimental findings \cite{wordsim}.

%Table \ref{table:study-eva-metrics} shows the evaluation method and performance metrics adopted by each of the primary studies. We see that 50\% (28/56) of the studies employ re-enactment based evaluation whereas $\approx$43\% (24/56) studies involve human developers in their evaluation process. Only 7\% (4/56) studies adopt both the evaluation methods. In the case of performance metrics, Mean Recall (MR), Hit@K, Query Effectiveness (QE) and Mean Average Precision (MAP) are the most popular metrics among the primary studies. About 30\%--38\% of the studies use them to evaluate their performance. Mean Precision (MP) and Mean Reciprocal Rank (MRR) are also used by 25\% of our selected studies. Estimation of time or effort spent (ETS) is another widely used metric in the qualitative evaluation, which was adopted by 18\% of the studies. The remaining metrics (\eg\ LS, NDCG) are used by about 5\% of the primary studies. Fig. \ref{fig:eval-dimensions} shows the box plot on the frequency of performance metrics employed by our primary studies. We see that about 50\% studies use at least \emph{two} metrics and 25\% studies use \emph{three} or more metrics for evaluating their performance. About 5\% of the studies also employ five or more metrics for their evaluation and validation.         

\begin{table}[!t]
	\centering
	\caption{Frequently used subject systems for experiment in various code searches}\label{table:context-subject-system-overlap}
	\vspace{-.3cm}
	\resizebox{5.5in}{!}{%
		\begin{threeparttable}
			\begin{tabular}{l|p{4in}}
				\hline
				%\rowcolor{LightGray}
				\textbf{Code search} & \textbf{Frequently used systems} \\
				\hline
				\hline
				Concept location & Adempiere (4), 
				JEdit (4), 
				ECF (3), 
				FileZilla (3), 
				eclipse.jdt.core (3), 
				WinMerge (3), 
				aTunes (3), 
				eclipse.pde.ui (3), 
				Rhino (2), 
				eclipse.jdt.ui (2), 
				Log4j (2), 
				Tomcat70 (2), 
				iReport (2), 
				eclipse.jdt.debug (2), 
				Sando (2), 
				Jajuk (2), 
				Sling (2), 
				JavaHMO (2), 
				Eclipse (2), 
				JBidwatcher (2)\\
				\hline
				Feature location & CAF (4), JEdit (2) \\
				\hline
				Bug localization & AspectJ (4), 
				Lucene (3), 
				ECF (3), 
				Pig (3), 
				eclipse.jdt.ui (3), 
				Solr (3), 
				Derby (3), 
				eclipse.jdt.core (3), 
				Tomcat70 (3), 
				OpenJPA (3), 
				ZooKeeper (3), 
				BookKeeper (3), 
				eclipse.jdt.debug (3), 
				Mahout (3), 
				eclipse.pde.ui (3), 
				Tika (3), 
				ZXing (2), 
				JodaTime (2), 
				SWT (2), 
				Chrome (2), 
				Eclipse (2) \\
				\hline
				Internet-scale code search & GitHub (9), 
				Stack Overflow (8), 
				SourceForge (4), 
				OSChina (2), 
				Java2s (2), 
				Javadb (2), 
				KodeJava (2), 
				OpenHub (2), 
				Google (2),  
				JEdit (2)\\
				\hline
			\end{tabular}	
		\end{threeparttable}
		\vspace{-.4cm}
	}
\end{table}

\begin{figure}[!t]
	\centering
	\includegraphics[width=4in]{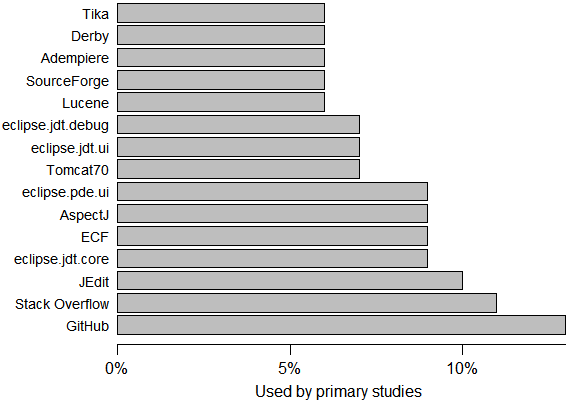}
	\vspace{-.2cm}
	\caption{Top-15 frequently used subject systems by the primary studies}
	\vspace{-.4cm}
	\label{fig:popular-subject-system}
\end{figure}

\textbf{(c) Subject systems:} Each of our primary studies uses one or more subject systems to evaluate their reformulated queries. Table \ref{table:context-subject-system-overlap} shows the popular subject systems used in four different working contexts.
In case of concept location, we see that eight systems including \texttt{Adempiere}, \texttt{JEdit}, \texttt{ECF}, and \texttt{FileZilla} are selected by at least three primary studies. However, a total of 20 subject systems were used by at least two primary studies to evaluate their reformulated queries and to locate their desired concepts in the software code. In case of feature location, only two systems (\texttt{CAF} and \texttt{JEdit}) are selected by two or more primary studies where \texttt{JEdit} is common. On the other hand, in the case of bug localization, we see that 20+ systems are selected by at least two or more primary studies on average. It should be noted that the primary studies from concept location, feature location, and bug localization share several subject systems (e.g., \texttt{ECF}, \texttt{JEdit}, \texttt{Eclipse}). This might be partially explained by their methodological overlap, as demonstrated in Table \ref{table:context-method-overlap}.
On the contrary, the primary studies from Internet-scale code search mostly use large-scale open-source software platforms (e.g., GitHub, SourceForge, OpenHub, OSChina) and popular programming Q\&A websites (e.g., Stack Overflow, KodeJava, Java2s) for their evaluation and validation. That is, they use hundreds if not thousands of subject systems from these platforms in their experiment.

Fig. \ref{fig:popular-subject-system} further demonstrates the Top-15 popular subject systems (or software platforms) used by our primary studies across different working contexts. We see that \texttt{GitHub} and \texttt{Stack Overflow} are the most popular choices due to their large collection of open-source projects and millions of question-answering threads containing code examples. \texttt{JEdit}, a Java-based IDE, has also been a popular subject system among the primary studies from multiple code search contexts. Besides, several other systems such as \texttt{eclipse.jdt.core},  \texttt{ECF}, and \texttt{AspectJ} have been used by $\approx$10\% of the primary studies. The use of diverse subject systems is essential to mitigate the threats to external validity of the experimental findings \cite{wordsim}.

%The majority of primary studies \cite{gayg,shepherd,sisman,sisman-jsep,refoqus,saner2017masud} adopt one or more subject systems to evaluate their query reformulation performance in the context of software changes (\eg\ bug localization, concept location). These systems generally span a limited number of application domains (\eg\ software development, multimedia, networking). Almost all of them are taken from the open source software domain except a few (\eg\ BSH, CAF) \cite{trace-query-transformation,flt-model-qr}. 
%Existing studies on Internet-scale code search \cite{testcase-cs-ist,iman-tse,lemos-type-qr} use extra-large software repositories (\eg\ GitHub, SourceForge, Google Code) containing thousands of open source systems for evaluating their algorithms. These systems span across numerous application domains, development platforms and programming languages. They also contain millions of LOC written by thousands of software developers. Thus, they pose as a great challenge during the retrieval of relevant code segments.

\begin{figure}[!t]
	\centering
	\includegraphics[width=3.6in]{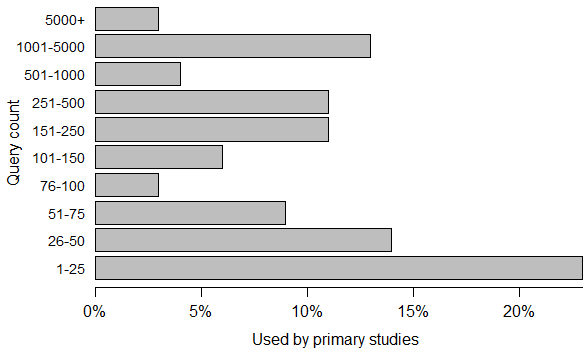}
	\vspace{-.2cm}
	\caption{Number of search queries used by primary studies}
	\vspace{-.2cm}
	\label{fig:query-count-stats}
\end{figure}

\begin{table}[!t]
	\centering
	\caption{Search queries used for experiment in various code searches}\label{table:context-query-count}
	\vspace{-.3cm}
	\resizebox{4in}{!}{%
		\begin{threeparttable}
			\begin{tabular}{l|c|c|l}
				\hline
				%\rowcolor{LightGray}
				\textbf{Code search} & \textbf{Query count} & \textbf{\#Study} & \textbf{Frequent reformulation} \\
				\hline
				\hline
				\multirow{3}{*}{} & 1--25 & 5  & \multirow{3}{*}{} \\
				Concept location &1000--5000& 5 & Query expansion \\
				&25--50 & 2 & \\
				\hline
				\multirow{3}{*}{} & 1--25 & 3 & \multirow{3}{*}{} \\
				Feature location &25--50  & 2 &  Expansion + reduction \\
				&150--250 & 2 & \\
				\hline
				\multirow{3}{*}{} & 1000--5000  & 4  & \multirow{3}{*}{} \\
				Bug localization & 250--500  & 2  & Query expansion \\
				&500--1000 & 2 & \\
				\hline
				\multirow{3}{*}{} & 1--25 & 8  & \multirow{3}{*}{} \\
				Internet-scale code search &25--50 & 5 & Query expansion \\
				&50--75 & 5 & \\
				\hline
			\end{tabular}	
		\end{threeparttable}
		\vspace{-.5cm}
	}
\end{table}

\textbf{(d) Search queries and human participants:} Each of our primary studies selects a set of search queries to evaluate their query reformulation techniques. Ideally, all queries (to be reformulated) should be collected from human developers. However, involving humans is costly and thus, 
only a few studies \cite{refoqus,trace-query-transformation,gayg,collab-feature-location} collect actual queries from the developers.
 %for their evaluation. Although such queries are the ideal choice, collecting them could be costly in terms of time and efforts. Hence, 
 The majority of primary studies (1) use historical artifacts (\eg\ past bug reports) \cite{refoqus,sisman,saner2017masud}, 
(2) reuse queries from earlier datasets \cite{iman-tse,qr-log-mining}, or simply (3) construct their queries from secondary sources (\eg\ programming Q\&A forums, tutorial sites) \cite{rack,icsme2018masud}. These sources often deliver a good enough representative of actual search queries from human developers. Furthermore, they can provide a relatively large dataset at low cost, which is essential to conduct a large-scale empirical evaluation. 

The number of used queries is an important aspect of evaluation. We collect the number of queries used by each of our primary studies and summarize our analysis in Fig. \ref{fig:query-count-stats} and Table \ref{table:context-query-count}. From Fig. \ref{fig:query-count-stats}, we see that 23\% (16/70) of the studies use 25 or less queries whereas 14\% (10/70) of them select 26 to 50 queries for their evaluation. On the other hand, 16 studies select 150 to 500 search queries whereas 14 other studies evaluate their reformulation techniques using 500+ search queries. We further breakdown our analysis across different working contexts. From Table \ref{table:context-query-count}, we see that the primary studies from concept location and bug localization use the maximum number of queries, which might be partially explained by the availability of thousands of bug reports or change requests in the open-source projects. On the other hand, the number of used queries in the Internet-scale code search is generally small. Unlike that of above working contexts, these queries cannot be easily extracted from the historical artifacts
and are often collected from human developers, which might partially explain their small numbers.

\begin{figure}[!t]
	\centering
	\includegraphics[width=3.4in]{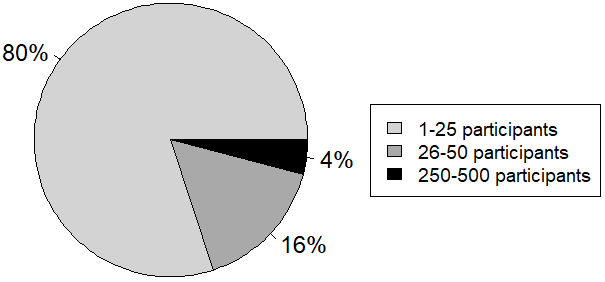}
	\vspace{-.2cm}
	\caption{Number of participants involved by primary studies in their evaluation}
	\vspace{-.1cm}
	\label{fig:participant-count-stats}
\end{figure}

Human participation is an important aspect of search query evaluation. We found that 34\% (24/70) of our primary studies involve human participants in their evaluation. Fig. \ref{fig:participant-count-stats} shows the statistics on human participation in query evaluation. We see that the majority of these studies (80\%) involve 25 or less participants whereas 16\% involves up to 50 developers. On the other hand, only one primary study involves 250+ human participants in a large-scale developer survey. Furthermore, we found that 58\% (14/24) of the human participation is associated to the primary studies from Internet-scale code search. However, the number of developers involved in each study is small on average (i.e., 1--25). 
%human participation as opposed to 25\% and 33\% for 
%60\% of the human participation for their evaluation. 
Involving human participants is always a great idea to complement an empirical evaluation. However, this involvement is often costly, which possibly explains the small number of participants recruited by the majority of our primary studies.

\begin{table}[!t]
	\centering
	\caption{Frequently used validation targets for comparison in various code searches}\label{table:context-validation-target}
	\vspace{-.3cm}
	\resizebox{5.4in}{!}{%
		\begin{threeparttable}
			\begin{tabular}{l|p{4in}}
				\hline
				%\rowcolor{LightGray}
				\textbf{Code search} & \textbf{Frequently used validation targets} \\
				\hline
				\hline
				Concept location & Rocchio's method \cite{rocchio} (4), 
				Baseline (4), 
				Pessimistic constant \cite{qperf} (2), 
				RSV \cite{rsv} (2), 
				Optimistic constant \cite{qperf} (2), 
				Random classification \cite{qperf} (2), 
				\citet{shepherd} (2), 
				Query reduction \cite{refoqus} (1), 
				Eclipse IDE-based search (1), 
				Dice expansion \cite{refoqus} (1), 
				Conquer \cite{hilltool} (1),
				\citet{sisman} (1), 
				Manual search (1), 
				Text-based search engine (1), 
				GES \cite{shepherd} (1), 
				FLAT3 \cite{flat3} (1), 
				Logistic regression \cite{logistic-reg} (1), 
				Refoqus \cite{refoqus} (1), 
				\citet{kevic} (1), 
				GREP (1), 
				ELex (1), and
				\citet{hillicse09} (1) \\
				\hline
				Feature location & 
				Baseline (6),
				Rocchio's method \cite{rocchio} (2), 
				Dice expansion \cite{refoqus} (2), 
				RSV \cite{rsv} (2), 
				Query reduction \cite{refoqus} (2), 
				LDA \cite{lda-blei} (2), 
				Eclipse search (1), 
				LSI \cite{ir-salton-book} (1), 
				SWUM \cite{ehill} (1), 
				\citet{shepherd} (1), 
				Manual search (1), 
				VSM \cite{tfidf} (1),  and
				\citet{semantictool} (1) 
				 \\
				\hline
				Bug localization & 
				Baseline (6), 
				BugLocator \cite{buglocator} (4), 
				BLIZZARD \cite{fse2018masud} (2), 
				Rocchio's method \cite{rocchio} (1), 
				BRTracer \cite{brtracer} (1), 
				BLUiR \cite{saha} (1), 
				Lucene \cite{lucene} (1), 
				Lobster \cite{stacktrace} (1), 
				\citet{sisman} (1), 
				rVSM \cite{buglocator} (1), 
				Manual search (1), 
				BLIA \cite{blia} (1), 
				AmaLgam+ \cite{versionhistoryjsep} (2), 
				Relevance model \cite{relevance-model} (1), 
				MULAB \cite{multi-abstraction-vsm} (1), 
				Observed behaviour \cite{observed} (1), 
				\citet{versionhistory} (1), 
				Genetic algorithm (1), and
				STRICT \cite{saner2017masud} (1) 
				 \\
				\hline
				Internet-scale code search &  
				Google search (6), 
				GitHub search (5), 
				BM25 \cite{bm25} (5), 
				QECK \cite{tsc2016} (4), 
				TF-IDF \cite{tfidf} (4), 
				Stack Overflow search (3), 
				OpenHub search (3), 
				CodeHow \cite{codehow} (3), 
				Portfolio \cite{portfolio} (3), 
				Baidu search (2), 
				Baseline (2), 
				OSChina search (2), 
				Manual search (2), 
				RACK \cite{rack} (2), 
				VF (VSM + Item set mining) (2), 
				SourceForge search (2), 
				\citet{pwordnet} (2), 
				Rocchio's method \cite{rocchio} (1), 
				Ohloh search (1), 
				NQE \cite{neural-query-expansion} (1), 
				Dice expansion \cite{refoqus} (1), 
				SWordnet \cite{swordnet} (1),
				Wordnet \cite{wordnet} (1), 
				FIM \cite{qr-stackoverflow} (1), 
				HRED-qs \cite{qr-stackoverflow} (1), 
				GooglePS \cite{qr-stackoverflow} (1),
				word2vec \cite{word2vec} (1), 
				K-NN Linear (1), 
				\citet{feature} (1), 
				Codota search \cite{codota} (1), 
				Bing search (1), 
				Cocabu \cite{cocabu} (1), 
				BIKER \cite{biker} (1),
				\citet{iman-tse} (1),
				RSV \cite{rsv} (1), 
				NLP2API \cite{icsme2018masud} (1),
				KBCS \cite{lemos-type-qr} (1), 
				Jungoloid \cite{jungloid} (1), 
				NCS \cite{neural-query-expansion} (1), and
				LDA \cite{lda-blei} (1)  \\
				\hline
			\end{tabular}	
		\end{threeparttable}
		\vspace{-.3cm}
	}
\end{table}

\begin{figure}[!t]
	\centering
	\includegraphics[width=4.2in]{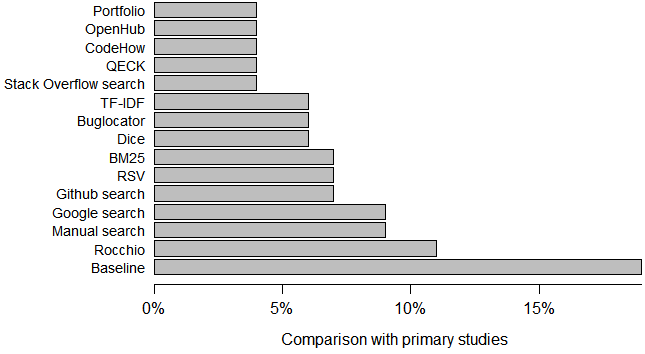}
	\vspace{-.2cm}
	\caption{Top-15 frequently used validation targets by primary studies}
	\vspace{-.4cm}
	\label{fig:popular-validation-target}
\end{figure}

\textbf{(e) Validation targets:} Comparison with closely related existing studies (a.k.a., validation targets) is an essential step to place any work in the literature. About 99\% of our primary studies validate their experimental findings by comparing with one or more existing alternatives. We collect the validation targets from each study and summarize our analysis in Table \ref{table:context-validation-target} and Fig. \ref{fig:popular-validation-target}. 

From Table \ref{table:context-validation-target}, we see that \emph{Baseline} and \emph{Rocchio's method} \cite{rocchio} have been frequently selected for comparison by the primary studies across three different search contexts -- \emph{concept location}, \emph{feature location}, and \emph{bug localization}. Baseline search query refers to the initial version of a query that 
%Baseline search queries are the initial versions that 
gets reformulated by the primary studies. On the other hand, Rocchio's method is a popular technique for query expansion borrowed from the Information Retrieval domain \cite{rocchio,qsurvey}. About 33\% of our primary studies compare their reformulated queries with that from Baseline and Rocchio's methods. Besides them, other techniques such as RSV, Dice expansion, and query reduction have also been frequently selected for comparison by the primary studies across concept location and feature location.
However, the primary studies that reformulate queries to localize software bugs compare with several bug localization techniques such as BugLocator \cite{buglocator}, AmaLgam+ \cite{versionhistoryjsep}, BLIZZARD \cite{fse2018masud}, BLUiR \cite{saha}, and \citet{sisman}. On the other hand, the primary studies from the Internet-scale code search frequently compare their approaches with several well-established general-purpose (e.g., Google search) or specialized search engines (e.g, GitHub search, Stack Overflow search)  and a few established search algorithms such as BM25 \cite{bm25} and TF-IDF \cite{tfidf}.
Furthermore, several primary studies such as QECK (S34), CodeHow (S15), and RACK (S39) were also chosen by the later studies to validate their findings.

%\small
%\input{subjects-query.tex}
%\normalsize

%Table \ref{table:system-n-query} shows the subject systems and the number of queries employed by each of the primary studies for their evaluation. We see that 90\%+  systems are taken from open source domain except a few (\eg\ BSH, CAF). \texttt{JEdit}, \texttt{Eclipse} and \texttt{eclipse.jdt.core} and \texttt{Adempiere} are the most popular subject systems among such studies
%that deal with local code searches (\eg\ concept location, bug localization).
%On the contrary, \texttt{GitHub}, \texttt{Stack Overflow} and \texttt{SourceForge} are the most frequently used Internet-scale
%software/code repositories in evaluating the query reformulations for  
%by the literature on query reformulation targeting 
%Internet-scale code search. 
%Fig. \ref{fig:eval-dimensions} shows box plots on the frequency of subject systems and search queries used by the primary studies. We see that each primary study employs two systems on average while 15\% of the studies make use of five or more systems for evaluation. About 50\% of the studies use more than 74 queries whereas 25\% use 300+ queries for their evaluation and validation.    

Fig. \ref{fig:popular-validation-target} further demonstrates the top 15 validation targets selected by the primary studies across different search contexts. We see that reformulated queries are often compared with their initial, non-reformulated versions (a.k.a., Baseline) to make sure that the reformulation is useful. About 18\% (13/70) of our primary studies perform such a comparison. 
Several traditional alternatives such as Google search, GitHub search, and Stack Overflow search have also been popular. Furthermore, six existing approaches including RSV, Dice, TF-IDF have also been used for comparison by 5\%+ of the primary studies on average. Comparison with appropriate existing alternatives is essential to claim the technical superiority of a proposed technique. Unfortunately, due to the unavailability of replication packages, such a comparison might not always be possible. For example, only $\approx$30\% of the primary studies were selected for comparison by the later studies, which might be partially explained by our findings on the replication packages in Table \ref{table:quality-ass}.

\begin{table}[!t]
	\centering
	\caption{Statistical tests conducted by primary studies}\label{table:statistical-test}
	\vspace{-.3cm}
	\resizebox{5.2in}{!}{%
		\begin{threeparttable}
			\begin{tabular}{l|p{1.4in}|p{3in}}
				\hline
				%\rowcolor{LightGray}
				\textbf{Category} & \textbf{Statistical test} & \textbf{Overview}\\
				\hline
				\hline
				& Mann-Whitney Wilcoxon & The test determines whether any two samples are drawn from the same distribution or not. It does not assume any specific distribution for its samples.\\
				\hhline{~--}
				& Wilcoxon Signed Rank & The test determines if two related, dependent samples are significantly different or not from each other. Both samples should have equal number of elements. \\
				\hhline{~--}
				& Friedman & The test determines if three or more samples have identical population distributions or not. The data can be either quantitative or ordinal. \\
				\hhline{~--}
				& Kendall's $\tau$ & The test determines the strength and direction of association between any two measured quantities.\\
				\hhline{~--}
				Non-parametric & Quade & The test is an extension of Wilcoxon Signed Rank test for three or more related samples. \\
				\hhline{~--}
				& Shapiro-Wilk & The test determines whether a sample is drawn from a normal distribution or not. \\
				\hhline{~--}
				& Cliff's $\delta$ & The test quantifies the difference between any two samples beyond their \emph{p-value} interpretation. \\
				\hhline{~--}
				& Vargha and Delaney's $A$ & The test quantifies the difference between any two samples using a probabilistic measure. It has a linear relationship with Cliff's $\delta$.\\
				\hhline{~--}
				& Holm's correction & The test is used to counteract the problem of multiple comparisons and to control the family-wise error rate.\\
				\hline
				\hline
				& Paired t & The test determines whether two related, dependent samples are significantly different from each other or not. It also assumes t distribution for its samples. \\ 
				\hhline{~--}
				& ANOVA & The acronym stands for Analysis of Variance. The test determines whether three or more independent populations are significantly different from each other, with the assumption that their distributions are normal. \\
				\hhline{~--}
				Parametric & Chi-squared & The test determines the difference between any two categorical variables in the same population. It assumes 
				$\chi^2$ distribution for the population, and shows how the two variables might be connected to each other. \\
				\hhline{~--}
				& Cohen's $D$ & The test quantifies the difference between any two samples with an assumption of normality and homogeneity in their variance. It often accompanies t-test and ANOVA. \\
				\hhline{~--}
				& Asymptotic General Independence & The test determines whether any two random variables are asymptotically independent of each other. Here, asymptotic aspect refers to a very large sample size for each variable. \\
				\hline				
			\end{tabular}	
		\end{threeparttable}
		\vspace{-.6cm}
	}
\end{table}

\begin{figure}[!t]
	\centering
	\includegraphics[width=4.6in]{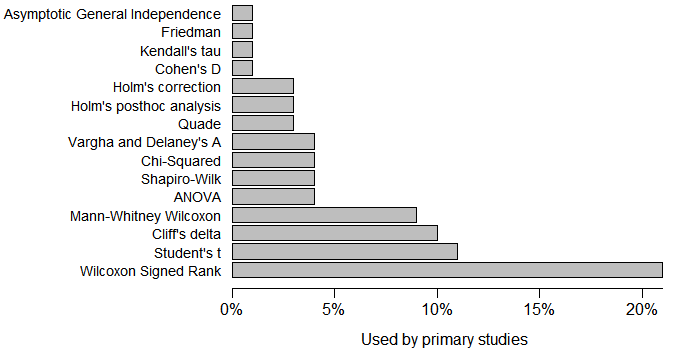}
	\vspace{-.2cm}
	\caption{Statistical tests used by primary studies}
	\vspace{-.2cm}
	\label{fig:statistical-test}
\end{figure}

\begin{figure}[!t]
	\centering
	\includegraphics[width=4in]{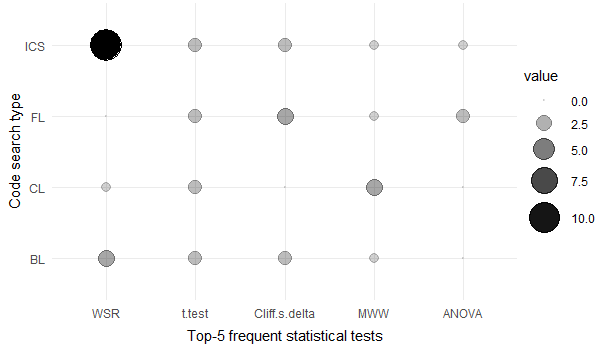}
	\vspace{-.1cm}
	\caption{Adoption of 5 frequently used tests by different search types}
	\vspace{-.1cm}
	\label{fig:wc-stat-test}
\end{figure}

%\small
%\input{developer-participant.tex}
%\normalsize   

\textbf{(f) Statistical tests:} The use of statistical tests is an essential part of testing any hypothesis and drawing any meaningful conclusions \cite{statistical-test}. About half of the primary studies (51\%, 36/70) conduct one or more statistical tests to draw their conclusions (e.g., superiority over the existing alternatives). Table \ref{table:statistical-test} provides an overview of the conducted tests and Figures \ref{fig:statistical-test}, \ref{fig:wc-stat-test} show their usage statistics.

From Fig. \ref{fig:statistical-test}, we see that Wilcoxon Signed Rank is the most frequently used statistical test. It has been adopted by 20\%+ (15/70) of our primary studies. It is a non-parametric paired test for two related, dependent samples, which makes it suitable for comparison between a proposed technique and a competing alternative. Student's t is another paired test that assumes normality for its samples. It has been used by 11\% of the primary studies. On the other hand, Cliff's $\delta$ is a non-parametric test to determine the effect size between any two samples, which is often accompanied by another test -- Mann-Whitney Wilcoxon. Both these tests are 
suitable for the samples with unknown distribution and are used by 8\%--10\% of our primary studies. Besides these four, 
there have been 11 other statistical tests (e.g., ANOVA, Shapiro-Wilk,  Chi-squared) each of which has been used 
%and each of  ANOVA, Shapiro-Wilk, Vargha and Delaney's A, and Chi-squared has been used
by $\approx$5\% of our 70 primary studies.

We also investigate how the top 5 frequent tests (according Fig. \ref{fig:statistical-test}) were used by the primary studies in different working contexts (e.g., bug localization, Internet-scale code search). From Fig. \ref{fig:wc-stat-test}, we see that Wilcoxon Signed Rank (WSR) test has been more popular in the Internet-scale code search than in the local code searches. On the other hand, t-test has been used in all four working contexts: bug localization, concept location, feature location, and Internet-scale code search. Both tests above involve paired comparison, which is often used to validate a proposed technique against existing baselines. It also should be noted that the primary studies from bug localization and feature location made use of at least four statistical tests whereas the studies from Internet-scale code search adopted all top 5 statistical tests in their experiments.

%Similar conclusion might be drawn for the effect size - cliff's delta.
%Fig. \ref{fig:eval-dimensions} shows box plots on the frequency of existing works replicated and developer participants engaged by our selected primary studies. We see that 50\% of these studies compare with at least two existing works and 25\% studies compare with three or more works from the literature. On the contrary, 47\% of the primary studies compare with one existing work only. From Table \ref{table:validation-participant}, we see that about 38\% of the primary studies involve human developers in their experiments.
%As shown in Fig. \ref{fig:eval-dimensions}, 50\% of these studies involve at least 16 software developers and 25\% of them involve 20+ developers in their evaluation/validation. A few studies \cite{multi-rec,cexchange} also deploy their tool on the web, evaluate their code search performance in the wild, and then collect feedback from hundreds of developers. Although this should be the ideal practice, very few studies adopt this validation approach due its high costs. It also should be noted that only 46\% of the primary studies conduct statistical significance tests (\eg\ Mann-Whitney Wilcoxon) during their comparisons (Table \ref{table:validation-participant}).

\begin{frshaded}
\noindent
\textbf{Summary of RQ$\mathbf{_2}$:} About \textbf{62\%} of our primary studies use re-enactment-based evaluation, 29\% of them involve human participants, and only 7\% of the studies combine both types of evaluation.
About 80\% of our studies that involve human participants involve less than 25 developers each.
We found that the use of performance metrics (e.g., Hit@K, MAP, MRR) might vary across evaluation types or working contexts.
%Almost half of our studies select at least two popular performance metrics each for their evaluation whereas
%30 different metrics have been used by all the studies.
%a total of 30 different metrics have been used by all studies.
%They also select more than 100 subject systems for their experiments. 
We also found that 70 primary studies selected more than 100 subject systems for their experiments. The studies from
%the studies from concept location and bug localization have used more queries than those from 
Internet-scale code search use a smaller number of queries than the others for their evaluation.
About 30\% of our primary studies were also selected for comparison by the later studies.
Our primary studies use a total of 15 statistical tests where Wilcoxon Signed Rank (WSR) test has been adopted by more than 20\% of the studies.
%They use \textbf{two} popular performance metrics on average. About \textbf{50\%} of the studies make use of at least \textbf{two} subject systems and at least \textbf{70} search queries for their evaluation. About \textbf{38}\% of the studies involve human developers in their evaluation process and about \textbf{53}\% studies compare with more than one existing work.     
\end{frshaded}
 
\begin{table}[!t]
	\centering
	\caption{Challenges and limitations of the primary studies}\label{table:challenge}
	\vspace{-.3cm}
	\resizebox{3.4in}{!}{%
		\begin{threeparttable}
			\begin{tabular}{l|l}
				\hline
				%\rowcolor{LightGray}
				\textbf{ID} & \textbf{Challenge \& limitation}\\
				\hline
				\hline
				CH1 & Injection of noisy keywords in the original query\\
				\hline
				CH2 & Extra cognitive burden on the developers \\
				\hline
				CH3 & Lack of generalizability \\
				\hline
				CH4 & Issues towards practical adoption \\
				\hline
				CH5 & Human bias and lack of rigour in evaluation \\
				\hline
				CH6 & Inappropriate use of external tools and dependencies \\
				\hline
				CH7 & Lack of a sound theory \\
				\hline
				CH8 & Miscellaneous \\
				\hline
			\end{tabular}	
		\end{threeparttable}
		\vspace{-.2cm}
	}
\end{table}

\begin{figure}[!t]
	\centering
	\includegraphics[width=4.2in]{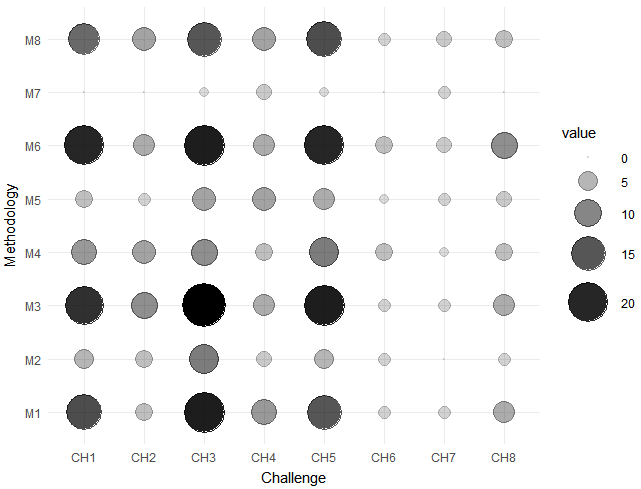}
	\vspace{-.1cm}
	\caption{Methodologies associated to different challenges encountered by the primary studies}
	\vspace{-.2cm}
	\label{fig:method-challenge-overlap}
\end{figure}

\subsection{Answering RQ$\mathbf{_3}$: Challenges and limitations in automated query reformulations for source code search}\label{sec:challenge}
Search queries often consist of one or more keywords that are unstructured. On the contrary, software code is full of structures (e.g., syntactic relations) and has an unrestricted vocabulary  \cite{devanbu-fse}. Besides, search queries and source code might be written by different persons using their own vocabularies that have a little chance of being the same \cite{vocaprob,semantictool}.
Due to these inherent differences between a query and the code, constructing the right search query is challenging.
Although each of our primary studies attempts to mitigate these challenges, they might not be entirely successful.
We critically examine the methodology, evaluation, validation, and threats to validity sections from each study, and summarize their reported or implied limitations. Similar to RQ$_1$, we use Grounded Theory approach (Section \ref{sec:gtheory}) to analyze the primary studies and detect their challenges and issues. Tables \ref{table:challenge} and \ref{table:study-challenge} show the key challenges and limitations of our primary studies. We also determine the association between methodologies and challenges by combining the findings from RQ$_1$ and RQ$_3$, and Fig. \ref{fig:method-challenge-overlap} summarizes our investigation. In this section, we discuss each of the key challenges and their mitigation strategies in detail as follows.     

%the underlying methodology and evaluation/validation approaches of each primary study, and record the  limitations and threats to validity either \textbf{reported} or \textbf{implied} by their \textbf{original authors}.
%and then determine the common challenges and limitations. 
%Similar to RQ$_1$, we select each of the primary studies and summarize their limitations using a few important natural language phrases. We then apply \emph{Grounded Theory} \cite{grounded-theory} on these phrases (Section \ref{sec:gtheory}), perform three types of coding--\emph{open coding}, \emph{axial coding} and \emph{selective coding}, and finally identify seven common key challenges. Table \ref{table:challenge} shows how each primary study suffers from each of these limitations/challenges. 

%As shown in Table \ref{table:challenges}, we focus on each of the challenges and discuss how they limit the existing studies 

\begin{table}[!t]
	\centering
	\caption{Primary studies, their challenges and limitations}\label{table:study-challenge}
	\vspace{-.3cm}
	\resizebox{4.6in}{!}{%
		\begin{threeparttable}
			\begin{tabular}{l|p{4.2in}| c}
				\hline
				%\rowcolor{LightGray}
				\textbf{ID} & \textbf{Primary studies} & \textbf{Total}\\
				\hline
				\hline
				CH1 & S1, S5, S7, S14, S15, S16, S19, S20, S23, S25, S26, S27, S28, S29, S35, S36, S38, S39, S40, S41, S42, S43, S44, S45, S46, S50, S51, S53, S54, S55, S57, S58, S59, S63, S64, S70 &  \textbf{36} (51\%)\\
				\hline
				CH2 & S3, S12, S13, S17, S24, S27, S28, S29, S30, S31, S33, S35, S43, S47, S48, S49, S53, S56 & 18 (26\%)\\
				\hline
				CH3 & S1, S2, S3, S4, S7, S8, S10, S11, S13, S14, S15, S16, S19, S22, S23, S24, S25, S27, S28, S30, S31, S32, S33, S34, S35, S36, S37, S39, S40, S41, S42, S43, S45, S46, S47, S49, S50, S51, S52, S54, S55, S56, S57, S59, S60, S64, S65, S66, S69 & \textbf{49} (70\%) \\
				\hline
				CH4 & S2, S3, S6, S22, S27, S29, S30, S32, S33, S34, S42, S44, S48, S50, S53, S54, S61, S62, S63, S64, S66, S67, S68 & 23 (33\%)\\
				\hline
				CH5 & S1, S4, S6, S11, S12, S13, S15, S16, S17, S18, S21, S22, S24, S26, S27, S28, S30, S31, S34, S35, S36, S37, S38, S39, S42, S43, S44, S45, S46, S47, S51, S52, S58, S60, S63, S64, S65, S66, S69 & \textbf{39} (56\%) \\
				\hline
				CH6 & S35, S37, S48, S51, S53, S56, S61 & 7 (10\%) \\
				\hline
				CH7 & S9, S14, S35, S44, S55, S62, S65 & 7 (10\%)\\
				\hline
				CH8 & S8, S14, S17, S26, S27, S32, S36, S38, S45, S49, S51, S55, S57, S61, S63 & 15 (21\%) \\
				\hline
			\end{tabular}	
		\end{threeparttable}
		%\vspace{-.4cm}
	}
\end{table}

\textbf{CH1--Injection of noisy keywords in the original query:} One of the major challenges of code search is the vocabulary mismatch between a query and the source code. In essence, each of our primary studies attempts to overcome this challenge. Unfortunately, almost half of them (51\%)  might fail due to their adoption of inappropriate methodologies and selection of noisy keywords.
For example, several studies \cite{api-book,codehow,concept-ase,qsynthesis,req-qr} rely on lexical similarity between a query and the API documentation to detect the relevant APIs and to expand the query.
However, textual similarity alone might not be sufficient to overcome the vocabulary mismatch problems \cite{vocaprob,vocaprob2}, and thus noisy keywords (e.g., false-positive API classes) 
 %such a similarity alone might not be sufficient and 
 %false-positive API classes (a.k.a., noise) 
 could be added to the original query \cite{qeffect,qperf}.
%enough since the relevant APIs could also be lexically dissimilar to the query. 
%might not be  lexically similar. 
%identify the relevant API classes (or methods) and to expand the NL query. 
%Thus, the subsequent steps of their code search approach are negatively affected by the noisy queries.
Many studies \cite{pwordnet,shepherd,multi-rec,lemos-type-qr,thesaurus-qr-idcs} adopt English language thesauri such as WordNet \cite{wordnet} to replace or expand a query with synonyms or semantically similar words. 
However, WordNet is based on natural language texts rather than source code. As a result, the keywords suggested by WordNet might not be appropriate for detecting relevant source code. An existing evidence \cite{semantictool} also suggests that the same word could have two different meanings when used in natural language texts and in source code respectively. In other words, the keywords from WordNet can introduce noise to a search query. Thus, the above studies on query reformulation might 
%Thus, WordNet-based query suggestion might not be effective for the source code search \cite{semantictool}. 
%In short, these above studies 
%might suffer from noisy queries and 
not be able to overcome the vocabulary mismatch problem during their code search.  

Several primary studies \cite{cocabu,icsme2018masud,tsc2016,cexchange,qecc-intent,ase2016masud,rack,iman-tse,twkraft-jsep} expand a search query using relevance feedback mechanisms (details in Section \ref{sec:background}) to mitigate the vocabulary mismatch problem. 
However, such an expansion might also hurt the initial query if the good keywords are not selected from the relevance feedback documents \cite{term-independence,query-drift}.
%by introducing noisy keywords if optimal set of keywords 
\citet{tsc2016} expand a query by collecting important keywords from relevant Q\&A threads of Stack Overflow using pseudo-relevance feedback. \citet{cocabu} extract important program elements (\eg\ method invocations) from such threads. Similarly, \citet{icsme2018masud} extract important API classes from the Q\&A threads of Stack Overflow to expand a query. All these studies apply various term weighting methods (\eg\ TF-IDF \cite{tfidf}) to selecting their keywords. Unfortunately, the term weighting method alone might not be able to discard the noisy keywords if the proxy of term importance is not appropriate \cite{emse2021masud}.
Several primary studies \cite{fse2018masud,kim-bl-qr,finelocator} use a bug report or a feature request as a query and collect candidate terms from relevant source code to reformulate the query. However, these approaches might also suffer from noisy keywords as the issue reports and the source code might have different vocabularies and the relevant code could not be retrieved \cite{vocaprob,verbose}.
%the bug and source code might different vocabularies \cite{vocaprob}
%are unable to capture appropriate candidate terms from the code.
 %are poor themselves. 
 \citet{cexchange} capture frequent identifier names from the source code retrieved by a given query and use them to expand the query. 
\citet{qecc-intent} mine frequently changed identifier names from the code change history to determine the intent of a search query.
%code change history and expand a search query with such identifier names that were frequently changed in the code apparently relevant to the query. 
\citet{twkraft-jsep} also expand a search query by repeating its structured terms (\eg\ identifier names) multiple times during feature location.    
All these studies above greatly rely on the identifier names to expand their queries.  However, the identifier names could be poor, generic, and less descriptive \cite{codes, lawrie-identifier}, which can hurt the original search queries.

A few primary studies \cite{iman-tse,icsme2018masud} learn word embeddings \cite{wembedding} from large corpora (e.g., Stack Overflow, GitHub). They attempt to mitigate the vocabulary mismatch problem by determining semantic relevance between a query and the source code. However, to learn word embeddings, 
 %They strongly assume that 
 both query keywords and code terms need to be present in the same corpus \cite{fasttext,word2vec,skip-gram}, which might not always hold true. \citet{rack} 
mine frequent keyword-API co-occurrences from Stack Overflow Q\&A threads and use the frequent API classes to expand a search query. However, their sole reliance on co-occurrence might also provide noisy keywords (e.g., false-positive API classes) that can hurt the original query during code search \cite{icsme2018masud}. 

From Fig. \ref{fig:method-challenge-overlap}, we note that the primary studies facing this challenge (i.e., CH1) frequently use three different methodologies: use of term weights and relevance feedback (M1), extraction of semantic relations from co-occurrences and thesauri (M3), and API recommendation using data mining (M6). 
While these methodologies are popular, they could be limited by their underlying assumptions, as discussed above. For example, frequency or co-occurrence might always not be the indication of a keyword's importance \cite{emse2021masud,qperf-mills-tosem}. API recommendation can also return false-positive API classes or methods due to the lack of appropriate contexts \cite{icsme2018masud}. Thus, these methodologies might inject noisy keywords into the original search query as a part of query reformulations. 
%On the other hand, primary studies not relying on these methodologies are less likely to introduce noisy keywords in their reformulated queries. 

\textit{Mitigation strategy.} Frequency or co-occurrence might not always be a reliable proxy to keyword importance, which could lead to the selection of noisy keywords. They could be carefully combined with other available proxies (e.g., ambiguity, parts of speech, position of a keyword) to avoid false-positives during keyword selection.

\textbf{CH2--Extra cognitive burden on the developers:} Several primary studies \cite{shepherd,multi-rec,hillicse09,cexchange,gayg,twkraft-jsep} impose extra cognitive burden on the developers during query reformulation. \citet{shepherd} first introduce an interactive approach namely \emph{Find-Concept} to support query reformulation in the context of concern location. Although their approach suggests relevant keywords from multiple sources (\eg\ Verb-DO pairs, synonyms), a developer is ultimately responsible to choose the right keywords and to construct the final query. \citet{hillicse09} extend this work with the phrasal concept analysis. \citet{multi-rec} further extend Find-Concept by recommending keywords before and after the code search.
%with multi-step recommendations of search keywords
 %(\eg\ pre-search recommendation, post-search recommendation). 
 All these studies above argue in favour of developer involvement rather than a complete automation in query construction.
 Thus, in essence, they warrant a certain level of expertise or experience from the developers and might not be suitable for novice developers.
 %significant level of skill/experience from the developers during query construction. Hence, novice developers might not find these approaches useful for query construction during the code search.
\citet{gayg} capture explicit feedback on a query from a developer and expand the query with the feedback to support concept location.
%the developers on a given query in the context of concept location tasks. They analyse such feedback and then reformulate the given query 
%Rocchio's method \cite{rocchio}. 
\citet{cexchange} implement a similar idea to improve queries for Internet-scale code search. 
Although the developer's feedback could be useful, capturing it regularly could be infeasible and costly. 
Several primary studies also warrant significant learning and cognitive efforts from the developers
such as generation of test cases \cite{testcase-cs-ist, lemos-type-qr}, input code examples \cite{query-by-example,precise-scalable-icpc,auto-query} and controlled search queries \cite{nlsupport,querythecode}. To use the approach of \citet{testcase-cs-ist}, a developer needs to express the information need as a complex test case. \citet{query-by-example} expect an input code example that can be transformed into an AST to retrieve the similar code examples. \citet{nlsupport} also expect a semi-structured query from the developers rather than a free-form query during code search.
\citet{multi-faceted-flt} provide multi-faceted information within IDE to support an interactive query reformulation during feature location. However, their technique warrants a certain level of domain expertise so that the developers can recognize the relevant program elements and can make important decisions during feature location \cite{multi-faceted-flt}. 

\textit{Mitigation strategy.} Given the significant learning curves of the above techniques, software practitioners might be reluctant to adopt them. Future work can focus on reducing the learning curves by capturing and leveraging the existing knowledge or expertise of the developers. For example, analysis of authored or reviewed code from the version control history might reveal the expertise of a developer, which can be leveraged to customize the above techniques.

%which has the potential to overwhelm the developers with information overload.
%could be overwhelming for the developers as well. 

\textbf{CH3--Lack of generalizability:} About 70\% (49/70) of our primary studies
suffer from a lack of generalizability due to their limited datasets, queries, and evaluation.
The majority of these studies \cite{time-aware-term-weighting,kevic,saner2017masud,fse2018masud, multi-faceted-flt,gayg} evaluate their approaches using only Java-based subject systems. According to an existing evidence \cite{bluirplus}, findings from Java-based systems might not always generalize for other systems such as C/C++ based systems.
%generalize for the systems from other programming platforms (\eg\ C/C++). 
Almost all primary studies also rely on open-source systems to conduct their experiments. 
Only a few studies \cite{flt-model-qr,trace-query-transformation,genqr-tse-perez} make use of proprietary dataset for their experiments. 
\citet{genqr-tse-perez} collect railway software systems from their industry partner (CAF) to evaluate their work. Other studies analyze proprietary search logs from Bing search engine \cite{swim} and Stack Overflow \cite{qr-stackoverflow} to generate their reformulated queries.
According to existing literature \cite{oss-css,rigbyOSSpeer}, open source and closed source systems could be different in various aspects (e.g., development practices, bug resolution, code reviews) and thus findings from these two types of systems might not always align with each other. 
%There exists a little evidence suggesting that the findings from open-source systems will generalize for the proprietary software systems.
%subject systems for their experiments. 
%Their findings could not be guaranteed to generalize for the closed source, industry applications \cite{flt-model-qr}.
%since most of the experiments are performed on 
%majority of the subject systems are taken from open source domain. 

Other studies \cite{gayg,refoqus,saner2017masud,fse2018masud,tsc2016,qperf} use a single code search engine (\eg\ Lucene)
to evaluate their reformulated queries.
However, the same query might not deliver the same performance across multiple search engines \cite{trconfig}. 
A number of studies \cite{shepherd,gayg,qr-log-mining,active-cs,kevic,lemos-type-qr,codehow,swordnet,query-by-example,qecc-intent} also make use of a limited set of queries (\eg\ $\le$50) for their evaluation.
A few studies \cite{kevic,qr-interactive-clarification} use only one subject system. Thus, their findings might 
suffer from the lack of generalizability as well. Several primary studies \cite{api-book,time-aware-term-weighting,qsynthesis,nlsupport} might not perform well for all types of queries due to their unique requirements for a given query (e.g., textual similarity to API classes \cite{api-book}, presence of nouns \cite{time-aware-term-weighting}, syntactic correctness \cite{nlsupport}), which cannot be always satisfied.
%\cite{qperf-mills-tosem}

From Fig. \ref{fig:method-challenge-overlap}, we note that the primary studies suffering from this limitation (i.e., CH3) frequently adopt three methodologies - M1, M3, and M6. However, unlike the above ones, the lack of generalizability is closely associated to the dataset or evaluation mechanism rather than the methodologies adopted in query reformulations.

\textit{Mitigation strategy.} To improve the generalizability of findings, future work should focus on designing comprehensive experiments. For example, subject systems should come from multiple programming languages (e.g., Java, C/C++, Python, Javascript) \cite{bluirplus}, development practices (e.g., open-source, closed-source) \cite{oss-css}, and application domains. More subject systems and queries should also be chosen to rigorously evaluate any proposed technique. Furthermore, multiple search engines should be used to evaluate the original and reformulated queries to avoid spurious findings \cite{trconfig,refoqus}.

%determine the benefits of reformulated queries.
%That is, they might return good reformulated queries only if the given queries are carefully designed in the first place,
%perform well only with carefully designed 
 %in the first place, 
%which, in essence, defeats the overall purpose of automated query reformulations. 

\textbf{CH4--Issues towards practical adoption:} About 33\% (23/70) of our primary studies suffer from 
%Several studies \cite{qr-log-mining,testcase-cs-ist,verbose,gayg,shepherd,query-by-example,irmarcus,twkraft-jsep} suffer 
major issues that might prevent their adoption by software developers. Several studies \cite{irmarcus,qr-log-mining,req-qr,low-cost-crowd-qr}
make use of Latent Semantic Indexing (LSI) and term-document co-occurrence matrix to reformulate their search queries.
Although the LSI might be useful to mitigate the vocabulary mismatch problem \cite{vocaprob}, it suffers from the \emph{curse of dimensionality problem}\footnote{http://mlwiki.org/index.php/Vector\_Space\_Models}.
That is, a large, sparse matrix is needed to store the semantic space, which can be both costly and inefficient.
Besides, the semantic space needs to be updated frequently for each incoming query before its execution, which is time-consuming and undesirable.
Several approaches require non-trivial items such as test cases \cite{testcase-cs-ist}, code examples \cite{query-by-example, auto-query}, and salient keywords \cite{shepherd,multi-rec} from developers, which could make the developers reluctant to use these approaches.
Several studies \cite{cmills-icsme2018,verbose,lawrie-qr-bl} make use of ground truth information to identify appropriate search keywords from a bug report. Although these studies might be useful to provide a proof of concept, they are not suitable for a practical use since the ground truth is not generally known during query construction \cite{cmills-icsme2018}.
Several other studies involve costly computations such as genetic algorithms \cite{genqr-tse-perez,manq-qr}, frequent AST construction \cite{dl-qr-change-sequence}, or ontology construction \cite{qr-interactive-clarification}, which might outweigh the benefit of query reformulation.
%warrant significant manual  from developers such as test case generation 
%writing code examples , and the selection of salient keywords  . 
A few studies \cite{time-aware-term-weighting,twkraft,twkraft-jsep} need significant configuration management or parameter tuning, whereas the others use proprietary, publicly unavailable datasets (e.g., Bing search logs \cite{swim}, Stack Overflow search logs \cite{qr-stackoverflow}).
%external resources (\eg\ Bing search logs) \cite{swim} just to get started. 
There also exist other approaches \cite{gayg,verbose,cexchange} that involve the developers in costly trials and errors. 
Furthermore, 47\% of our studies do not have a replication package and thus might not be replicated properly.
All these constraints and issues might prevent the developers from using $\approx$33\% of our primary studies to reformulate their search queries.

\textit{Mitigation strategy.} Future work can take several steps to mitigate this threat. First, any cost-benefit analysis through human studies will inform the usability of a technique. Developer frequently used general-purpose search engines (e.g., Google) for code search that support traditional, natural language queries \cite{google-for-code-search}.
%are mostly comfortable with natural language queries, and 
Thus they should not be forced to design non-traditional queries (e.g., test cases, example code) for code search. The designed tools should be easily integrated into the working context of a developer \cite{surfexample}. Furthermore, the replication packages should be available to foster reuse and adoption.

\textbf{CH5--Human bias and lack of rigour in evaluation:} About 56\% (39/70) of our primary studies
might suffer from subjective bias and the lack of rigour in their evaluation. Several studies \cite{qr-log-mining,cexchange,iman-tse,lemos-type-qr,testcase-cs-ist,query-by-example} evaluate their approaches involving 
undergraduate or graduate students in a laboratory setting. These participants neither represent the real users (\eg\ professional developers) nor are adequate in number. For example, 80\% of our studies that engage humans have less than 25 participants each (Fig. \ref{fig:participant-count-stats}).
These participants were also responsible for both the ground truth preparation and the query evaluation.
Thus, there exists a good chance of subjective bias in the evaluation since the appropriate measures (e.g., inter-rater agreement, pooling method \cite{iman-tse}) were not always taken. 
 Replication of an existing work might also introduce implementation bias especially if the replication package is not available. A few studies \cite{rack,icsme2018masud,fse2018masud,hillicse09,active-cs,refoqus} might suffer from this bias.  
Many primary studies \cite{time-aware-term-weighting,twkraft-jsep,sisman,active-cs,kevic,ase2017masud} report their best performances using their well-tuned configurations against certain subject systems (Table \ref{table:context-subject-system-overlap}). 
%such configuration settings and tool parameters that are well tuned against their subject systems. 
Unfortunately, the same performance might not be guaranteed for other subject systems with the same configurations.
Several studies \cite{genqr-tse-perez,low-cost-crowd-qr,lemos-type-qr} do not make use of any frequently adopted metrics whereas the others \cite{neural-query-expansion,alteration-intent,kevic,qr-interactive-clarification,qr-stackoverflow} do not compare with established baselines from the literature. Besides, 49\% of the primary studies do not perform any statistical test (e.g., significance tests), which is an important step of rigorous validation.

From Fig. \ref{fig:method-challenge-overlap}, we note that the studies suffering from the above limitation or challenge (i.e., CH5) frequently adopt four methodologies: M1, M3, M6, and M8. Unlike many of the above ones, this challenge is closely associated to the experiment design rather than the methodologies adopted in search query reformulations.

\textit{Mitigation strategy.} Future work should involve appropriate participants (e.g., software practitioners) and a sufficient number of participants to mitigate the threats from human studies. The ground truth should be also carefully constructed using either polling method or inter-rater agreement analysis \cite{iman-tse}. The source code, experimental data, and other details should be made publicly available to facilitate smooth replication by third parties \cite{replication}. Furthermore, the conducted experiments should use appropriate performance metrics and statistical tests based on evaluation type and working contexts (check RQ2 for details). 

%There also exist other studies \cite{qr-log-mining,api-book,flt-model-qr,kevic,req-qr,ccmapping} that suffer from weak evaluation and validation approaches. 

\textbf{CH6--Inappropriate use of external tools and dependencies:} Over the last two decades, many tools, methods and technologies from Information Retrieval (IR) and Natural Language Processing (NLP) domains have been used to solve Software Engineering problems. Although they work with their default configurations, they need to be configured properly for Software Engineering problems for an optimal performance. They also need to be used with caution \cite{refoqus,trconfig}.
For example, several studies make use of WordNet \cite{shepherd,pwordnet,qsynthesis,time-aware-term-weighting} and
Stanford CoreNLP POS tagger \cite{api-book,rack} in the context of Software Engineering problems. 
Both these tools were trained on regular English texts (\eg\ Wall Street Journal) rather than the source code. 
Thus, the semantic inferences and POS tag predictions made by these tools could be either noisy or even inaccurate, as confirmed by the existing evidence \cite{semantictool,semantic-tool-code}. Many primary studies \cite{gayg,refoqus,tsc2016,concept-ase, active-cs,api-book, flt-model-qr, kevicdict} also use TF-IDF \cite{tfidf} as a 
%adopt TF-IDF \cite{tfidf} as a 
\emph{de facto} term weighting method to select important keywords from source code. 
However, TF-IDF was originally designed for regular texts (\eg\ news articles, legal documents) rather than source code. 
Regular texts and source code differ in their syntax, structures, semantics, and vocabulary \cite{devanbu-fse}. Thus, the adoption of TF-IDF to select keywords from source code might not be appropriate. Several studies heavily depend on third-party items such as code change history \cite{qecc-intent,kim-bl-qr}, semantic web technologies (e.g., SPARQL) \cite{nlsupport,qr-interactive-clarification}, tags from Stack Overflow \cite{tsc2016}, and input code examples \cite{auto-query,query-by-example}, and thus might be negatively affected by their absence.

\textit{Mitigation strategy.} Future work can take several steps to mitigate this threat. For example, the established tools from other domains (e.g., WordNet \cite{wordnet}, Word2Vec \cite{word2vec}) need to be carefully re-trained using software engineering data (e.g., source code, bug reports). Traditional techniques for keyword selection (e.g., TF-IDF \cite{tfidf}) should be adapted and tuned to software engineering data. Besides, the future techniques should depend on open-source, well-maintained dependencies to avoid any potential integration issues.

%It should be noted that regular texts and source code differ in their syntax, structures and semantics. Unlike regular texts, source code is scarce in vocabulary but rich in structures \cite{devanbu-fse}. Thus, the adoption of TF-IDF for keyword selection from the source code might not be appropriate. A few studies \cite{ase2017masud,fse2018masud} make use of a novel method namely \emph{CodeRank} for keyword selection from the source code, and attempt to overcome the limitations of TF-IDF. However, an exhaustive comparison between TF-IDF and CodeRank in keyword selection from the source code has been long due.

%\small
%\input{challenges.tex}
%\vspace{-.2cm}
%\normalsize

\textbf{CH7--Lack of a sound theory:} Several primary studies \cite{observed,qecc-intent,thesaurus-qr-idcs,flt-model-qr} make counter-intuitive or unsubstantiated theoretical assumptions in their approach.
%adopt unverified or weak theoretical assumptions in their query reformulation. 
\citet{qecc-intent} attempt to determine a query's intent using the code change history.
First, they mine the code change history of 631 GitHub projects and construct a database that contains thousands of $<$\texttt{code segment}, \texttt{changes}$>$ pairs. Then they analyze the code segment retrieved by a query, identify its frequently changed identifier names
using the database as query intent, and then reformulate the query with the intent.
%made to the segment using the database, and then use them to reformulate the query.
However, the idea of capturing a query's intent with the changed code 
%Although the reported experimental results look promising, their idea 
lacks a strong theoretical justification. A few other studies \cite{alteration-intent,dl-qr-change-sequence} might also have similar issues. 
%Especially, their assumption around the \emph{query intent} is vague and counter-intuitive.
According to \citet{observed}, an ideal bug report contains three parts- (a) observed behaviour (OB), (b) expected behaviour (EB) and (c) steps to reproduce (S2R) the bug. Using an empirical study, \citeauthor{observed} demonstrate that OB can be used as a reformulated query for bug localization.
However, how OB could be less noisy than the S2R was demonstrated neither theoretically nor empirically. According to a recent work \cite{oscar-duplicate}, S2R often contains terms that better indicate what the bug is about and includes key terms. Thus,  solely relying on OB might lead to the removal of important search keywords from a bug report \cite{cmills-icsme2018,oscar-duplicate}.
%the idea of using OB over EB or S2R lacks a sound theoretical justification. 
%\citeauthor{observed} experimentally show that OB texts could be a good reformulated version of the given query (bug report). However, such finding has little theoretical justification. 
%\citet{cmills-icsme2018} also could not reproduce the findings associated with OB. 
According to \citet{thesaurus-qr-idcs}, WordNet could be effective for query expansion during interface driven code search (IDCS) rather than keyword based code search (KBCS). However, such a claim was also not theoretically justified.

\textit{Mitigation strategy.} Future work can take a few steps to mitigate this threat. For example, the intuition or logical reasoning behind any proposed technique should be clearly articulated before presenting any experimental evidence. Comprehensive understanding of the relevant literature can help develop such intuition. Then any proposed theory should be also substantiated by strong empirical evidence to place it in the literature.

\textbf{CH8--Miscellaneous:} About 21\% (15/70) of our primary studies suffer from miscellaneous issues and limitations as follows. \emph{(a) Chronology of bug reports is not respected.} \citet{finelocator} do not respect the chronology of bug reports through their use of 10-fold cross validation. That is, they might have identified previous bugs from the history
by training their approach on the later bug reports, which is impractical. \emph{(b) Underlying semantics are either overlooked or misused.} The primary studies on query quality prediction \cite{qperf,qperf-mills-tosem} make use of lexical and syntactic properties of a query but mostly overlook the semantic aspect. They also target binary classification (e.g., good query and bad query), which might not be sufficient since the search queries can have non-binary quality levels (e.g., very good, good, bad, very bad). \citet{dl-qr-change-sequence} consider two code segments similar if they have similar AST nodes. However, their approach does not guarantee that both code segments have similar semantic, i.e., implement the same programming task. \citet{qecc-intent} also assume that a query's intent can be better expressed with the changed identifier names rather than focusing on the semantics of the query, which is counter-intuitive.
\emph{(c) Corner-cases are not considered.} \citet{anne} extend the vocabulary of source code by annotating program entities (e.g., \texttt{int arr[]}) with their equivalent concepts from natural language texts (e.g., integer array). However, their approach, as the authors report, might fall short for the nested program entities with subsumption problems. \citet{kevicdict} reformulate natural language  query (i.e., change request) into source code elements by mining version history. However, their approach might be restricted to existing code elements and thus might be insufficient for change requests where new software features are requested. 

\textit{Mitigation strategy.} To mitigate the above threats, future work can take several steps including performing time-aware training with bug reports, leveraging query semantics, and considering corner-cases in their techniques.

\begin{frshaded}
\noindent
\textbf{Summary for RQ$\mathbf{_3}$:} 
%Primary studies on automated query reformulations suffer from eight major challenges and limitations. 
Almost \textbf{half} of our primary studies might add unexpected noise to a query during its reformulation whereas 26\% of them impose extra cognitive burden on the developers. The majority of these studies also suffer from a lack of generalizability due to their limited datasets, queries, and evaluation. About 33\% of our studies suffer from the issues that might prevent their adoption by software developers whereas more than \textbf{half} of them could have subjective bias in their evaluation. Furthermore, several studies heavily depend of third-party items for their query reformulations and the others make counter-intuitive assumptions in their approaches.
%targeting code search experience \textbf{seven} key challenges including \textbf{unsolved vocabulary mismatch problem} (\textbf{30}\%), \textbf{lack of generalizability} ($\approx$\textbf{84}\%) and lack of proper rigour in dataset preparation or conducted experiments (\textbf{50}\%). They also impose \textbf{extra cognitive burdens} on the developers (\textbf{30}\%) during query reformulations/code search and \textbf{lack direct applicability} in practice (\textbf{20}\%). Many of these primary studies employ less appropriate tools or libraries (\eg\ WordNet) (9\%) and also incorporate \textbf{unsubstantiated ideas} (11\%) into their approach for query reformulation.  
\end{frshaded}

\subsection{Answering RQ$\mathbf{_4}$: Evidence of existing researches on automated query reformulation}
\label{sec:sevidence}
\begin{figure}[!tbh]
	\centering
	\includegraphics[width=3.8in]{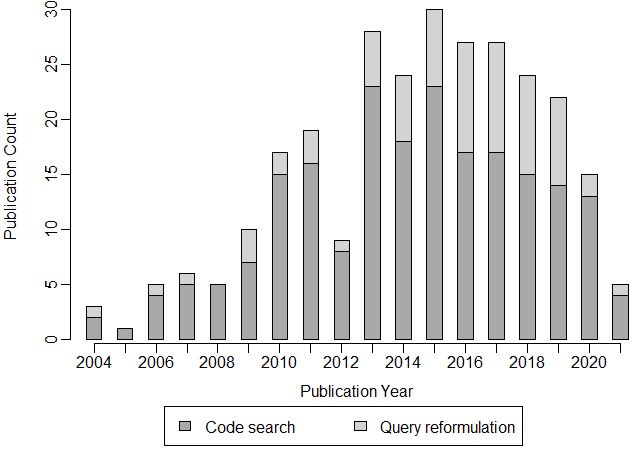}
	\vspace{-.3cm}
	\caption{Publications on code search and query reformulation}
	\label{fig:year-cs-qr}
	\vspace{-.4cm}
\end{figure}

We collect publication year, venue, and authors of each primary study to construct the evidence of 
%and determine the evidence of 
 existing researches on automated query reformulations. 
 Such an evidence not only demonstrates the interest and enthusiasm around the research topic 
 but also shows the quality of the work being produced and published.
Tables \ref{table:year-venue-stat}, and \ref{table:venue-year-stat} and Fig. \ref{fig:year-cs-qr}
present various statistics on the conducted studies over the last 15+ years.

%\begin{figure}[!tb]
%	\centering
%	\includegraphics[width=3.5in]{rqs/rq4/study-year-stat}
%	\vspace{-.2cm}
%	\caption{Publication counts across different calendar years}
%	\label{fig:year-stat}
%	\vspace{-.2cm}
%\end{figure}

Table \ref{table:year-venue-stat} shows the frequency of publications on automated query reformulations per calendar year. 
%It should be noted our primary studies target not only local code searches (\eg\ concept location, bug localization, feature location) but also Internet-scale code search. 
We see that the first work on automated query reformulation got published back in 2004. \citet{irmarcus} first introduced query reformulation in the context of concept location where they used word similarity derived from \emph{latent semantic space} to automatically reformulate a given search query (\eg\ \emph{``font style"}).
From 2005 to 2012, there had been a few studies targeting query reformulations, and on average 2-3 papers got published every year. However, the query reformulation received much attention as a research topic since 2013. We believe that such an attention or enthusiasm could be partially attributed to a few contemporary doctoral theses on query reformulations and code searches \cite{shaiduc-thesis,ehill-thesis}. Since 2013, there has been a significant increase in the research activities and publication counts. For example, during the last six years (2015--2021), 47 studies on query reformulations got published where 2016 and 2017 were the most productive years (e.g., 10 studies per year). It also should be noted that $\approx$80\% of our selected primary studies were conducted during the last nine years (2013--2021). 

\begin{table}[!t]
	\centering
	\caption{Existing publications on query reformulation targeting source code search}\label{table:year-venue-stat}
	\vspace{-.35cm}
	\resizebox{3.8in}{!}{%
		\begin{threeparttable}
			\begin{tabular}{l|p{3in}|c}
				\hline
				\textbf{Year} & \textbf{Published studies} & \textbf{Total} \\
				\hline
				\hline
				2004 & S50 (WCRE) &  1 \\
				\hline
				2006 & S18 (LNCS) & 1 \\
				\hline
				2007 & S43 (ASOD) & 1\\
				\hline
				2009 & S12 (ICSE), S30 (ICSM), S55 (LNCS) & 3 \\
				\hline
				2010 & S48 (ICSE), S51 (ASE) & 2\\
				\hline
				2011 & S3 (IST), S31 (ICPC), S47 (ASE) & 3 \\
				\hline
				2012 & S8 (ASE) & 1 \\
				\hline
				2013 & S6 (MSR), S10 (ICSE), S24 (ICSE), S28 (MSR), S52 (ASE) & 5\\
				\hline
				2014 & S1 (IST), S4 (ASE), S11 (ICSE), S27 (EMSE), S42 (MSR), S45 (MSR) & 6\\
				\hline
				2015 & S13 (SCAM), S14 (ASE), S15 (ASE), S33 (ICSME), S36 (SANER), S46 (ICSE), S53 (SANER) & 7 \\
				\hline
				2016 & S5 (Internetware), S17 (VLC), S23 (SPLC), S29 (ICSE), S34 (TSC), S37 (Internetware), S38 (ASE), S39 (SANER), S41 (ICSE), S56 (ASE) & 10 \\
				\hline
				2017 & S2 (ICCIT), S20 (TSE), S21 (JSEP), S22 (ASE), S26 (ASE), S32 (TOSEM), S40 (SANER), S44 (ICSME), S49 (WSDM), S54 (JSEP) & 10 \\
				\hline
				2018 & S7 (EMSE), S9 (DKE), S16 (FCS), S19 (ICSME), S25 (FSE), S35 (SPE), S60 (Access), S67 (ICSME), S68 (ICSME) & 9 \\
				\hline
				2019 &  S57 (IST), S58 (MAPL), S59 (AUSE), S61 (SAC), S63 (SPE), S65 (Access), S69 (TSE), S70 (ICPC) & 8 \\
				\hline
				2020 & S62 (IST), S64 (Access) & 2 \\
				\hline
				2021 & S66 (ICSE) & 1 \\
				\hline
				\multicolumn{2}{c}{~} & \textbf{70} \\
				\hline 
				
			\end{tabular}		
		\end{threeparttable}	
	\vspace{-.3cm}
	}
\end{table}

%\begin{figure}[!t]
%	\centering
%	\includegraphics[width=4.6in]{rqs/rq4/study-venue-stat}
%	\vspace{-.3cm}
%	\caption{Publication counts across different venues}
%	\label{fig:venue-stat}
%	\vspace{-.4cm}
%\end{figure}

\begin{table}[!t]
	\centering
	\caption{Existing Publications on Query Reformulation Targeting Source Code Search}\label{table:venue-year-stat}
	\vspace{-.35cm}
	\resizebox{3.8in}{!}{%
		\begin{threeparttable}
			\begin{tabular}{l|p{3in}|c}
				\hline
				\textbf{Venue} & \textbf{Published studies} & \textbf{Total} \\
				\hline
				\hline
				%& & & & & & & & & & & &
				ASE & S4 (2014), S8 (2012), S14 (2015), S15 (2015), S22 (2017), S26 (2017), S38 (2016), S47 (2011), S51 (2010), S52 (2013), S56 (2016) & 11 \\
				\hline
				ICSE & S10 (2013), S11 (2014), S12 (2009), S24 (2013), S29 (2016), S41 (2016), S46 (2015), S48 (2010), S66 (2021) & 9 \\
				\hline
				ICSM/ICSME & S19 (2018), S30 (2009), S33 (2015), S44 (2017), S67 (2018), S68 (2018) & 6\\
				\hline
				WCRE/SANER & S50 (2004), S36 (2015), S39 (2016), S40 (2017), S53 (2015) & 5 \\
				\hline
				IST & S1 (2014), S3 (2011), S57 (2019), S62 (2020) & 4 \\
				\hline
				MSR & S6 (2013), S28 (2013), S42 (2014), S45 (2014) & 4\\
				\hline
				Access & S60 (2018), S64 (2020), S65 (2019) & 3 \\
				\hline
				Internetware & S5 (2016), S37 (2016) & 2 \\
				\hline
				SPE & S35 (2018), S63 (2019) & 2\\
				\hline
				LNCS & S18 (2006), S55 (2009) & 2\\
				\hline
				JSEP & S21 (2017), S54 (2017) & 2 \\
				\hline
				EMSE & S7 (2018), S27 (2014) & 2 \\
				\hline
				TSE & S20 (2017), S69 (2019)  & 2 \\
				\hline
				ICPC & S31 (2011), S70 (2019) & 2 \\
				\hline
				VLC  & S17 (2016) & 1 \\
				\hline
				WSDM  & S49 (2017) & 1\\
				\hline
				AUSE  & S59 (2019) & 1\\
				\hline
				FSE  & S25 (2018) & 1\\
				\hline
				ASOD  & S43 (2007) & 1\\
				\hline
				SCAM  & S13 (2015) & 1\\
				\hline
				DKE  & S9 (2018) & 1 \\
				\hline
				TOSEM  & S32 (2017) & 1\\
				\hline
				TSC  & S34 (2016) & 1\\
				\hline
				ICCIT  & S2 (2017) & 1\\
				\hline
				SPLC  & S23 (2016) & 1\\
				\hline
				MAPL  & S58 (2019) & 1\\
				\hline
				SAC  & S61 (2019) & 1\\
				\hline
				FCS  & S16 (2018) & 1\\
				\hline
				\multicolumn{2}{c}{~} & \textbf{70} \\
				\hline 
				
			\end{tabular}
		\end{threeparttable}	
	}
\end{table}

%Fig. \ref{fig:venue-stat} shows the publication count for each venue. 
%for the researches on automated query reformulation targeting code search. 
Table \ref{table:venue-year-stat} shows how many primary studies got published at each venue in different calendar years.
We see that 50\% (35/70) of our selected studies got published in the top venues of Software Engineering
such as ICSE, ASE, ICSME, SANER, and MSR.
%(\eg\ ICSE, ASE, ICSME, MSR). 
About 28\% of them (20/70) also made into the flagship conferences such as ICSE (A*) and ASE (A*). Many of these 70 studies were extended later and got published in the reputed journals such as EMSE and JSEP.
 %(\eg\ EMSE, JSEP). We did not consider the extended works in our analysis unless there was a significantly novel contribution. 
There were also a number of Journal-First studies on query reformulations that got published in several high-quality journals such as TSE, TOSEM, EMSE, JSEP, SPE and TSC. While ASE and ICSE communities pioneered this research topic, other communities (\eg\ ICSME, MSR, SANER) have also been catching up to speed especially in the last few years.
%We also see such an evidence of significant research activities from the Table \ref{table:year-venue-stat}.

We also investigate the comparative research interest in query reformulation and in code search. It should be noted each primary study is related to code search but each study on code search might not be related to query reformulation. Thus, we repeat the relevant part of our study selection process and identify the studies on code search as well. Fig. \ref{fig:year-cs-qr} summarizes our analysis using a bar plot.
According to our investigation, ~34\% (70/207) of the code search studies focus on query reformulations.
As shown in Fig. \ref{fig:year-cs-qr}, a significant fraction of code search studies target query reformulation every year. For example, a maximum of 23 code search studies were conducted in 2015, and 7 of them focused on search query reformulations.

%\begin{figure}[!t]
%	\centering
%	\includegraphics[width=4.6in]{rqs/rq4/author-cloud-gs}
%	\vspace{-.3cm}
%	\caption{Word cloud of author names using publication counts in automated query reformulations}
%	\label{fig:author-cloud}
%	\vspace{-.4cm}
%\end{figure}

We also identify the authors from each of our 70 primary studies and calculate their contribution frequency.
%design a word cloud using their first and last names.
We found a total of 172 Software Engineering researchers who contributed to the research of automated query reformulations 
to improve code search. Appendix \ref{appendix:E} shows the authors who contributed to at least two primary studies.
%In Fig. \ref{fig:author-cloud}, the size of each of the names indicates their relative publication counts on automated query reformulations.

The demand for low-cost, accurate, and easy-to-use solutions for code search is likely to grow in the coming years due to
the constant growth of open-source repositories (\eg\ GitHub, BitBucket) and the increasing complexity or size of modern software systems.
%have been constantly growing and software systems are becoming even more complex gradually. . 
Reformulation of a query is an inevitable step of code search regardless of the search context. 
Thus, automated query reformulation is likely to be an active, interesting research topic for the years to come.
%However, given our identified challenges and limitations (Section \ref{sec:threats}), we suggest that the community should focus on designing more practical solutions that can be easily integrated by software developers in their regular workflows.
 %going to be even more attractive to the researchers as a problem area. Consequently, the research outcomes are also going to be even more relevant and appealing to the software practitioners (\eg\ developers).         
\begin{frshaded}
	\noindent
	\textbf{Summary of RQ$\mathbf{_4}$:} To date, according to our investigation, a total of \textbf{70} primary studies  
	on automated query reformulations supporting code search have been conducted. About \textbf{80\%} of them were conducted in the last nine years. This trend not only indicates the timeliness or relevance but also the attention or enthusiasm around this research topic. A large group \textbf{172} researchers contributed to this topic. About 50\% of the primary studies got published at the top venues of Software Engineering such as ICSE, ASE, ICSME, MSR, and SANER.
	%from all over the world either had or have been working on this research topic. The researches on automated query reformulation regularly got published in the top Software Engineering venues such as \textbf{ICSE}, \textbf{ASE}, \textbf{ICSME}, \textbf{MSR},  \textbf{EMSE} and \textbf{JSEP}.  
	Given the increasing size and complexity of modern software systems, the research activities on automated query reformulations and code searches are likely to increase in the coming years.
\end{frshaded}

\subsection{Answering RQ$\mathbf{_5}$: A closer look at local code search and Internet-scale code search}
\label{sec:qualitative}
%\textbf{Query Reformulations for Local Code Search (LCS) \& Internet-scale Code Search (ICS):} 
Local code search such as bug localization, concept location, or feature location 
attempts to find the location of a software bug, a concept, or a feature respectively within the code of a software system.
On the other hand, Internet-scale code search attempts to find out relevant, reusable code segments from thousands of software projects stored at Internet-scale code repositories (\eg\ GitHub). More specifically, while the former detects code locations that need to be modified, the later finds code segments that can be reused.
%against a programming task from  \emph{Ohloh}, \emph{SourceForge}). 
%Primary studies target either one of these two searches.  
Although these two searches differ in their working context and end goal, they could overlap in several other aspects. 
In this systematic literature review, we analyze 70 primary studies where 58\% of them reformulate queries for local code searches and the remaining 42\% studies attempt to improve Internet-scale code search through query reformulations (e.g., Fig. \ref{fig:study-wc-type}).
We refer to these two groups of studies as QR$_{LCS}$ and QR$_{ICS}$ respectively.
%that reformulate queries for both searches.
%As shown in , 58\% of the studies target local code searches where
A closer look at the query reformulations supporting these two searches 
%which target both these code searches. According to our investigation, query reformulation studies intended for local code searches and Internet-scale code search 
%overlap significantly in their underlying methodologies, evaluation metrics and experienced challenges. 
%For example, primary studies on these two types of query reformulations \cite{tsc2016,refoqus,icsme2018masud,iman-tse,sisman,active-cs} often borrow methodologies and algorithms (\eg\ Rocchio \cite{rocchio}) from each other. Consequently, they also share common limitations (Table \ref{table:challenges}) and strengths. 
%However, there is a marked lack of research that investigates and contrasts between these two types of query reformulations/code searches. An in-depth comparison between query reformulation for local code searches (hereby \textbf{QR-LCS}) and query reformulation for Internet-scale code search (hereby \textbf{QR-ICS}) 
can provide valuable insights.  In this section, we compare and contrast between these two groups of primary studies based on 
%which could encourage novel, more effective solutions. 
%We contrast between these two types of query reformulations in terms of 
their underlying methodologies, algorithms, evaluation methods, challenges, and limitations as follows.

\begin{figure}[!t]
	\centering
	\includegraphics[width=5.5in]{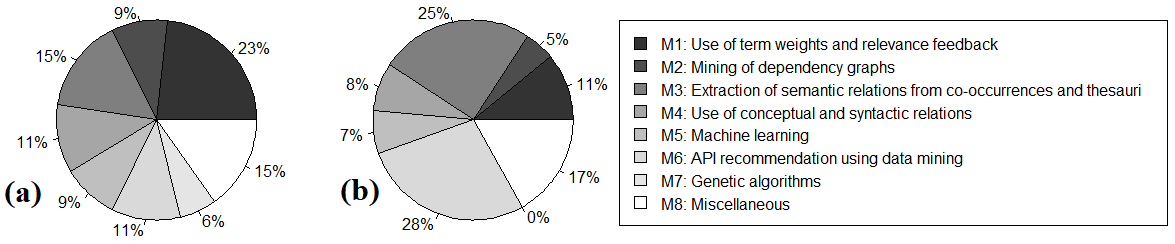}
	\vspace{-.2cm}
	\caption{Use of methodologies by the primary studies supporting (a) local code search, and (b) Internet-scale code search}
	\label{fig:compare-methodology}
	\vspace{-.3cm}
\end{figure}

\begin{figure}[!t]
	\centering
	\includegraphics[width=3.6in]{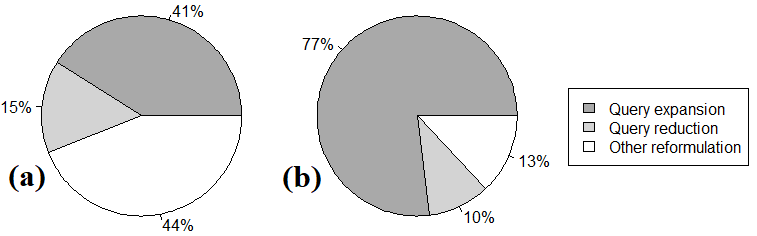}
	\vspace{-.2cm}
	\caption{Types of reformulation provided by the primary studies supporting (a) local code search, and (b) Internet-scale code search}
	\label{fig:compare-qtype}
	\vspace{-.5cm}
\end{figure}

\begin{figure}[!t]
	\centering
	\includegraphics[width=3.8in]{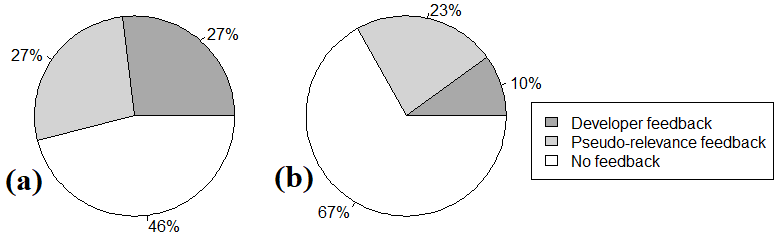}
	\vspace{-.2cm}
	\caption{Relevance feedback by the primary studies supporting (a) local code search, and (b) Internet-scale code search}
	\label{fig:compare-rf}
	\vspace{-.5cm}
\end{figure}

\textbf{Underlying methodologies and algorithms in query reformulation.} Among 70 primary studies, 40 studies (QR$_{LCS}$) 
reformulate queries to support local code searches (\eg\ bug localization, concept location) whereas the remaining 30 studies (QR$_{ICS}$) target Internet-scale code search. Fig. \ref{fig:compare-methodology} shows the methodologies adopted by these two groups of primary studies.
We see that term weighting and relevance feedback mechanism are frequently used to select keywords and to reformulate queries by the primary studies associated to local code searches. About 23\% of them adopt this methodology whereas the ratio is only 11\% for the studies from QR$_{ICS}$.
In local code search, there are several software artifacts such as bug reports and feature requests from which 
the keywords can be chosen based on their weight or appropriateness, which explains the popularity of term weighting methods.
On the other hand, in the Internet-scale code search, query keywords are often chosen by the developers on the fly according to their information need.
Semantic relations, word co-occurrences, and thesauri have also been used by 15\% and 25\% of the primary studies from these two groups respectively.
However, the most popular methodologies for query reformulations that support Internet-scale code search are data mining and API recommendation.
About 28\% of the studies from QR$_{ICS}$ adopt this methodology whereas the ratio is only 11\% for the counterpart.
In Internet-scale code search, search queries are often short (e.g., 2--3 keywords) and thus they require an expansion to retrieve the desired code \cite{koderlog,stolee2015}. Many primary studies \cite{iman-tse,rack,icpc2019rodrigo,icsme2018masud,biker} expand these queries with relevant API classes or methods mined from large software repositories (e.g., GitHub, Stack Overflow), which explains the popularity of API recommendation methodology (i.e., M6).
Several studies \cite{lemos-type-qr,correlation-cs,pwordnet,thesaurus-qr-idcs,anne} also expand these short queries with synonyms or similar words
from well-known thesauri such as WordNet \cite{wordnet}. This explains a high usage ratio (25\%) of thesaurus based methodology (i.e., M3) by the studies from QR$_{ICS}$ 
(Fig. \ref{fig:compare-methodology}-(b)).

Fig. \ref{fig:compare-qtype} shows the types of query reformulations performed by primary studies supporting both local and Internet-scale code searches.
We see that query expansion is very common for both groups (e.g., 41\%--77\%). However, queries are more frequently expanded in the Internet-scale code search due to their short length \cite{koderlog-msr}. Query reduction is less common for both groups of studies. Interestingly, in local code search, a significant number  of queries (e.g., 44\%) receive other types of reformulations such as the combination of query expansion, reduction, or replacement.

Fig. \ref{fig:compare-rf} shows how relevance feedback is used to reformulate queries by the primary studies.
We see that query reformulations in the local code search use comparatively more feedback. About 54\% of them use either direct feedback from developers or pseudo-relevance feedback from a search engine. On the other hand, only 33\% of studies from the Internet-scale code search (QR$_{ICS}$) use any form of feedback to reformulate their queries. It should be noted that these QR$_{ICS}$ studies capture less feedback 
%from any source to reformulate their queries 
although they involve human developers more frequently than their counterpart (check Section \ref{sec:validation}-(d)).
%We also see that QR-LCS studies collect more feedback from the developers than QR-ICS studies do (\ie\ 27\% vs. 4\% DF). 
Any local code search takes place on a single software project whereas the Internet-scale code search deals with thousands of cross-domain software projects. 
Thus, the results from Internet-scale code search might be noisier, and capturing relevance feedback against them could be costly and challenging. 
%Search results from multiple, cross-domain projects can be noisier than that from a single project.
%Capturing relevance feedback against cross-domain results could be both challenging and costly.
%developer feedback during Internet-scale code search could be  
%costlier than with local code searches. 
Such a high cost might have discouraged any form of relevance feedback. According to Fig. \ref{fig:compare-rf}, 46\%--67\% of studies from both groups did not use any form of relevance feedback to reformulate their search queries.

%. According to our investigation, 
%46\%  and 67\% of the studies from our two groups above did not use any relevance feedback.

%73\% QR-ICS studies apply no relevance feedback at all during their query reformulations (Fig. \ref{fig:compare-qtype}). 

\begin{figure}[!t]
	\centering
	\includegraphics[width=4.2in]{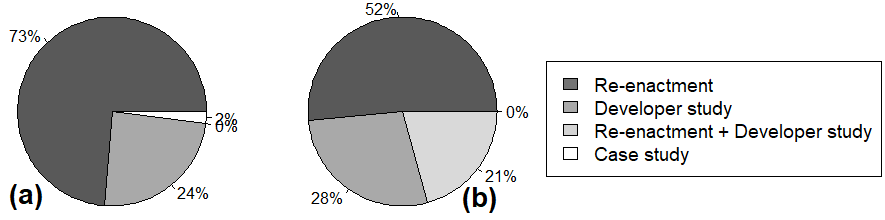}
	\vspace{-.2cm}
	\caption{Evaluation methods used by the primary studies supporting (a) local code search, and (b) Internet-scale code search}
	\label{fig:compare-eval-type}
	\vspace{-.4cm}
\end{figure}

  \begin{figure}[!t]
  	\centering
  	\includegraphics[width=4.2in]{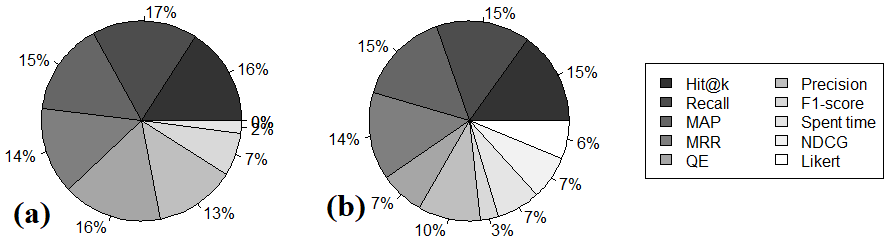}
  	\vspace{-.2cm}
  	\caption{Popular evaluation metrics used by the primary studies supporting (a) local code search, and (b) Internet-scale code search}
  	\label{fig:compare-metric}
  	\vspace{-.4cm}
  \end{figure}

%studies adopt re-enactment based evaluation (RBE), 43\% (24/56) studies adopt developer-oriented evaluation (DOE) and the remaining 7\% (4/56) studies employ both types of evaluations. According to our investigation, 67\% studies from local code searches employ re-enactment based evaluation and 30\% of them employ developer-oriented evaluation (Tables \ref{table:study-tfc}, \ref{table:study-eva-metrics}). On the contrary, such measures are 26\% and 61\% respectively for Internet-scale code searches.

\textbf{Evaluation and validation in query reformulation.} Among 70 primary studies, 
45 (64\%) studies use re-enactment-based evaluation, 18 (26\%) studies involve human participants whereas 6 (9\%) studies use a combination of both (see Fig. \ref{fig:empirical-developer-both}).
%On the other hand, the remaining 6 studies conduct case studies to evaluate their approaches.
Fig. \ref{fig:compare-eval-type} shows a breakdown of these statistics for two groups of primary studies.
We see that 73\% of the studies supporting local code searches evaluate their queries through \emph{re-enactment} using historical artifacts (e.g., bug reports, version control history). On the contrary, 52\% of the studies supporting Internet-scale code search 
adopt the same where they make use of existing benchmarks and programming Q\&A sites (e.g., Stack Overflow).
However, human participants were more frequently involved in the Internet-scale code searches. For example, 
%their popular evaluation method is developer study. 
$\approx$50\%  (28\% + 21\%) of the studies from the Internet-scale code search involve human developers in their evaluation. Such a ratio is only 24\% for the counterpart. Existing primary studies from local code search
 %Query reformulation studies on local code searches (\eg\ bug localization, feature location) 
 often construct their dataset and ground-truth from the version control history (e.g., bug-fixing commits) of open source projects that are widely available. On the contrary, historical data from Internet-scale code search such as search engine logs are proprietary artifacts that might not be publicly available. Thus, these studies often involve human developers to evaluate their queries.
 
 %Thus, re-enactment based evaluation is not widely used in the studies on Internet-scale code search. Developer-oriented evaluation (DOE) is rather a viable choice for evaluating studies on this code search. However, construction of test queries and corresponding ground truth is not only time-consuming but also subject to human bias and errors.     

Fig. \ref{fig:compare-metric} further contrasts between two groups of our primary studies 
based on their use of popular performance metrics.
We see that Hit@K, Recall, MAP and MRR are generally popular among both groups. However, the studies from local code search are more interested about query effectiveness (QE).
%traditional metrics (\eg\ Hit@K, MRR, MAP, MP, MR) are frequently employed by the query reformulation studies from both searches. We also see that Query Effectiveness (QE) is adopted by 33\% of QR-LCS studies whereas such ratio is 26\% with QR-ICS studies. 
In local code searches such as concept location, bug localization, or feature location, the quick identification of the first correct result is important. The underlying idea is that once the first correct document is located, 
the other relevant documents can be easily located through their structural dependencies (\eg\ call graphs) within a software project \cite{shaiduc-thesis}. The same might not be true for the Internet-scale code search and 7\% of their studies use QE.
Their search results are retrieved from thousands of cross-domain software projects and static dependencies might not be applicable beyond a single project. We also see that unlike QR$_{LCS}$,  QR$_{ICS}$ studies are more interested in developer-oriented metrics such as spent time, NDCG, and Likert that capture a developer's impression on the results retrieved by a query.

%The higher adoption ratio of MAP by QR-ICS studies could be explained by the notion that Internet-scale code search is focused on delivering all the correct results at the top positions of the result list rather than just the first correct result only. The same explanation also applies with another state-of-the-art metric for rank evaluation, NDCG. 
%We also see that Likert Scale (LS) and other ad hoc metrics (\eg\ ETS, efforts and time spent) are more frequently used in evaluating the Internet-scale code search approaches due to their developer-centric experiments and lack of historical ground truth.

  \begin{figure}[!t]
	\centering
	\includegraphics[width=5.4in]{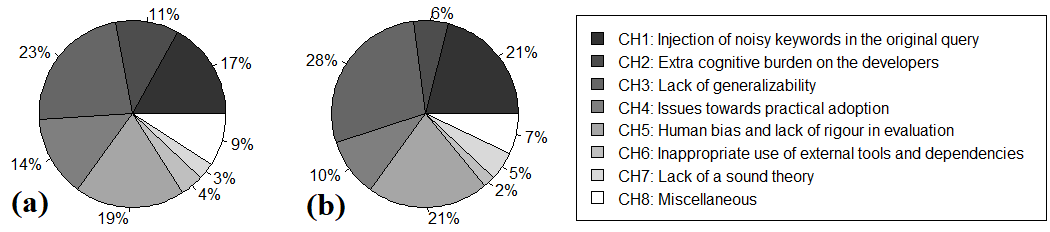}
	\vspace{-.2cm}
	\caption{Challenges and limitations of query reformulations targeting (a) local code search, and (b) Internet-scale code search}
	\label{fig:compare-challenge}
	\vspace{-.2cm}
\end{figure}

\textbf{Challenges and limitations in query reformulation.} Since the studies supporting local code search (QR$_{LCS}$) and the studies supporting Internet-scale code search (QR$_{ICS}$) overlap in their methodologies (check Fig. \ref{fig:compare-methodology}), 
%both local code searches (\eg\ bug localization) and Internet-scale code search overlap in their methodologies and algorithms,
they might suffer from similar types of challenges or limitations. Fig. \ref{fig:compare-challenge} 
summarizes our analysis in this aspect. We see that the studies from both groups are similarly affected by several challenges and limitations (Table \ref{table:challenge}). About 17\% of QR$_{LCS}$ studies could suffer from noise in their queries whereas the corresponding statistic is 21\% for the QR$_{ICS}$ studies. In local code search, developers can select keywords from existing artifacts (e.g., bug reports) whereas they need to select keywords on the fly during Internet-scale code search. Thus, the later queries need more extension (Fig. \ref{fig:compare-rf}) and might have a higher chance of getting noisy keywords.
%shows the major challenges/limitations experienced by our selected primary studies. We see that 48\% studies on Internet-scale code search suffer from vocabulary mismatch problem (\ie\ CH$_1$) as opposed to 18\% studies on local code searches. 
%During software bug localization, developers often make use of bug report texts with little no modifications in constructing a search query. On the contrary, during Internet-scale code search, the developers guess a few important keywords that best describe a programming task. Thus, queries in Internet-scale code search are comparatively shorter which need frequent expansions. Besides, it is extremely hard for a developer to guess the right keywords that are used in thousands of software projects \cite{vocaprob}. Thus, \emph{vocabulary mismatch problem} is a major challenge for Internet-scale code search. Such observation is also supported by the higher adoption rate of thesaurus-based expansion (using synonyms) in the QR-ICS studies (Fig. \ref{fig:compare-methodology}).
The existing query reformulation approaches in the local code search impose more cognitive burden on the developers than their counterpart. Changes to wrong code often have major negative consequences (\eg\ subsequent bugs) \cite{vcc,vcc-2}. Thus, precisely locating the source code of interest is a major step in the local code searches. On the contrary, developers might accept the partially relevant code examples during Internet-scale code search, which they can customize based on their needs. 
Both groups of studies also suffer from a lack of generalizability, which is the most prevalent issue. The findings from 
23\%--28\% of the studies from both groups might not be generalizable, as acknowledged or implied by their authors.
Although existing software artifacts (e.g., software, bug reports, version control history) are extensively used to evaluate query reformulations in the local code search, the majority of them target Java-based systems. According to an existing evidence \cite{bluirplus}, the findings from Java-based systems might not generalize for other systems such as C/C++ systems.
On the other hand, the datasets used to evaluate query reformulations in the Internet-scale code search are either small or proprietary in nature. For example, 60\% of the QR$_{ICS}$ studies use 75 or less number of queries each in their evaluation.
A few studies make use of proprietary items such as search logs from Bing \cite{swim} and Stack Overflow search engine \cite{qr-stackoverflow}, which might not be available for a public reuse and replication.  From Fig. \ref{fig:compare-challenge}, we also see that both groups suffer similarly with the remaining issues including human bias, weak evaluation (CH5, Table \ref{table:challenge}), extreme external dependencies (CH6, Table \ref{table:challenge}), and a lack of sound theoretical justification (CH7, Table \ref{table:challenge}). 

%While local code searches can be extensively evaluated using the dataset constructed from version control history of open source projects, Internet-scale code search does not have such leverage. The code search logs of various search engines (\eg\ GitHub, Google) are often proprietary artifacts which are not publicly available. Besides, search logs alone might also not be sufficient enough for the their ground truth construction. 
%According to our investigation, the median query count of each primary study on Internet-scale code search is 54 whereas such number for the local code search study is $\approx$four times higher, 209. 
%The construction of Internet-scale corpora (\eg\ \emph{Sourcerer} \cite{sourcerer}, \emph{IJaDataset} \cite{iman-codesearch,iman-tse}) for laboratory experiments could be even more challenging and time-consuming. The studies on local code searches collect their test queries from only two Java-based systems (Fig. \ref{fig:eval-dimensions}). Thus, 76\% of them also suffer from generalizability issues. 

\begin{table}[!t]
	\centering
	\caption{Comparison between query reformulations for local code search and Internet-scale code search}\label{table:compare-lcs-ics}
	\vspace{-.35cm}
	\resizebox{5.4in}{!}{%
		\begin{threeparttable}
			\begin{tabular}{p{1.2in}|p{2.2in}|p{2.2in}}
				\hline
				\textbf{Dimension} & \textbf{Local code search (LCS)} & \textbf{Internet-scale code search (ICS)}\\
				\hline
				\hline
				\multicolumn{3}{c}{\textbf{Query \& search intent}}\\
				\hline
				Initial query (ad hoc) & Preprocessed version of a bug report, a change request, or a feature request & Developer chosen keywords \\  
				\hline
				Initial query length & Comparatively long & Short (\eg\ 1-3 keywords \cite{koderlog}) \\
				\hline
				Keyword selection & Performed with term weighting and relevance feedback (M1) & Performed with data mining and API recommendation (M6) \\  
				\hline
				Query reduction & Used by 15\% of the studies & Used by 10\% of the studies \\ 
				\hline
				Operational context & Software change tasks (e.g., bug fixing) & General-purpose programming tasks \\
				\hline
				Search intent & Precisely locating an entity of interest & Locating the most relevant code example to a programming task \\
				\hline
				End goal & Making changes to the existing code & Implementing a programming task \\
				\hline
				Relevance feedback & Used by 54\% of the studies & Used by only 33\% of the studies \\ 
				\hline
				\multicolumn{3}{c}{\textbf{Corpus, ground truth, \& evaluation}}\\
				\hline
				Corpus documents & Source documents (e.g., classes, methods) from a single software project (\eg\ JEdit \cite{gayg}) & Source documents from thousands of software projects (\eg\ Sourcerer \cite{sourcerer}) \\
				\hline
				Ground truth construction & Commits from the version control history \cite{fse2018masud,refoqus} & Relevance judgement from developers using pooling method \cite{iman-tse}\\   
				\hline
				External resources & Past bug reports, bug-fixing history, author history, knowledge graph \cite{concept-ase} & WordNet, Stack Overflow, GitHub \\  
				\hline
				Ranking priority & First correct result should rank high & All correct results should rank high\\
				\hline
				Popular metrics & Hit@K, Recall, MAP & Hit@K, Recall, MAP \\ 
				\hline
				\multicolumn{3}{c}{\textbf{Miscellaneous}}\\
				\hline
				Major challenges & Lack of generalizability, developer bias, weak evaluation, and noisy queries & Lack of generalizability, developer bias, weak evaluation, and noisy queries\\
				\hline
				Experience required & Domain knowledge (\eg\ design, architectures) of a system under study & Relevant API libraries \\
				\hline
				Research evidence & 58\% of the primary studies &  42\% of the primary studies \\
				\hline
			\end{tabular}
		\end{threeparttable}
	}
\vspace{-.3cm}
\end{table}

%From Fig. \ref{fig:compare-challenge}, we also see that studies on Internet-scale code search suffer more from the lack of generalizability than their counterparts on local code searches (\ie\ 91\% vs. 76\%, CH$_3$). Such issue of primary studies from Internet-scale code search can be partially explained by their use of small datasets. 

%needs to precisely map the reported concern/bug to one or more specific source code entities (\eg\ methods, classes). Code changes to wrong entities might result into unresolved bugs or even new bugs. 
%,and then collects a list of (partially) relevant code examples. 
%That is, appropriate keyword selection is more crucial for the local code searches since they need more precise mapping to the code. 

Table \ref{table:compare-lcs-ics} further compares and contrasts between the query reformulation approaches supporting local code search and Internet-scale code search. We see that they are different from each other in terms of the majority of  dimensions such as query length, keyword selection methodology, operational context, search intent, and even in the ranking priority.
For example, in a local code search, developers attempt to precisely locate the source code entity that needs a change.
Changing the wrong code entity can have serious consequences \cite{vcc,vcc-2}. On the contrary, in the Internet-scale code search, developers attempt to locate the most relevant code example that can be reused to implement a programming task.
Despite these differences, the two groups of studies are also surprisingly similar across several dimensions such as their performance metrics (e.g., Hit@K, Recall, MAP) and experienced challenges or limitations (e.g., lack of generalizability).

 \begin{frshaded}
 	\noindent
 	\textbf{Summary of RQ$\mathbf{_5}$:} 
 	About \textbf{58\%} of our primary studies support local code searches (e.g., concept location, bug localization) whereas the remaining \textbf{42\%} focus on the Internet-scale code search. The query reformulation approaches supporting these searches share their methodologies, algorithms, performance metrics, and even suffer from similar challenges and limitations. Appropriate keyword selection is an essential step of any query reformulation.
 	In local code search, keywords are often selected using 
 	%The studies supporting local code search rely on 
 	various term weighting methods (e.g., TF-IDF) whereas data mining and API recommendations are adopted
 	in the context of Internet-scale code search. The primary studies from these two searches also
 	%the counterparts from Internet-scale code search use data mining to select appropriate keywords.
 	 %They also 
 	 differ in their reformulation types, evaluation methods, and human participation. However, both groups of studies suffer from a common set of issues such as a lack of generalizability, weak evaluation, and noisy queries. 
 \end{frshaded}  

\subsection{\textbf{Answering RQ$\mathbf{_6}$:} Recommended practices in query reformulations for code search}\label{sec:recommended-practice}
 Query reformulation is an integral part of source code search regardless of working context. Our systematic review using 70 primary studies informs the frequently adopted methodologies (RQ1) and evaluation methods (RQ2), and identifies their common challenges and limitations (RQ3). While query reformulation practices could be significantly different for local and Internet-scale code searches (RQ5),
there are several recommended, well-established practices that should be followed.
%several practices are recommended regardless of the working context. 
In this section, we discuss several recommended practices in query reformulations and code search based on experimental evidence and the general understanding of community.

\textbf{(a) Capture search queries from human users.} In local code search, the \emph{title} and \emph{description} of a change request
have been frequently used as baseline queries assuming that they need a reformulation \cite{buglocator,versionhistoryjsep,saha,ase2017masud,saner2017masud}. On the other hand, in the Internet-scale code search, several primary studies \cite{li2016,api-book,concept-ase,codehow,cocabu} use the title of a question from Stack Overflow as baseline query. 
However, these queries might not always represent an actual user's queries \cite{refoqus}. Thus, although capturing search queries from secondary sources might be necessary for an exhaustive evaluation, queries should be also collected from actual users (e.g., professional developers).  

\textbf{(b) Choose appropriate granularity for corpus documents.} Unlike regular texts (e.g., news article), source code has different granularity levels: package, file, class/module, method, and line. To search for code using IR methods, the source code of a project needs to be indexed at an appropriate granularity level. Many primary studies \cite{buglocator,versionhistoryjsep,saha,fse2018masud} choose files and classes to index their corpus as a part of bug localization. Since software bugs might be scattered through multiple methods \cite{parnindebug}, a granularity level such as \emph{file} or \emph{class} might be appropriate. On the other hand, most studies from concept or feature location \cite{refoqus,saner2017masud,hillicse09} use methods to index their corpus. They identify the relevant entity in the code that needs a change \cite{refoqus}. Choosing a small granularity such as \emph{method} is likely to help the post-search analysis by developers who can leverage their cognitive abilities \cite{hillicse09,multi-faceted-flt}. In the Internet-scale code search, developers often look for short, working code examples \cite{nasehi}, which also makes method as a feasible choice for corpus indexing \cite{surfexample,portfolio-tosem}. 

\textbf{(c) Apply standard preprocessing to source code.} Corpus documents (e.g., classes, methods) need to be preprocessed as a part of indexing. Traditionally, three preprocessing steps have been applied to source code: \emph{stop word removal}, \emph{token splitting}, and \emph{stemming}. First, stop words are frequently used words, but they convey very little semantics. Our primary studies often use a standard list of stop words \cite{stopword} and programming keywords \cite{java-keyword} to preprocess their corpora. Several studies \cite{saner2017masud,refoqus} remove such words that have a length of less than three characters or that occur in more than 25\% of the corpus documents. Second, source code is full of structured entities such as camel-case tokens, snake-case tokens, method invocations, field invocations, and so on. Our primary studies \cite{observed,saner2017masud,ase2017masud,refoqus} split these entities using appropriate regular expressions and advanced algorithms (e.g., Samurai \cite{samurai}). Third, stemming is a widely used preprocessing step that extracts the root of any given word. Two stemming algorithms --\emph{Porter} \cite{porter-stemming} and \emph{Snowball} \cite{snowball-stemming} -- have been frequently used by the primary studies. However, the impacts of stemming have been mixed according to a few studies \cite{kevic,saha,stemming}. For example, stemming might have a strong impact on short queries but negligible impact on large queries \cite{stemming}. Thus, it should be used selectively based on the type or size of queries and the search task at hand (e.g., bug localization).

\textbf{(d) Choose appropriate reformulations to improve a query.} Queries can be reformulated in three different ways: expansion, reduction, and replacement (Section \ref{sec:background}). Given an initial query, appropriate reformulation techniques should be chosen to reformulate the query. Developers frequently use short queries to search for reusable code in the Internet \cite{koderlog,stolee2015,google-for-code-search}. Similarly, they choose a few keywords from a change request as a query 
and execute the query to detect the code of interest (e.g., faulty code, software feature). If a query is short, it is expanded with relevant, complementary keywords \cite{emse2018masud, hillicse09}. On the other hand, long queries (e.g., preprocessed bug report) are reduced by discarding noisy, irrelevant keywords from them \cite{verbose,refoqus}. Similarly, hard, difficult queries are replaced with better alternatives \cite{hard-trace-ase,carmel}. Sometimes, a combination of multiple reformulation techniques might be required where machine learning can be employed to suggest the best reformulated query \cite{refoqus,ase2017masud}.

\textbf{(e) Use multiple search engines to evaluate the same query.} To retrieve relevant source code, a query needs to be executed with a search engine. In a local code search (e.g., bug localization, concept location), primary studies frequently use two search engines -- \emph{Lucene} \cite{lucene} or \emph{Indri} \cite{saha}  -- to execute their queries. Lucene is a Vector Space Model (VSM)-based search engine that combines Boolean search and BM25 algorithm \cite{bm25}. On the other hand, Indri combines probabilistic language modelling and BM25 algorithm \cite{saha}. These engines can be installed against a local code repository and are easy to use. Several studies \cite{emse2018masud, codehow,icsme2018masud,icpc2019rodrigo} also use \emph{Lucene} to execute their queries in Internet-scale code search. A few studies make use of well-established search engines such as Google \cite{emse2018masud}, GitHub search \cite{cocabu}, and Stack Overflow search \cite{cocabu,li2016}. According to existing literature \cite{trconfig,refoqus}, the same query can perform differently with different search configurations, which could pose a threat to the generalizability of findings. Thus, the same reformulated query needs to be evaluated with multiple search engines to mitigate the threat. 

\textbf{(f) Choose appropriate performance metrics for evaluation.} Queries need to be evaluated using appropriate performance metrics \cite{threats-to-validity}.
Our investigation (Fig. \ref{fig:year-pop-metric-freq}) suggests that several metrics such as Hit@K, MAP, MRR, QE, precision, and recall have been frequently used by the primary studies for the last 15+ years. While precision and recall are traditional metrics, Hit@K, MAP, MRR, and QE have received significant attention since 2013. We found that these metrics have been widely used in both local and Internet-scale code searches (Fig. \ref{fig:context-metric-overlap}). However, they are more prevalent in the re-enactment based evaluation than in the developer studies (Fig. \ref{fig:pop-metric-eval-type}). Thus, future studies can choose them for their re-enactment based evaluation. On the other hand, likert, spent time can be used for developer study-based evaluation.

\textbf{(g) Choose appropriate dataset for experiments.} Our analysis suggests that the use of experimental dataset (e.g., subject systems) could vary based on the context of code search (Table \ref{table:context-subject-system-overlap}). For example, primary studies frequently use individual systems in local code searches (e.g., bug localization, concept location) whereas large-scale software repositories (e.g., GitHub, SourceForge, Stack Overflow) are frequently used in Internet-scale code search. The widely used subject systems can be found in Fig. \ref{fig:popular-subject-system}, which could be a good starting point for the future work.
%and Appendix \ref{appendix:D}, 
%which could be a good. 
We also note that only a few studies evaluate their query reformulation techniques using proprietary systems (e.g., CAAF \cite{flt-model-qr}) whereas the majority use open-source software systems, which could lead to threats to generalizability (e.g., CH3, Table \ref{table:challenge}) \cite{oss-css,open-close}. Thus, the future work should use appropriate systems from both development practices to achieve generalizability \cite{threats-to-validity}.

\textbf{(h) Compare with the state-of-the-art techniques.} Comparison with the state-of-the-art techniques is essential to place one's work in the literature \cite{compare-sota}. Table \ref{table:context-validation-target} records all the techniques used by our primary studies for their validation. We see that validation targets could be different based on working context. However, a few techniques (e.g., BM25 \cite{bm25}, Rocchio \cite{rocchio}, RSV \cite{rsv}) were used for validation across multiple code search contexts. Fig. \ref{fig:popular-validation-target} shows the frequently used validation targets, which could be a good starting point for the future work.

\textbf{(i) Use appropriate statistical tests to answer research questions.} To test any null hypothesis, statistical tests are essential \cite{statistical-test,threats-to-validity}. According to our investigation, 51\% (36/70) of our primary studies conduct one or more statistical tests. Table \ref{table:statistical-test} outlines the use cases of various tests whereas Figures \ref{fig:statistical-test}, \ref{fig:wc-stat-test} focus on their uses by the primary studies. From Fig. \ref{fig:statistical-test}, we see that three non-parametric tests such as Wilcoxon Signed Rank (WSR), Mann-Whitney Wilcoxon (MWW), and Cliff's delta, and one parametric test namely Student's t have been frequently used by the primary studies. Non-parametric tests are used when the probability distributions of test samples are not assumed whereas parametric tests should be chosen when the samples have a normal distribution \cite{param-non-param}. All these items could be a reference point for the future work for any statistical analysis. 

\textbf{(j) Discuss limitations and threats to validity.} Threats or limitations in a research
could emerge due to various choices made by the researchers \cite{threats-to-validity,threat-to-validity}. According to our investigation, primary studies in query reformulations suffer from eight different challenges or limitations (Table \ref{table:challenge}). Interestingly, several challenges (e.g., lack of generalizability (CH3), injection of noise into the original query (CH1), human bias (CH5)) are common in both local and Internet-scale code searches (Fig. \ref{fig:compare-challenge}). As done by earlier work, future work should also mitigate and discuss their challenges and threats, which is key to advance the current state of knowledge.

\textbf{(k) Provide replication package for reuse.} Among 70 primary studies, we found that 24 studies provide complete replication package including dataset and executable code, 13 studies offer only partial information (e..g, dataset only), whereas 33 studies do not provide any replication package (Appendix \ref{appendix:A}). Replication of any research work could be the key to its validity and long-term impacts \cite{replication,replication2}. Thus, making replication package publicly available is one of the recommended practices to foster extended work and reuse.

\begin{frshaded}
	\noindent
	\textbf{Summary of RQ$\mathbf{_6}$:} Although query reformulation practices could be significantly different between local and Internet-scale code searches, there are a set of practices that have been well-accepted and recommended by the community. For example, collecting search queries from professional developers and selecting both open and closed-source systems are recommended for designing an experiment. Choosing appropriate granularity levels and preprocessing steps is essential to index the corpus. The same reformulated query should be executed against multiple search engines to ensure a fair evaluation.
	%and appropriate performance metrics will ensure a sound evaluation. 
	On the other hand, comparison with state-of-the-art techniques and conducting statistical tests are required to validate a proposed technique. Furthermore, threats and limitations should be discussed and the replication package should be shared publicly to advance the current state of knowledge.
\end{frshaded}

\section{\textbf{Answering RQ$\mathbf{_7}$:} Future Research Directions}
\label{sec:future-work}
 Our selection criteria and personal knowledge led to a set of 70 primary studies being identified since 1998 supporting code search through query reformulations.
%To date, 70 primary studies supporting code search through query reformulations have been conducted. 
Among them, 40 studies focus on local code searches that involve software change tasks (\eg\ bug localization, feature location, concept location). On the other hand, 30 studies reformulate queries to support the Internet-scale code search.
%provide query reformulation supports for Internet-scale code search. 
Despite these significant number of studies, we believe that there is still room for future work and many novel dimensions are yet to be explored. Based on our empirical and qualitative analyses, we present a list of future research directions that can
improve query reformulations and support code search as follows.
%on query reformulation targeting the source code search as follows:  

\subsection{Adapt traditional keyword selection algorithms to bug reports for bug localization}
Information Retrieval (IR) has been adopted by at least 20 Software Engineering tasks including bug localization \cite{refoqus}. Bug localization is a type of local code search where bug reports serve as a major source of queries.
A few recent studies \cite{parninireval,bias-bug-loc-lo,bl-misoo} point out potential biases and limitations in this local code search. According to them, IR-based localization is only good when the bug reports contain localization hints (\eg\ program entity names). However, these empirical studies use the whole texts of a bug report as a query, which are often noisy, and overlook the potential of optimal search queries. \citet{cmills-icsme2018} recently conduct a large-scale empirical study using Genetic Algorithms and present  positive evidence for both bug reports and IR-based localization. They suggest that the bug reports contain sufficient keywords to return the faulty source code at high ranks using Information Retrieval.
%the top-most position of the result list.
 In particular, the IR-based localization can succeed 67\%--88\% of the times even if their bug reports do not contain any localization hints \cite{cmills-icsme2018}. Similar finding has been reported by a recent work of \citet{emse2021masud}.
 Thus, the real challenge becomes the extraction of right search keywords from a bug report regardless of its localization hints.
  %for bug localization, \ie\ local code searches. 
  
\textbf{Future work.} Appropriate term weighting algorithms can play a major role in selecting the search keywords. 
To date, existing 
term weighting studies exploit each term's frequency \cite{refoqus,kevic,tfidf,rocchio}, statistical relations and syntactic dependencies \cite{fse2018masud,saner2017masud}, or occurrence probability \cite{sisman,sisman-jsep,carmelbook} as a proxy to its relative importance (a.k.a., term weight).
%literature \cite{,,} employ three types of algorithms such\emph{frequency-based}, \emph{graph-based} and \emph{probabilistic}--for keyword selection. 
However, many of these algorithms were designed for regular texts in the Information Retrieval domain and 
%Each of these term weighting algorithms was borrowed from Information Retrieval domain that mostly deals with regular texts. Hence, these algorithms 
might not be well-configured for bug reports containing a mix of regular texts and structured entities (\eg\ stack traces, test cases, program elements, commit diffs). Future studies can focus on better adapting the IR algorithms to these structured entities. For example, rather than treating these entities as regular texts, they can be represented as graphs and thus can be better leveraged with graph-based algorithms (e.g., PageRank algorithm \cite{pagerank}) for keyword selection.

%For example, besides frequency, contextual information (e.g., keyword position, scope, dependencies) can be leveraged to determine the salience of a keyword, which can lead to a high-quality query for IR-based bug localization.
%and to leverage their strengths in keyword selection during bug localization.

%Thus, more sophisticated and efficient term weighting algorithms are required to improve the search queries in IR-based bug localization. 

\subsection{Design practical GA-based query reformulations for IR-based bug localization} 
\citet{cmills-icsme2018} demonstrate that Genetic Algorithms (GA) are capable of generating optimal or near-optimal queries from a bug report (that can localize faulty code) when the ground truth information is provided. Unfortunately, in practice, the ground truth might not be known when a search query is constructed to find a bug. Thus, the ground truth should not be a part of fitness calculation when designing a GA-based query reformulation technique. In other words, we need a fitness function that can identify the better one from two different search queries without needing their ground truth, which is a major challenge.
 %That is, designing an appropriate \emph{fitness function} for query reformulation with Genetic Algorithms is a major challenge. In particular, given two candidate search queries, the fitness function should be able to identify the better one without executing them. 
 
 \textbf{Future work.} Several studies \cite{qperf,refoqus,qperf-mills-tosem,trconfig,ase2017masud} make use of query difficulty metrics (\eg\ specificity, coherency \cite{qsurvey}) and machine learning algorithms to predict a query's success against the search engines.
  %predict the best one from a list of queries. 
  However, these metrics have non-linear relationships with a query's real performance according to recent studies \cite{qperf-mills-tosem,emse2021masud}.
  %with the query quality, which warrants special care during their use. 
  One possible way to address this is to formulate the query construction as a multi-objective optimization problem, as demonstrated by \citet{manq-qr}. They design a complex fitness function that attempts to optimize 15 objectives while reformulating a bug report into an optimal query for IR-based bug localization.
 % and thus require a machine learning algorithm to be effective.
  %and generally work in collaboration with machine learning algorithms. 
% Thus, they might not be an ideal choice to design a fitness function. 
Future work can design one or more fitness functions that can be used in practice for GA-based keyword selection from bug reports.
For example, difficulty of a query could be a part of its fitness calculation where the easy queries will be prioritized over the difficult ones during evolutionary search.

%Future works should focus on designing more appropriate fitness functions since Genetic Algorithm has the potential for delivering the optimal queries from the bug reports for bug localization \cite{cmills-icsme2018}. 
%GA-based query reformulation from the bug reports. Such a fitness function could be considered as the \emph{holy grail} for the GA-based query construction. 
%IR-based localization has not been widely adopted by the software practitioners due to its inherent limitations (e.g., low accuracy) \cite{parninireval}. However, we believe that IR-based bug localization equipped with GA-based optimal queries 
%can make a difference in this regard. 

%be a preferable alternative to the developers as opposed to the traditional costly localization processes.  

\subsection{Leverage word embeddings in query reformulation \& code search}
\label{sec:fw3} 
Several primary studies \cite{pwordnet,thesaurus-qr-idcs,hillicse09,multi-rec} use English language thesauri such as WordNet \cite{wordnet} to expand their query with synonyms or semantically similar words.
However, \citet{semantictool} suggest that the same word can have two different meanings
when used in source code and when used in regular texts. Since WordNet is based on regular English texts, its suggested keywords might not be appropriate for source code search.
%Thus, English language thesaurus might be neither appropriate nor sufficient for query expansion intended for source code search. 
Thus, several studies \cite{ccmapping,infer,swordnet,wordsim} construct software-specific thesauri from the mining of software repositories (e.g., open-source projects, programming Q\&A site).
%(\eg\ source code, Stack Overflow Q\&A threads). 
However, construction and update of these thesauri are costly and their effectiveness in the query reformulation is not well tested. 
Recently, several studies \cite{wembedding,iman-tse,icsme2018masud,spain-paris} make use of word embedding technology
to determine semantic similarity or relevance between any two software specific words. 
A few of them \cite{iman-tse,icsme2018masud} use embeddings to reformulate queries for code search and report positive findings.
However, the true potential of word embeddings in query reformulations might not have been explored yet.

%These studies also leverage the word embeddings 
%in reformulating queries for source code search, and report positive evidence.

\textbf{Future work.} Word embedding technology approximates the meaning of a word using a high dimensional numeric vector \cite{word2vec}. The vector places each word as a single coordinate within a high dimensional semantic space (a.k.a., semantic hyperspace).  Such a geometric approximation of word semantics can be leveraged for various tasks in both \emph{local} and \emph{Internet-scale code searches}. For example, by placing all keywords from a document (e.g., bug report, feature request, source code) in the semantic hyperspace, important keywords could be detected using various geometric theories (e.g., Hopkins statistics \cite{hopkins}). By visualizing the semantic hyperspace using appropriate tools (e.g., t-SNE \cite{t-SNE}), we could make a better sense of (a) relevance between a query and its target code, or (b) the difference between observed behaviour and expected behaviour of a reported bug. 

%In particular, various geometric theories could be applied to the semantic space 
%reduces various natural language processing tasks (\eg\ synonym detection, semantic distance calculation) into simple algebraic or geometric operations. Thus, we believe that this technology has lots to offer to various text retrieval tasks including query reformulation, code search, and even bug understanding. For example, the semantic hyperspace constructed using word embeddings could be useful to (a) understand erroneous and expected software behaviours from a bug report \cite{observed}, 
%might even encourage various geometric theories into the text processing tasks of Software Engineering (\eg\ observed/expected behaviour detection \cite{observed}).
%In essence, it  
%not only in query reformulation or code search but also in other text retrieval tasks of Software Engineering. In particular, 
 %The semantic hyperspace constructed from the embedding vectors of a large vocabulary might also help design (1) 
%(b) design appropriate fitness functions for the GA-based query reformulations \cite{cmills-icsme2018} and (c) design more effective keyword selection algorithms to overcome the vocabulary mismatch problems.

\subsection{Complement term weighting algorithms with contextual information} 
Determining importance of a term within a body of texts (\eg\ bug report, source document) has been 
%long been recognized 
as a major challenge \cite{kevic,twkraft}. TF-IDF \cite{tfidf} is a term weighting algorithm that has been extensively used both in Information Retrieval and in Software Engineering. The algorithm determines a term's importance without considering its contextual information (\eg\ surrounding terms). However, a term's semantics are often determined by its contexts \cite{wordsim,wordnet}. Besides, several primary studies demonstrate the benefits of incorporating context in their term weighting algorithms. To date, several contextual items such as spatial code proximity \cite{sisman}, positional relevance \cite{sisman-jsep,fse2018masud}, term co-occurrences \cite{hillicse09,saner2017masud,shepherd,multi-rec,ase2017masud}, syntactic dependencies \cite{saner2017masud}, time-awareness \cite{time-aware-term-weighting}, and structural awareness \cite{fse2018masud,twkraft,twkraft-jsep} have been incorporated in term weighting. However, their use was ad hoc and their roles in query reformulations were not investigated exhaustively.

\textbf{Future work.} While context can deliver useful information in term weighting, it should be used carefully to avoid noise.
Future work can investigate how multiple contexts can be incorporated in the term weight calculation.
%Future work can investigate how combining these contextual items might benefit the 
%traditional term weighting algorithms such as TF-IDF \cite{tfidf}.
Genetic Algorithms can be used to optimize the relative weights of multiple contexts during their combination \cite{wordsim}. Machine learning can also be used to determine complex, non-linear relationships between these contexts and a term's importance \cite{kevic}.

%, machine learning algorithms could be used to design even more complex, non-linear relationships between these contexts and a term's importance. 

\subsection{Leverage Stack Overflow in query reformulation for local code search} 
%Despite innovative attempt to represent queries as code examples \cite{query-by-example} or test cases \cite{testcase-cs-ist,semantic-cs,stolee2016}, 
Software developers frequently use natural language queries to find relevant code on the web \cite{sourcerer,koderlog}. 
However, their queries are often short and thus fail to
%Each of their queries contains a few keywords that might not be able to 
capture their information need \cite{koderlog}.
%They choose a list of keywords that best describe a programming task and then submit them to a search engine (\eg\ GitHub code search \cite{github-code-search}). Unfortunately, these queries do not work well since they often lack necessary information required for the task. 
Existing findings \cite{koderlog,icsme2018masud,iman-tse, lemos-type-qr} suggest that inclusion of relevant API classes or methods in a query can improve its chance of retrieving the relevant code.
%consistently improves the code search performance. 
A few studies \cite{rack,cocabu,tsc2016} analyze the Q\&A threads from Stack Overflow 
to expand a natural language query and to support the Internet-scale code search. 
For example, \citet{rack} analyze co-occurrences of query keywords and API classes in question title and accepted answer respectively
%of query keywords in the title of a question and the API classes in the corresponding accepted answer, 
and expand a query with frequently co-occurred API classes. However, the true potential of Stack Overflow in query reformulation might not have been explored yet.

\textbf{Future work.} Several studies \cite{swim,neural-query-expansion,qr-log-mining} perform machine translation where they translate a natural language query into relevant API classes or methods. However, they make use of proprietary search logs (e.g., Bing logs) that might not be publicly available for reuse. Interestingly, Stack Overflow can be leveraged to generate such logs where keywords and API classes can be extracted from question titles and accepted answers respectively. Given the size of Stack Overflow, millions of logs can be captured at low costs. Then the machine translation techniques based on these logs can be used to suggest (a) relevant API classes for a new feature request \cite{librec} or (b) potentially buggy API classes for a bug report.

\subsection{Minimize query worsening during reformulation}
Query reformulations have both benefits and costs. Existing studies suggest that automated query reformulation
might improve the performance of a query up to 20\% during code search \cite{time-aware-term-weighting, emse2018masud}. However, several studies \cite{hillicse09,flt-model-qr,query-drift} also question the idea of a complete automation in reformulation.
%in query reformulation. 
Without human intervention, automated reformulations might add noisy keywords that could drift a query away from its original topic \cite{query-drift}. Thus, we need a mechanism to 
%such solutions that can 
maximize the benefits and minimize the costs of automated query reformulations.

\textbf{Future work.} Several studies \cite{qperf,refoqus,ase2017masud,emse2021masud} use query difficulty metrics and machine learning to automatically identify the difficult queries and then reformulate them.
%\citeauthor{refoqus} and colleagues \cite{qperf,refoqus} determine the quality of each given query using 28 query difficulty metrics \cite{qsurvey} and reformulate the poor queries to improve concept location tasks. Later studies \cite{ase2017masud,emse2021masud} also adopt the query difficulty metrics to separate good queries from poor ones using machine learning. However, any classifier could be prone to false-positives, and any wrong classifications will lead to costly, inappropriate reformulations to a search query.
%. A few studies \cite{refoqus,ase2017masud} employ query difficulty metrics and machine learning to deliver the best reformulation for a given query. In order 
%To minimize the cost, \citet{ase2017masud} even return the original query using machine learning if the reformulated queries are not good %enough. 
However, unless the models are 100\% accurate, there will be always false-positive results. Thus, the risk of query worsening due to reformulations is still prevalent. According to \citet{trace-query-transformation}, human developers might perform well in removing irrelevant terms from a search query due to their cognitive abilities. On the contrary, automated techniques perform well in adding the relevant terms to a query \cite{trace-query-transformation}. Thus, query reformulation task can be considered as a multi-step changes to a query (e.g., query expansion, query reduction) where human and machine capabilities are combined together.
%as suggested by earlier studies, human cognitive abilities can be leveraged 
%should be incorporated into the query reformulation activities.
%%we thus believe that human cognitive power could be leveraged in this case. According to \citet{trace-query-transformation}, human developers might perform well in removing irrelevant terms from a search query, but could perform poorly in adding the new relevant terms. Relevant keywords could be hidden within thousands of identifier names (\eg\ classes, methods) of a system's codebase. Future work can combine the strengths of both human developers and automated tools to minimize the cognitive burdens on the developers and to reduce the costs of inappropriate query reformulations. 

\subsection{Apply pseudo-relevance feedback (PRF) to Internet-scale code search}  
Relevance feedback is an important mechanism to reformulate a search query \cite{rf-revisited, relevance-model}.
However, capturing developers' feedback on queries by analyzing their results could be costly and sometimes impractical.
Many primary studies \cite{refoqus,ase2017masud,ase2016masud,sisman} thus use pseudo-relevance feedback (PRF) (Section \ref{sec:background}) as a feasible choice to reformulate their queries. PRF has been reported to improve the queries during code search \cite{ase2017masud,refoqus}. However, the majority of studies except a few \cite{cexchange,qecc-intent} focus on local code searches such as concept location, bug localization, and feature location.
%, not many studies except a few  use PRF in the Internet-scale code search.
% has not been well studied. 
%That is, they naively assume the top K documents
%the Top-K documents 
%retrieved by a given query as relevant and then extract important keywords from them 
%to reformulate the query.
%for query reformulation. 
%These approaches have been reported to improve upon the given queries \cite{expand-reduce,refoqus}. However, such a feedback mechanism might not help much if the given queries are already poor \cite{fse2018masud}. Then the retrieved documents are likely to be \emph{irrelevant}, which will lead to noisy reformulated queries. While PRF has been frequently used in 

\textbf{Future work.} In the Internet-scale code search, the results are retrieved from thousands of open source projects, which could be noisier and less homogeneous than that from local code search. PRF naively assumes the top few results retrieved by a query as \emph{relevant} \cite{prf}, which might not apply to the Internet-scale search. Thus, future work can adapt the PRF mechanism to mitigate the noise and to support the query reformulations in Internet-scale code search.

\subsection{Leverage PageRank algorithm in term weighting and source code retrieval}
\citeauthor{saner2017masud} adapt PageRank \cite{pagerank}, a popular algorithm from Information Retrieval,
to reformulate queries in several Software Engineering tasks
%, and use it for search query reformulation in various problem contexts 
such as bug localization \cite{fse2018masud}, concept location \cite{saner2017masud,ase2017masud} and Internet-scale code search \cite{icsme2018masud}. Other studies \cite{portfolio,sourcerer-coderank,pagerank-link-analysis} make use of PageRank algorithm to rank code snippets during their code search. However, the true potential of PageRank algorithm in query reformulation and code search might not have been explored exhaustively.

\textbf{Future work.} PageRank operates on a graph with nodes and edges, implements a notion of voting, and then identifies the most important nodes from the graph using a recursive computation \cite{rada,blanco}. Source code 
has an unrestricted vocabulary \cite{devanbu-fse} and is full of structures, entities, and
dependencies among them (e.g., control dependency, data dependency), which can be represented as a graph \cite{ase2017masud}.
Thus, unlike TF-IDF, PageRank is a natural choice for analyzing the source code. While there have been a few attempts using static relationships \cite{ase2017masud,fse2018masud}, future work can use PageRank algorithm to exploit the code structures to support query reformulations and code search. For example, PageRank can be used to (a) detect meaningful terms from PRF documents for query expansion and (b) shortlist terms from source code for corpus indexing.

% along this line,  further work is warranted to understand the true potential of PageRank algorithm in various IR-based code retrieval tasks (e.g., ).

%Then they use these patterns and natural language features (\eg\ n-grams, POS tags) to detect the missing components (\eg\ S2R) in a bug report.    
            
\subsection{Standardize query reformulation in IR-based bug localization}
\citet{antoniol} first use Vector Space Model (VSM) to recover traceability links. 
\citet{buglocator} later refine VSM as rVSM and incorporate past bug reports in the IR-based bug localization. \citet{saha} make use of structures both from bug reports and from source code documents to localize the bugs. \citet{brtracer} boost up the bug-proneness score of a source document based on the stack traces found in a bug report. \citet{kak-version-history} and \citet{locus} incorporate version control history in the IR-based bug localization. Finally, \citet{versionhistoryjsep} incorporate five major items -- past bug reports, structures, stack traces, version history, and author history -- from the literature, and outperform the earlier approaches on IR-based localization. Thus, existing literature often adopt an incremental approach and  include more and more external artifacts in their approaches. Although these artifacts have positive impact upon the localization performance, their inclusion makes the proposed approaches \emph{less scalable} and \emph{less usable} Besides, these approaches become subject to external dependencies, which is not an ideal choice from the design point of view.
Such limitations might partially explain the reluctance of software practitioners in adopting the IR-based bug localization techniques in their workflow \cite{bias-bug-loc-lo,tse-bug-dev-study,parninireval,parnindebug}.

\textbf{Future work.} Several approaches equipped with query reformulation \cite{refoqus,fse2018masud,ase2017masud,observed} make an effective use of primary resources available to practitioners (e.g., \emph{bug report} and \emph{source code}), and  
localize the bugs with a competitive accuracy and at low cost. Given that Information Retrieval has been adopted by at least 20 software engineering tasks, query reformulation has the potential to significantly improve them.
%can be an integral part of them. 
For example, several tasks such as bug localization \cite{observed}, duplicate bug report detection \cite{oscar-duplicate}, bug triaging and bug report summarization can be benefited from query reformulations.

%Thus, we believe that query reformulation can be an important part of at least 20 IR-based 
%Software Engineering tasks including  Future work can investigate the impacts of query reformulations upon these 20 tasks and 
%develop more appropriate tools to supports them.
%and incorporate query reformulations in these IR-based solutions.

\begin{frshaded}
	\noindent
	\textbf{Summary of RQ$\mathbf{_7}$:} Despite significant research, there is still room for further work in automated query reformulations and code searches (e.g., bug localization, concept location, feature location, Internet-scale code search).
	%a significant number of existing studies, many works are yet to be done on automated query reformulations intended for code search or software artifact retrieval. 
	Existing findings suggest that up to \textbf{88\%} of bug reports contain appropriate keywords that make good queries and can find the bugs using Information Retrieval methods.
	%\textbf{contain} the appropriate keywords to localize the buggy source documents. 
	Unfortunately, many existing techniques are not effective enough to identify these keywords. The selection of keywords could be further improved by adding more contexts (e.g., time-awareness), designing an appropriate fitness function (for GA-based solutions), leveraging the structures from source code, or using neural language modeling (e.g., word embeddings). The adverse effects of automated query reformulations could be mitigated by combining both human cognitive power and tools' strengths. Large software repositories (e.g., GitHub) and programming Q\&A site (e.g., Stack Overflow) have a lot to offer to the research of query reformulations and code search. They store invaluable technical knowledge curated by a large technical crowd, which can be leveraged to support not only code searches but also other software development and maintenance activities.
	%, which warrants further studies. \textbf{Keyword selection} could be improved by adding more contexts or neural language modelling such as word embeddings. \textbf{Graph-based} term weighting approaches (\eg\ PageRank) have more potential than frequency-based alternatives (\eg\ TF-IDF) for keyword selection from the source code. \textbf{Genetic Algorithms} equipped with appropriate fitness functions have also \textbf{enormous potential} for improving IR-based bug localization. \textbf{Stack Overflow} could greatly help in improving the search queries for Internet-scale code search. Future works should also target to \textbf{minimize} both the costs of automated query reformulations (\eg\ query worsening) and the adverse effects of 
	%pseudo-relevance feedback (\eg\ topic drift). 
\end{frshaded}

\section{Threats to Validity}\label{sec:threats}
We identify several threats to the validity of our findings in this work. While we did our best to mitigate the majority of these threats, a few of them can be addressed in the future surveys on automated query reformulations. We classify our identified threats into several standard categories and discuss them as follows.

\subsection{Threats to internal validity}
These threats are associated with experimental or internal errors and human biases \cite{wordsim}. Selection of primary studies and their classification into certain categories could be a source of such threats. However, we take careful steps to mitigate them. 

First, we choose 70 primary studies using a well established, systematic approach (Section \ref{sec:methodology})
that involves (a) generating keywords from research questions using PIO criterion, (b)
extensive searching against 11 widely used publication databases (e.g., IEEE Xplore, ACM Digital library), and (c) multi-level filtration based on specific criteria. We first collect a large set of 2,970 studies that match our search keywords, and then carefully select 70 studies from them using six levels of noise filtration. Despite these careful, time-consuming steps, relevant studies might be omitted due to sub-optimal queries or the limitations of 
%sub-optimal quality of our selected keywords and 
 search engines provided by the publication databases (Section \ref{sec:study-selection}). 
To mitigate this threat, we also added a few studies that were relevant (according to our prior work experience) but were accidentally missed by the study selection process. However, we still cannot rule out the possibility of missing a few relevant studies on the topic of interest.

Second, we use the Grounded Theory approach (Section \ref{sec:gtheory}) 
to classify the primary studies based on their adopted methodologies, algorithms (Section \ref{sec:qr-techniques}), and experienced challenges (Section \ref{sec:challenge}).
%in the classification of the primary studies based on either their adopted methodologies/algorithms (Table \ref{table:study-category}) or their experienced challenges (Table \ref{table:challenges}). 
Grounded Theory has been a popular choice for qualitative analysis in 
Software Engineering  research for decades \cite{gt-survey,observed,grounded-theory}. Thus, the threats identified above might be mitigated.  

Third, local code search and Internet-scale code search differ in their working context and end goals. While former targets software change tasks (\eg\ bug localization, feature location), the latter one focuses on code reuse or general-purpose programming tasks. One might argue about the design of our survey since the query reformulation studies from both searches were captured. However, according to our investigation, query reformulation approaches from these two searches have a significant overlap in their reformulation techniques (Table \ref{table:context-method-overlap}, Fig. \ref{fig:compare-methodology}), evaluation methods (Figures \ref{fig:compare-eval-type}, \ref{fig:compare-metric}), and even strengths or limitations (Fig. \ref{fig:compare-challenge}). 
Thus, analysis restricted to only either local code searches or Internet-scale code search misses a significant number of strongly related studies from the literature. Besides, collecting both set of studies allows us to 
compare and contrast between them, which has led us to meaningful insights (Section \ref{sec:qualitative}) and novel ideas for future work (Section \ref{sec:future-work}). 
%and derive meaningful insights for future works. 
There also exists a large body of work on query reformulations in the Information Retrieval domain \cite{qsurvey,carmelbook}. Many of our primary studies adapt their ideas from them. However, since we deal with Software Engineering problems such as code search, those studies from IR literature were not included in our survey.   

Fourth, our systematic review is restricted to only such studies that deliver either automated or semi-automated
 tool supports to reformulate queries for code search. 
 %in the query reformulation for code searches. 
 That is, it does not include empirical studies, developer surveys and sometimes even the simple incremental works. Such a choice was made to construct a homogeneous set of primary studies that solve a single research problem, i.e., automated query reformulation to support code search.

\subsection{Threats to external validity}
These threats relate to the generalizability of any reported findings \cite{wordsim}. Our primary studies are carefully chosen from a large population of 2,970 results that were retrieved from 11 popular publication databases (Fig. \ref{fig:study-selection}). Thus, these primary studies are likely to represent the core studies on query reformulation supporting code search in the literature.
%query reformulation studies on source code search from the literature. 
Consequently, our findings and insights derived from these studies could also generalize for the remaining, accidentally missed studies from the literature. Furthermore, we leverage the Grounded Theory approach for our qualitative analysis, which has been a standard practice for years \cite{observed,missing-comp,grounded-theory,gt-survey}. 
Thus, the threats to external validity might be mitigated.

\subsection{Threats to construct validity}
These threats relate to the appropriateness of evaluation methodology used in a study \cite{wordsim}. Since our selection of primary studies involves certain quality control mechanism, such threats may arise. We select 70 studies through a well-established approach of extensive database search and well-documented filtration criteria (Fig. \ref{fig:study-selection}). Furthermore, we control the quality of these studies by asking 10 common questions about their quality (Section \ref{sec:assessment}). While the majority of these studies pass the quality check, the remaining few do not meet certain quality standards (\eg\ lack of rigorous evaluation and validation). However, we retain them in our analysis due to their strong relevance to our survey topic-- query reformulation in source code search. It also should be noted that three studies \cite{hard-trace-ase,trace-query-transformation,qr-service-discovery} on traceability link recovery were also included in our survey due to their solid contributions in query reformulation algorithms.  

\subsection{Threats to conclusion validity}
These threats arise when the relationships between two variables are not backed up by solid evidence or reliable data \cite{observed}. In our qualitative analysis (Section \ref{sec:qualitative}), we compare and contrast between local code searches and Internet-scale code search and make several statements about their characteristics, methodologies, strengths, and limitations. 
%We also contrast among three keyword selection approaches, and suggest the best course of actions during query reformulation/code search.
We also outline several future research directions based on our analyses.
Our claims and suggestions were backed up not only by substantial analytical data (RQ$_1$, RQ$_3$) but also by empirical evidence (RQ$_4$). 
%(\eg\ Tables \ref{table:study-category}, \ref{table:challenges}, \ref{table:study-eva-metrics}) that are derived from Grounded Theory.
Thus, threats to the conclusion validity of our systematic survey might also be mitigated. 

%Although local code searches and Internet-scale code search differ in their working contexts and end goals, they share a lot.      
%In particular, we assume that the given queries are written using natural language texts and  
%the developers issue natural language queries for local or Internet-scale code search,  

\begin{table}[!t]
	\centering
	\caption{Comparison between existing surveys/reviews and our systematic reviews}\label{table:related-work}
	\vspace{-.35cm}
	\resizebox{5.4in}{!}{%
		\begin{threeparttable}
			\begin{tabular}{l|p{1in}|l|c|c|c|>{\raggedright\arraybackslash}p{1.5in}}
				\hline
				\textbf{Survey} & \textbf{Topic} & \textbf{FQR} & \textbf{Period} & \textbf{\#Studies} & \textbf{SLR} & \textbf{Dimension}\\
				\hline
				\hline
				\citet{qsurvey} & Information retrieval & Yes & 1993-2009 & 27 & No & Performance limitations \\ 
				 \hline
				 \citet{dit-flt-survey} & Feature location & No &  1992-2011 & 89 & Yes & Type of analysis, user input, data sources, output, programming language support, evaluation methods, and software systems\\
				 \hline
				 \citet{rubin-toronto-flt-survey} & Feature location & No & 1995--2010 & 24 & No & Underlying technology, program representation, input, and user interactions  \\
				 \hline
				 \citet{bl-survey-wong-tse} & Fault localization & No & 1977--2014 & 331 & No & Underlying technology, subject programs, evaluation metrics, and critical aspects \\
				 %\hline
				 %\citet{br-survey-scis} & Bug report analysis & No & - & - & No & Bug report optimization, bug triage, and bug fixing \\
				 \hline
				 \citet{grundy-cs-survey} & Internet-scale code search & No & 2002--2020 & 81 & Yes &  Code search technique, evaluation, and publication trends  \\
				 \hline
				 Our systematic review & Concept location, feature location, bug localization, and Internet-scale code search & Yes & 2004--2021 & 70 & Yes & Underlying methodology, evaluation design, publication trends, limitations, and challenges \\
				 \hline
			\end{tabular}
		   \centering
		   \textbf{FQR}=Focus on query reformulation, \textbf{SLR}=Use of systematic literature review
		\end{threeparttable}
	}
\vspace{-.5cm}
\end{table}

\section{Related Work}\label{sec:related}
Source code search has two major steps -- (a) query construction and (b) retrieval of relevant code \cite{stolee2014,refoqus}.  
There have been several surveys  \cite{rubin-toronto-flt-survey,dit-flt-survey,bl-survey-wong-tse} that examine the evidence of existing researches on local code searches (\eg\ feature location, bug localization).
%categorize them into certain taxonomy, and then discuss the future research directions and open issues.
They mostly focus on the studies that deal with the second step of code search, i.e., retrieval of relevant code.
Unfortunately, to the best of our knowledge, there exists no systematic survey 
that deals with query reformulation approaches adopted in source code search.
The closely related work is a similar survey \cite{qsurvey} from the Information Retrieval (IR) domain. Although IR-based methods have been widely adopted by at least 20 Software Engineering tasks \cite{refoqus,cmills-icsme2018}, they experience unique challenges (Section \ref{sec:challenge}), which warrants further investigation \cite{parninireval,trconfig}. Our survey identifies the existing work on query reformulations supporting Software Engineering tasks, categorize them based on their underlying methodologies or limitations, and then discusses their
open issues.

%future works and open issues.    
 
\citet{dit-flt-survey} first provide a comprehensive survey of 89 studies on feature location techniques (FLT). They analyze each of these studies using seven different dimensions : (a) type of analysis, (b) user input, (c) data sources, (d) output, (e) programming language support, (f) evaluation method, and (g) software systems used. Based on an analysis, they also provide a detailed taxonomy on the FLT studies. \citet{rubin-toronto-flt-survey} perform a similar survey using 24 studies on feature location techniques. They provide necessary guidelines for software practitioners to choose the right feature location technique for their tasks. \citet{bl-survey-wong-tse} analyze 331 studies on software fault localization and provide a detailed classification of their studies. Unlike the above work, our survey deals with not only feature location or bug localization but also other types of code searches such as concept location and Internet-scale code search.
More specifically, we deal with a \emph{cross-cutting concept} namely query reformulation that concerns each of these code search activities.
%but also other techniques that target bug localization, concept location and Internet-scale code search.
%published between 1977 to 2014, and provide a detailed classification on the existing body of works. Unlike them, we capture a %broad range of studies that address the problem of source code searches in Software Engineering. 
Our survey also analyzes 70 primary studies on query reformulations and categorize them based on several aspects such as query reformulation methodology (RQ$_1$), evaluation method (RQ$_2$), and their experienced challenges or limitations (RQ$_3$).

%attempts to analyse and categorize the query reformulation approaches that are employed in various code searches.

\citet{br-survey-scis} present a survey on bug report analysis where they classify the existing work on bug-report optimization (\eg\ bug severity prediction), bug-report triage (\eg\ bug report assignment), and bug-fixing (\eg\ fixing time prediction). Similarly, we also deal with the primary studies that perform bug report analysis. However, our focus was to capture the studies that accept a bug report as an input, construct appropriate queries from the input, and then localize the bug using these queries.

%and then deliver appropriate search queries for bug localization, feature location or concept location.

A few other studies \cite{observed,cmills-icsme2018,oscar-duplicate,grundy-cs-survey} also investigate the existing work on query reformulations supporting code searches and other Software Engineering tasks. However, their analysis was 
limited and restricted to either local \cite{cmills-icsme2018,observed} or Internet-scale code search \cite{grundy-cs-survey}.
%limited and intended for local code searches only. 
Many of them also perform traditional literature review rather than a systematic review \cite{slr}, which
%On the hand, the traditional literature review 
might not guarantee that all eligible primary studies were captured from literature.
% due to a lack of structure in the data collection and subsequent analysis.
On the other hand, we conduct 
%Unlike a normal literature review, the 
a systematic literature review that was guided by a set of research questions and well-crafted search queries (e.g., PIO criterion). The quality of our primary studies was also controlled by clear exclusion/inclusion criteria and they were collected from 11 popular publication databases. Thus, our goal was to establish a comprehensive understanding of the existing research on query reformulations targeting code search. We also 
%Unlike them, we conduct a systematic literature review by (a) asking important research questions, (b) carefully selecting the primary studies from 11 popular publication databases, and (c) 
perform qualitative analysis using the Grounded Theory approach \cite{grounded-theory}, a widely used qualitative technique, to answer several of our research questions. No earlier survey addresses the same topic in this scale, which makes our work novel. Table \ref{table:related-work} also demonstrates how our survey fills in the gap of literature by focusing on query reformulations in the context of source code searches.

\section{Conclusion}\label{sec:conclusion}
Software maintenance can cost up to 80\% of total budget in modern software development. Searching for the code of interest (e.g., buggy code, reusable code example) is an integral part of software maintenance.
Developers often choose a few important keywords capturing their information need and attempt to find the code of interest either from a local software codebase or an Internet-scale code repository (\eg\ GitHub). Unfortunately, existing evidence suggests that the developers often perform poorly in choosing their keywords. As a result, they spend a significant amount of their time manually reformulating queries and analyzing the corresponding search results. Towards this goal, a number of studies attempt to support the developers through automated query reformulations. In this article, we conduct a systematic literature review using 70 primary studies on automated query reformulations intended for source code search. We summarize our quantitative and qualitative findings below and answer our research questions briefly as follows.
\begin{itemize}[noitemsep,topsep=1pt]
	\item \textbf{RQ$_1$:} Existing approaches for query reformulation adopt eight major methodologies
	including term weighting, relevance feedback, semantic relations, thesaurus lookup, and data mining to reformulate their queries and to support code searches in various working contexts.
	
	%/algorithms -- \emph{term weighting}, \emph{term-query co-occurrences}, \emph{thesaurus}, \emph{ontology model}, \emph{search log mining}, \emph{machine learning} and \emph{ad hoc heuristics}. 

	\item \textbf{RQ$_2$:} Existing studies evaluate their approaches through re-enactment, developer surveys, or using both. They use a total of 30 performance metrics including the popular ones (e.g., Hit@K, MAP).
	However, developer involvement is generally low, and the number of queries used to evaluate queries in the Internet-scale code search is often small. About 30\% of the previous approaches were selected for comparison by the later approaches.

	%employ two different evaluation methods -- \emph{re-enactment based evaluation} (50\%) and \emph{developer-oriented evaluation} (43\%). On average, each primary study makes use of \emph{two} subject systems and at least 70 search queries for the evaluation/validation.

	%About \textbf{54\%} of our primary studies use empirical evaluation, 33\% perform qualitative investigation, and only 9\% of the studies use both.
	%About 80\% of the studies that perform qualitative evaluation involve less than 25 developers each.
	%Almost half of our studies select at least two popular performance metrics each for their evaluation whereas
	%30 different metrics have been used by all the studies.
	%a total of 30 different metrics have been used by all studies.
	%They also select more than 100 subject systems for their experiments. 
	%We found that 70 primary studies have selected 100+ subject systems and the studies from
	%the studies from concept location and bug localization have used more queries than those from 
	%Internet-scale code search use less queries than the others for their evaluation.
	%About \textbf{30\%} of our primary studies were also selected for comparison by the later studies.
	%Our primary studies use a total of 15 statistical tests where Wilcoxon Signed Rank test has been adopted by more than 20\% of the studies.
	
	\item \textbf{RQ$_3$:} Existing query reformulation approaches suffer from eight major challenges and limitations including
	noisy keywords in their queries, vocabulary mismatch problem, lack of generalizability, human-induced biases, weak evaluation, and other prevailing issues that might prevent them from adoption by the software practitioners.

	% \emph{vocabulary mismatch problem} (30\%) and \emph{lack of external validity} (84\%). About 30\% of the studies also impose \emph{extra cognitive burdens} on the developers during query reformulation.

	\item \textbf{RQ$_4$:} Our selection criteria and personal knowledge led to a set of 70 primary studies from the last 15+ years
	%A total of 70 primary studies were conducted over the last 15 years 
	where 80\% of them were done in the last nine years. These studies appeared in the top Software Engineering venues such as ICSE, ASE, ICSME, TSE, TOSEM, and EMSE. Given the increasing size and complexity of modern software systems, the research 
	topics such as automated query reformulation and code search are likely to gain increased attention in the coming years.
	
	%Thus, researches on automated query reformulation targeting code searches are drawing much attention in the recent years.
	
	%To date, according to our investigation, a total of \textbf{70} primary studies  
	%on automated query reformulations supporting code search have been conducted. About \textbf{80\%} of them were conducted in the last nine years. This trend not only indicates the timeliness or relevance but also the attention or enthusiasm around this research topic. A large group \textbf{172} researchers contributed to this topic. About 50\% of the primary studies got published at the top venues of Software Engineering such as ICSE, ASE, ICSME, MSR, and SANER.
	%from all over the world either had or have been working on this research topic. The researches on automated query reformulation regularly got published in the top Software Engineering venues such as \textbf{ICSE}, \textbf{ASE}, \textbf{ICSME}, \textbf{MSR},  \textbf{EMSE} and \textbf{JSEP}.  
	%Given the increasing size and complexity of modern software systems, the research activities on automated query reformulations and code searches are likely to increase in the coming years.

	\item \textbf{RQ$_5$:} About 58\% of our primary studies perform query reformulations to support local code searches (e.g., concept location, bug localization, feature location) whereas the remaining 42\% focus on Internet-scale code search. 
	%In local code search, keywords are often selected using 
	%The studies supporting local code search rely on 
	%various term weighting methods (e.g., TF-IDF) whereas data mining and API recommendations are adopted
	%in the context of Internet-scale code search. 
	The primary studies from these two searches differ in their adopted methodologies, reformulation types, evaluation methods, and human participation. However, both groups of studies 
	%adopt similar methodologies for query reformulations and 
	suffer from a common set of issues such as a lack of generalizability, weak evaluation, and noisy queries. 
	
	%deal with local code searches (\eg\ bug localization) whereas the remaining 41\% studies target Internet-scale code search. The former group uses \emph{term weighting algorithms} extensively whereas the latter group greatly relies on various \emph{thesauri} (\eg\ WordNet) for query reformulation.
	
	\item \textbf{RQ$_6$:} Despite noticeable differences between local and Internet-scale code searches, there are a set of common, well-established practices in query reformulations targeting code search. These practices include selection of search queries from professional developers, selection of subject systems from both open and closed-source domains, execution of a query against multiple search engines, and comparison of a proposed technique with state-of-the-art techniques through replication.
	
	%Similarly, reformulated queries should be executed against multiple search engines to ensure a sound evaluation. 
	
	\item \textbf{RQ$_7$:} Constructing the right queries for code search has been a challenge for decades. Despite significant research, there is still room for further investigation. Future work can focus on 
	%Bug reports often contain appropriate keywords to localize the software bugs. Unfortunately, existing approaches are not efficient enough to identify them and construct the right queries.
	 adding more contexts (e.g., time-awareness) to their keyword selection algorithms, designing an appropriate fitness function for GA-based solutions, leveraging the structures from source code, or using neural language modeling (e.g., word embeddings) to deliver better queries for the code search.

	 %should focus on developing better algorithms to select appropriate reformulation keywords not only from bug reports but also from source code documents.
\end{itemize}

\section*{Acknowledgement}
This research was supported by Tenure-track startup grant, Dalhousie University, International Dean's Scholarship from University of Saskatchewan (2014--2017), Saskatchewan Innovation \& Opportunity Scholarship (2017--2018), and the Natural Sciences and Engineering Research Council of Canada (NSERC).

\newpage

\appendix

\section{Selected Primary Studies}\label{appendix:B}
Table \ref{table:selected-studies} shows the 70 primary studies selected for our systematic literature review.

\small
\begin{longtable}[!t]{l|p{1.2in}|p{3.6in}}
	\caption{Selected Primary Studies for Systematic Literature Review\vspace{-.35cm}}\label{table:selected-studies}\\
	%\vspace{-.35cm}
	%\resizebox{5.4in}{!}{%
	%\begin{threeparttable}
	%\begin{tabular}{l|l|p{4.8in}}
	%\hline
	\hline
	\textbf{ID} & \textbf{Primary Study} & \textbf{Description}  \\
	\hline
	\endfirsthead
	\multicolumn{3}{c}%
	{\tablename\ \thetable\ -- \small{Continued from previous page}} \\
	\hline
	\textbf{ID} & \textbf{Primary Study} & \textbf{Description} \\
	\hline
	\endhead
	\hline \multicolumn{3}{r}{\small{Continued on next page}} \\
	\endfoot
	\hline
	\endlastfoot
	S1 & \citet{time-aware-term-weighting} & Query reduction using noun term selection and time-aware term weighting for concept location \\
	\hline
	S2 & \citet{qr-log-mining} & Query expansion using keyword co-occurrences across the past queries from code search logs\\
	\hline
	S3 & \citet{testcase-cs-ist} & Ad hoc keyword selection from given test cases for semantic code search \\
	\hline
	S4 & \citet{active-cs} & Query expansion using modified Rocchio's method and the re-ranking of results for improved code search\\
	\hline
	S5 & \citet{api-book} & Query reduction by extracting nouns, verbs, adjectives and type information from a free-form query for code search \\
	\hline
	S6 & \citet{sisman} & Query reformulation using spatial code proximity for Information Retrieval-based bug localization \\
	\hline
	S7 & \citet{cocabu} & Query expansion using structured code entities from the relevant Q \& A threads of Stack Overflow\\
	\hline
	S8 & \citet{qperf} & Automatic performance prediction of a given search query using query difficulty analysis and machine learning without executing the query\\
	\hline
	S9 & \citet{flt-model-qr} & Query reformulation for feature location using Rocchio, RSV and Dice methods with software model families \\
	\hline
	S10 & \citet{refoqus} & Best query reformulation suggestion using Rocchio, RSV, Dice methods, query difficulty analysis and machine learning  \\
	\hline
	S11 & \citet{kevic} & Query reduction using advanced heuristics and logistic regression \\
	\hline
	S12 & \citet{hillicse09} & Semi-automated reformulation of NL queries with query keyword contexts and NL phrase extraction from method and field signatures\\
	\hline
	S13 & \citet{lemos-type-qr} & Query reformulation with NL thesaurus (\eg\ WordNet) in the context of interface-driven and keyword-based code searches \\
	\hline
	S14 & \citet{cexchange} &  Query reformulation using frequent identifier names from result files and other contextual information \\
	\hline
	S15 & \citet{codehow} & Query reformulation with relevant APIs by matching query texts and corresponding API documentations \\
	\hline
	S16 & \citet{correlation-cs} & Query expansion by using tag co-occurrences in the same and duplicate questions of Stack Overflow  \\
	\hline
	S17 & \citet{multi-rec} & Query reformulation using pre-search (\eg\ identifier and Verb-DO completion, term co-occurrences) and post-search recommendations (\eg\ synonym suggestion, spelling corrections) \\
	\hline
	S18 & \citet{qr-service-discovery} & Query reformulation using ontology models for web service discovery \\
	\hline
	S19 & \citet{icsme2018masud} &  Query reformulation using crowd generated knowledge and large-scale data analytics derived from Stack Overflow corpus\\
	\hline
	S20 & \citet{iman-tse} & Query reformulation using semantically related API classes for the improved Internet-scale code search\\
	\hline
	S21 & \citet{sisman-jsep} & Query reformulation using spatial code proximity, term ordering, Markov Random Field and query conditioning \\
	\hline
	S22 & \citet{ase2017masud} & Query reformulation using term weighting method, CodeRank, query difficulty analysis, and machine learning \\
	\hline
	S23 & \citet{req-qr} & Query reformulation of NL query using textual similarity with product requirements for feature location\\
	\hline
	S24 & \citet{multi-faceted-flt} &  Semi-automated query reformulation using structure, intent and dependency facets of search results\\
	\hline
	S25 & \citet{fse2018masud} & Query reformulation using bug report quality dynamics and graph-based term weighting algorithms\\
	\hline
	S26 & \citet{concept-ase} & Query reformulation using relevant API entities from RecoDec, conceptual knowledge, TF-IDF, and advanced heuristics\\
	\hline
	S27 & \citet{swordnet} & Query reformulation using semantically similar word pairs mined from comment-code contexts of source documents\\
	\hline
	S28 & \citet{ccmapping} & Query reformulation using semantically similar word pairs mined from leading comments and method signatures across projects\\
	\hline
	S29 & \citet{verbose} & Query reduction using gradual term removal with the help of query effectiveness\\
	\hline
	S30 & \citet{gayg} & Query reformulation using explicit relevance feedback from the developers and Rocchio's expansion method\\
	\hline
	S31 & \citet{precise-scalable-icpc} & Query reformulation with AST generation and XPath representation of example code snippet\\
	\hline
	S32 & \citet{qperf-mills-tosem} & Best query prediction using query difficulty analysis and machine learning for concept location and traceability recovery \\
	\hline
	S33 & \citet{query-by-example}  & Query reduction by subtree generation from the AST of input code \\
	\hline
	S34 & \citet{tsc2016} & Query expansion using Rocchio's method and pseudo-relevance feedback from Stack Overflow \\
	\hline
	S35 & \citet{qecc-intent} &  Query expansion based on intent mining from code change history and machine learning \\
	\hline
	S36 & \citet{pwordnet} & Query expansion using POS tagging and synonym addition from WordNet for improved concept location \\
	\hline
	S37 & \citet{li2016} & Query reformulation by term replacement with synonymous tags and by term expansion with related tags from Stack Overflow  \\
	\hline
	S38 & \citet{ase2016masud} & Query reformulation using frequently co-occurred keywords from source code and Stack Overflow threads \\
	\hline
	S39 & \citet{rack} &  Query reformulation using keyword-API co-occurrences in the Q \& A threads of Stack Overflow \\
	\hline
	S40 & \citet{saner2017masud} & Query reduction by selecting  important keywords using graph-based term weighting namely PageRank\\
	\hline
	S41 & \citet{swim} & Query reformulation using co-occurrence probability of keyword and API classes in the search engine logs\\
	\hline
	S42 & \citet{thesaurus-qr-idcs} & Query expansion using synonyms/antonyms from three thesauri-- WordNet, \citet{swordnet}, and type thesaurus\\
	\hline
	S43 & \citet{shepherd} & Query reformulation using Verb-DO extraction and their expansion using stemmed words, synonyms and similar usages\\
	\hline
	S44 & \citet{observed} & Query reduction using observed behaviour identification, Grounded Theory and natural language discourse analysis\\
	\hline
	S45 & \citet{kevicdict} & Query reformulation using NL-SCL mapping between changeset and IDE interaction history, and term weighting\\
	\hline
	S46 & \citet{qsynthesis} & Query reformulation with API method signatures using sentence parse tree extraction, uni-gram, and probabilistic context free grammar (PCFG) \\
	\hline
	S47 & \citet{querythecode} & Query reduction using POS tagging, selection of nouns and verbs and source code ontology\\
	\hline
	S48 & \citet{nlsupport} & Query reformulation using semantic web based ontology models, Ginseng, and SPARQL \\
	\hline
	S49 & \citet{anne} & Query/corpus reformulation using the mapping between programming concepts and their syntactic forms from Stack Overflow Q\&A threads\\
	\hline
	S50 & \citet{irmarcus} & Query reformulation using similar terms from the source code with Latent Semantic Indexing (LSI)\\ 
	\hline
	S51 & \citet{hard-trace-ase} & Query reformulation using web data mining and term weighting methods--domain term frequency, domain specificity, and concept generality\\ 
	\hline
	S52 & \citet{trace-query-transformation} & Query reformulation using trace query transformation rules generated by FP-growth algorithm\\
	\hline
	S53 & \citet{qr-interactive-clarification} & Query reformulation using ontology models across multiple repositories and candidate navigation path ranking \\
	\hline 
	S54& \citet{twkraft} & Query/corpus reformulation by weight boosting (\eg\ repeating) of structured entities from various contexts in the source code\\
	\hline
	S55 & \citet{sniff} & Query/corpus expansion by appending corresponding API documentations against method invocations. \\
	\hline
	S56 & \citet{auto-query} & Query suggestion by mining program dependency graphs from a given code segment\\
	\hline
	S57 & \citet{finelocator} & Query reformulation by expanding methods with neighboring methods based on semantic similarity, temporal proximity, and call dependency \\   
	\hline
	S58 & \citet{neural-query-expansion} & Query expansion using co-occurring API methods from the underlying corpus using encoder-decoder transformation and conditional probability analysis\\
	\hline
	S59 & \citet{collab-feature-location} & Query expansion using experts' descriptions of a feature, their confidence level, and Rocchio's method \\
	\hline
	S60 & \citet{interactive-qr-word-relations} & Query expansion using first and second order word co-occurrences, and object-oriented relations among the words \\ 
	\hline
	S61 & \citet{kim-bl-qr} & Query reformulation by adding attachment contents and by reducing noisy keywords from a bug report \\
	\hline
	S62 & \citet{manq-qr} & Query reformulation using Genetic algorithm and multi-objective optimization (e.g., query difficulty, OB, S2E, EB)\\
	\hline
	S63 & \citet{dl-qr-change-sequence} & Query expansion by deep learning change sequences and predicting the potentially changed terms for improved code search \\
	\hline
	S64 & \citet{low-cost-crowd-qr} & Query reformulation for feature location using Rocchio's method, relevance feedback, and expert's descriptions of a feature as the feedback documents \\
	\hline
	S65 & \citet{alteration-intent} & Query reformulation using old code examples as relevance feedback and then extracting expansion keywords (i.e., intent) from the change history of those code examples \\
	\hline
	S66 & \citet{qr-stackoverflow} & Query reformulation by learning to predict reformulated query from search logs of Stack Overflow using sequence-to-sequence learning and attention mechanism \\
	\hline
	S67 & \citet{lawrie-qr-bl} & Query selection for bug localization using Genetic algorithm and information need analysis as a fitness function \\
	\hline
	S68 & \citet{cmills-icsme2018} & Query reformulation for bug localization using Genetic algorithm and query effectiveness as a fitness function \\
	\hline
	S69 & \citet{genqr-tse-perez} & Query reformulation for feature location using Genetic algorithm using expansion, reduction, and selection as a mutation operation, and textual similarity as a fitness function\\
	\hline
	S70 & \citet{icpc2019rodrigo} & Query expansion for code example search using semantic similarity analysis, API suggestion, and term weighting (e.g., TF-IDF) \\
	\hline
\end{longtable}

%\vspace{-.8cm}
\normalsize

%\begin{appendices}
\newpage

\section{Quality Assessment of Primary Studies}\label{appendix:A}
Table \ref{table:quality-ass} shows answers for the questions outlined in Section \ref{sec:assessment}. We represent our responses as follows-- \pie{360} as ``Yes", \pie{0} as ``No", and \pie{270} or \pie{180} as ``Somewhat Yes".   

\small
\begin{longtable}[!t]{l|l|c|c|c|c|c|c|c|c|c|c}
	\caption{Quality Assessment of Primary Studies\vspace{-.35cm} \label{table:quality-ass}}\\
	%\vspace{-.35cm}
	%\resizebox{5.4in}{!}{%
	%\begin{threeparttable}
	%\begin{tabular}{l|l|p{4.8in}}
	%\hline
	\hline
	\textbf{ID} & \textbf{Primary Study} & \textbf{Q1} & \textbf{Q2} & \textbf{Q3} & \textbf{Q4} & \textbf{Q5} & \textbf{Q6} & \textbf{Q7} & \textbf{Q8} & \textbf{Q9} & \textbf{Q10}  \\
	\hline
	\endfirsthead
	\multicolumn{11}{c}%
	{\tablename\ \thetable\ -- \small{Continued from previous page}} \\
	\hline
	\textbf{ID} &\textbf{Primary Study} & \textbf{Q1} & \textbf{Q2} & \textbf{Q3} & \textbf{Q4} & \textbf{Q5} & \textbf{Q6} & \textbf{Q7} & \textbf{Q8} & \textbf{Q9} & \textbf{Q10} \\
	\hline
	\endhead
	\hline \multicolumn{11}{r}{\small{Continued on next page}} \\
	\endfoot
	\hline
	\endlastfoot
	S1 & \citet{time-aware-term-weighting} & \pie{360} & \pie{360} & \pie{360} &  \pie{360} & \pie{360}   & \pie{360} & \pie{360} & \pie{360} & \pie{360}  & \pie{180}\\
	\hline
	S2 & \citet{qr-log-mining} & \pie{360} & \pie{360}  & \pie{360} & \pie{360} & \pie{360} & \pie{360} & \pie{360} & \pie{0} & \pie{180} & \pie{360}   \\
	\hline
	S3 & \citet{testcase-cs-ist} & \pie{360} & \pie{360} & \pie{360} & \pie{180} & \pie{360} & \pie{360} & \pie{360} & \pie{360} & \pie{360} & \pie{360} \\
	\hline
	S4 &\citet{active-cs} & \pie{360} & \pie{360} & \pie{360}  & \pie{360}  & \pie{360} & \pie{360} & \pie{360} & \pie{360} & \pie{360} & \pie{0} \\
	\hline
	S5 & \citet{api-book} & \pie{360} & \pie{360}  & \pie{360} & \pie{360} & \pie{180} & \pie{360}  & \pie{180} & \pie{0} & \pie{180} & \pie{0}\\
	\hline
	S6 & \citet{sisman} & \pie{360} & \pie{360} & \pie{360}  & \pie{360}  & \pie{360}   & \pie{360} & \pie{360} & \pie{360} & \pie{360} & \pie{0}    \\
	\hline
	S7& \citet{cocabu} & \pie{360} &  \pie{360} & \pie{360}  & \pie{180} & \pie{360} & \pie{360} &  \pie{360} & \pie{360} & \pie{0} & \pie{360}\\
	\hline
	S8 & \citet{qperf} & \pie{360} & \pie{360} & \pie{360}  & \pie{0} & \pie{360} & \pie{360} & \pie{360} & \pie{360} & \pie{360}  & \pie{360}\\
	\hline
	S9 & \citet{flt-model-qr} & \pie{360} & \pie{360} & \pie{360} & \pie{360}  & \pie{360} & \pie{360} & \pie{360} & \pie{360} & \pie{0} & \pie{0}\\
	\hline
	S10 & \citet{refoqus} & \pie{360} & \pie{360}  &  \pie{360} & \pie{360}  & \pie{360}  & \pie{360} & \pie{360} & \pie{360} & \pie{180}  & \pie{180} \\
	\hline
	S11 & \citet{kevic} & \pie{360} & \pie{360}  & \pie{360}  &  \pie{360} & \pie{180} & \pie{360} & \pie{360} & \pie{0} & \pie{360} & \pie{360} \\
	\hline
	S12& \citet{hillicse09} & \pie{360} & \pie{360} & \pie{360} & \pie{360}  & \pie{360} & \pie{360} & \pie{360} & \pie{360} & \pie{360}  & \pie{180}\\
	\hline
	S13& \citet{lemos-type-qr} & \pie{360} & \pie{360} & \pie{360} & \pie{360}  & \pie{360} & \pie{360} & \pie{360} & \pie{360} & \pie{360}  &  \pie{360}   \\
	\hline
	S14& \citet{cexchange} & \pie{360} & \pie{360}  & \pie{360} & \pie{360} & \pie{360} & \pie{360} & \pie{360} & \pie{360} & \pie{360} & \pie{360} \\
	\hline
	S15& \citet{codehow} & \pie{360} & \pie{360} & \pie{360}  & \pie{360} & \pie{360} & \pie{360} & \pie{360} & \pie{360} & \pie{360} & \pie{0} \\
	\hline
	S16& \citet{correlation-cs} & \pie{360} & \pie{360} & \pie{360} & \pie{360} & \pie{360} & \pie{360} & \pie{360} & \pie{360} & \pie{360} & \pie{0}  \\
	\hline
	S17& \citet{multi-rec} & \pie{360} & \pie{360} & \pie{360} & \pie{360} & \pie{360} & \pie{360} & \pie{360} & \pie{360} & \pie{360} & \pie{360}\\
	\hline
	S18 & \citet{qr-service-discovery} & \pie{360} & \pie{360}  & \pie{360} & \pie{180} & \pie{180} & \pie{0} & \pie{180} & \pie{0} & \pie{360}  & \pie{0} \\
	\hline
	S19& \citet{icsme2018masud} & \pie{360} &\pie{360}  &\pie{360}  &\pie{360}  &\pie{360}   &\pie{360} &\pie{360} &\pie{360} &\pie{360}  &\pie{360}    \\
	\hline
	S20 & \citet{iman-tse} & \pie{360} & \pie{360} & \pie{360}  & \pie{360} &  \pie{360}  & \pie{360} & \pie{360} & \pie{360} & \pie{360}  &  \pie{180}\\
	\hline
	S21& \citet{sisman-jsep} & \pie{360} & \pie{360} & \pie{360} & \pie{360} & \pie{360}  & \pie{360} & \pie{360} & \pie{360} &  \pie{360} & \pie{180}\\
	\hline
	S22& \citet{ase2017masud} & \pie{360} & \pie{360} & \pie{360} &\pie{360}  &\pie{360}   &\pie{360} &\pie{360} & \pie{360} & \pie{360}  & \pie{360}   \\
	\hline
	S23& \citet{req-qr} & \pie{360} & \pie{360}  & \pie{360}  &  \pie{360}  &  \pie{270} & \pie{360} & \pie{360} & \pie{360} & \pie{360} &  \pie{0}   \\
	\hline
	S24& \citet{multi-faceted-flt} & \pie{360} & \pie{360} & \pie{360} & \pie{90} & \pie{360} & \pie{360} & \pie{360} & \pie{360} & \pie{360} & \pie{180}\\
	\hline
	S25& \citet{fse2018masud} & \pie{360} & \pie{360} & \pie{360} & \pie{360} & \pie{360}  & \pie{360}& \pie{360}&\pie{360} & \pie{360} & \pie{360}\\
	\hline
	S26&\citet{concept-ase} & \pie{360} & \pie{360}  & \pie{360}  & \pie{360}  & \pie{360}   & \pie{360}& \pie{360} & \pie{360} & \pie{360} &  \pie{0}   \\
	\hline
	S27&\citet{swordnet} & \pie{360} & \pie{360} & \pie{360}  & \pie{270} & \pie{360} & \pie{360} & \pie{360} & \pie{360} & \pie{360} & \pie{360}\\
	\hline
	S28&\citet{ccmapping} & \pie{360} & \pie{360} & \pie{360} & \pie{180} & \pie{360} & \pie{360} & \pie{360} & \pie{360} & \pie{360} & \pie{0}    \\
	\hline
	S29&\citet{verbose} & \pie{360} & \pie{360} & \pie{0} & \pie{180} & \pie{180}  & \pie{360} & \pie{360} & \pie{0} & \pie{180} & \pie{360} \\
	\hline
	S30&\citet{gayg} & \pie{360} & \pie{360} & \pie{360} & \pie{360}  & \pie{360} & \pie{360} & \pie{360} & \pie{360} & \pie{360} & \pie{180}   \\
	\hline
	S31&\citet{precise-scalable-icpc} & \pie{360} & \pie{360} & \pie{360}  & \pie{360}  &  \pie{360}  & \pie{360} & \pie{360} & \pie{360} &  \pie{360} & \pie{0}    \\
	\hline
	S32&\citet{qperf-mills-tosem} & \pie{360} & \pie{360}  & \pie{360} & \pie{0} &  \pie{360}  & \pie{360} & \pie{360} &  \pie{360} & \pie{360} & \pie{0}    \\
	\hline
	S33&\citet{query-by-example} & \pie{360} & \pie{360}  & \pie{360}  & \pie{360}  & \pie{360}   & \pie{360} & \pie{360} & \pie{0} &  \pie{360} &  \pie{0}   \\
	\hline
	S34&\citet{tsc2016} & \pie{360} & \pie{360} & \pie{360}  & \pie{360} & \pie{360} & \pie{360} & \pie{360} & \pie{360} & \pie{360} &  \pie{360}\\
	\hline
	S35&\citet{qecc-intent} & \pie{360} & \pie{360} & \pie{360}  & \pie{360}  & \pie{360}  & \pie{360} & \pie{360} & \pie{360} & \pie{360} & \pie{0}\\
	\hline
	S36&\citet{pwordnet} & \pie{360} & \pie{360}  &  \pie{360} &  \pie{360} & \pie{270} & \pie{360} & \pie{360} & \pie{0} & \pie{360} & \pie{0}\\
	\hline
	S37&\citet{li2016} & \pie{360} &  \pie{360} & \pie{360}  & \pie{360}  & \pie{360} & \pie{360} & \pie{360} & \pie{360} & \pie{360} & \pie{0}    \\
	\hline
	S38&\citet{ase2016masud} & \pie{360} & \pie{360}  & \pie{360}  & \pie{360}  & \pie{180} & \pie{360} & \pie{360} & \pie{0} & \pie{360} & \pie{180}\\
	\hline
	S39&\citet{rack} & \pie{360} & \pie{360} & \pie{360} & \pie{360} & \pie{360}  & \pie{360} & \pie{360} & \pie{360} & \pie{360}  & \pie{360}\\
	\hline
	S40&\citet{saner2017masud} & \pie{360} & \pie{360} &  \pie{360} & \pie{360}  & \pie{360}  & \pie{360} & \pie{360} & \pie{360} & \pie{360} &  \pie{360}   \\
	\hline
	S41&\citet{swim} & \pie{360} & \pie{360}  & \pie{360}  & \pie{360}  & \pie{360}  & \pie{360} & \pie{360} & \pie{0} &  \pie{360} &  \pie{0}   \\
	\hline
	S42&\citet{thesaurus-qr-idcs} & \pie{360} & \pie{360}  & \pie{360}  & \pie{360}  & \pie{360}   & \pie{360} & \pie{360} & \pie{360} & \pie{360}  &  \pie{180}   \\
	\hline
	S43&\citet{shepherd} & \pie{360} & \pie{360} & \pie{360}  & \pie{360}  & \pie{360}   & \pie{360} & \pie{360} & \pie{360} & \pie{360}  & \pie{180}    \\
	\hline
	S44&\citet{observed} & \pie{360} & \pie{360}  & \pie{360}  & \pie{360}  & \pie{360}  & \pie{360} & \pie{360} & \pie{360} & \pie{360} & \pie{360}    \\
	\hline
	S45&\citet{kevicdict} & \pie{360} & \pie{360}  & \pie{360}  & \pie{360}  & \pie{180} & \pie{360} & \pie{360} & \pie{0} & \pie{360}  &  \pie{360}   \\
	\hline
	S46&\citet{qsynthesis} & \pie{360} & \pie{360}  & \pie{360}  & \pie{180}  & \pie{180}  & \pie{360} & \pie{360} & \pie{0} & \pie{360}  & \pie{0}    \\
	\hline
	S47&\citet{querythecode} & \pie{360} & \pie{360}  & \pie{360} & \pie{360} & \pie{180}  & \pie{360} & \pie{360} & \pie{0}  & \pie{360}  & \pie{0}   \\
	\hline
	S48&\citet{nlsupport} & \pie{360} & \pie{360}  & \pie{360} & \pie{360} & \pie{360}  & \pie{360} & \pie{360} & \pie{0}  & \pie{360}  &  \pie{270}    \\
	\hline
	S49&\citet{anne} & \pie{360} & \pie{360}  & \pie{360}  & \pie{360}  & \pie{360}   & \pie{360} & \pie{360} & \pie{360} & \pie{360}  & \pie{360} \\
	\hline
	S50&\citet{irmarcus} & \pie{360} & \pie{360}  & \pie{360}  & \pie{180}  & \pie{360}   & \pie{360} & \pie{360} & \pie{0} & \pie{360} & \pie{0}    \\
	\hline
	S51&\citet{hard-trace-ase} & \pie{360} & \pie{360} & \pie{360}  & \pie{360}  & \pie{360}   & \pie{360} & \pie{360} & \pie{0} & \pie{360} & \pie{360}   \\
	\hline
	S52&\citet{trace-query-transformation} & \pie{360} &  \pie{360}  & \pie{360}  & \pie{360}  &  \pie{180}  &\pie{360} & \pie{360} & \pie{360} & \pie{360}  & \pie{0}    \\
	\hline
	S53&\citet{qr-interactive-clarification} & \pie{360} & \pie{360}  &  \pie{360} & \pie{360}  & \pie{360} & \pie{360} & \pie{360} & \pie{360}  & \pie{360} &  \pie{0}   \\
	\hline
	S54&\citet{twkraft-jsep} & \pie{360} &  \pie{360} & \pie{360}  & \pie{0} & \pie{360}  & \pie{360} & \pie{360} & \pie{360} &  \pie{360} &  \pie{180}   \\
	\hline
	S55&\citet{sniff} & \pie{360} & \pie{360}  & \pie{360} & \pie{180} & \pie{360}  & \pie{360} & \pie{360} &\pie{0}  & \pie{360}  &  \pie{0}   \\
	\hline
	S56&\citet{auto-query} & \pie{360} & \pie{360} & \pie{360} & \pie{360} & \pie{360} & \pie{360} & \pie{360} & \pie{360} & \pie{360} & \pie{0}  \\
	\hline
	S57&\citet{finelocator} & \pie{360} & \pie{360} & \pie{360} & \pie{360} & \pie{360} & \pie{360} & \pie{360} & \pie{360} & \pie{360} & \pie{0 } \\
	\hline
	S58&\citet{neural-query-expansion} & \pie{360} & \pie{360} & \pie{360} & \pie{360} &  \pie{360} & \pie{360} & \pie{360} &  \pie{0} & \pie{360} & \pie{0} \\
	\hline
	S59&\citet{collab-feature-location} & \pie{360} & \pie{360} & \pie{360} & \pie{360} & \pie{360} & \pie{360} & \pie{360} & \pie{360} & \pie{90} & \pie{0} \\
	\hline
	S60&\citet{interactive-qr-word-relations} &  \pie{360} &  \pie{360} & \pie{360} & \pie{360} & \pie{360} & \pie{360} & \pie{360} & \pie{360} & \pie{360} & \pie{0} \\
	\hline
	S61&\citet{kim-bl-qr} & \pie{360} & \pie{360} & \pie{360} & \pie{360} & \pie{360} & \pie{360} & \pie{360} & \pie{0} & \pie{180} & \pie{0} \\
	\hline
	S62&\citet{manq-qr} & \pie{360} & \pie{360} & \pie{180} & \pie{270} & \pie{180} & \pie{360} & \pie{180} & \pie{0} & \pie{270} & \pie{0} \\
	\hline
	S63&\citet{dl-qr-change-sequence} & \pie{360} & \pie{360} & \pie{360} & \pie{360} & \pie{360} &  \pie{360} & \pie{360} & \pie{360} & \pie{0} & \pie{0}\\
	\hline
	S64&\citet{low-cost-crowd-qr} & \pie{360} & \pie{360} & \pie{360} & \pie{360} & \pie{180} & \pie{360} & \pie{360} & \pie{360} & \pie{360} & \pie{0}\\
	\hline
	S65&\citet{alteration-intent} & \pie{360} & \pie{360} & \pie{360} & \pie{360} & \pie{180} & \pie{270} & \pie{360} & \pie{180} & \pie{360} & \pie{0} \\
	\hline
	S66&\citet{qr-stackoverflow} & \pie{360} & \pie{360} & \pie{360} & \pie{360} & \pie{360} & \pie{360} & \pie{360} & \pie{0} & \pie{360} & \pie{360} \\ 
	\hline
	S67&\citet{lawrie-qr-bl} & \pie{360} & \pie{360} & \pie{360} & \pie{180} & \pie{180} & \pie{360} & \pie{360} & \pie{0} & \pie{360} & \pie{180} \\
	\hline
	S68&\citet{cmills-icsme2018} & \pie{360} & \pie{360} & \pie{360} & \pie{180} & \pie{180} & \pie{360} & \pie{360} & \pie{360} & \pie{360} & \pie{360} \\
	\hline
	S69&\citet{genqr-tse-perez} & \pie{360} & \pie{360} & \pie{360} & \pie{360} & \pie{360} & \pie{360} & \pie{360} & \pie{360} & \pie{360} & \pie{360} \\
	\hline
	S70&\citet{icpc2019rodrigo} & \pie{360} & \pie{360} & \pie{360} & \pie{360} & \pie{360} & \pie{360} & \pie{360} & \pie{360} & \pie{360} & \pie{360} \\
	\hline
	
\end{longtable}

%%\vspace{-.8cm}
\normalsize

\newpage
\section{Evaluation Metrics}\label{appendix:C}
Table \ref{table:evaluation-metrics} defines the performance metrics frequently used by the primary studies.

\small
\begin{table}[H]
	\centering
	\caption{Metrics for performance evaluation}\label{table:evaluation-metrics}
	\vspace{-.35cm}
	\resizebox{5.4in}{!}{%
		\begin{threeparttable}
			\begin{tabular}{l|p{3in}|l}
				\hline
				\textbf{Metric} & \textbf{Overview} & \textbf{Formula} \\
				\hline
				\hline
				Hit@K & Percentage of all search queries ($Q$) for each of which at least one relevant result is returned within the top K results. & Hit@K = $\frac{\sum_{q\in Q}rel(q, ~K)}{|Q|}$\\
				\hline
				QE & Query Effectiveness (QE) refers to the rank of the first relevant result returned by a search query. &  \\ 
				\hline
				MRR & Reciprocal Rank is the multiplicative inverse of the rank of the first relevant result returned by a search query ($q \in Q$).
				Mean Reciprocal Rank (MRR) averages the reciprocal ranks for all search queries ($Q$). & MRR(Q) = $\frac{1}{|Q|}\sum_{q\in Q} \frac{1}{rank(q)}$ \\  
				\hline
				Precision & Precision refers to the fraction of retrieved results that are relevant. Here TP means true positive and FP means false positives. & P = $\frac{TP}{TP+FP}$\\
				\hline
				Recall & Recall refers to the fraction of ground truth results that are retrieved. Here, FN means false negatives. & R = $\frac{TP}{TP+FN}$\\
				\hline
				F1-score & F1-score calculates the harmonized mean of precision and recall measures. & F1-score = $\frac{2\times P\times R}{P+R}$\\
				\hline
				MAP & Precision@k calculates precision at every occurrence of a relevant result within a ranked list. Average Precision (AP) averages over all Precision@k measures for a given query $q$. & AP(q) = $\frac{\sum_{k=1}^{K}P_k}{|RR|}$\\
				\hhline{~~~}
				&Mean Average Precision (MAP) calculates the mean of AP for all queries ($Q$). Here $P_k$ refers to precision@k and  $RR$ means the relevant results retrieved by a query. & MAP(Q) = $\frac{\sum_{q\in Q}AP(q)}{|Q|}$\\
				\hline
				NDCG & Discounted Cumulative Gain (DCG) determines the quality of a ranking technique by rewarding when the highly relevant results are found at the top positions and by penalizing when the highly relevant results are found at low positions of ranked list.
				& DCG(q) = $\sum_{k=1}^{K}\frac{grel_k}{log_2(k+1)}$ \\
				\hhline{~~~}
				& Normalized Discounted Cumulative Gain (NDCG) calculates to the ratio between DCG and Ideal Discounted Cumulative Gain (IDCG) which is based on ground truth information. & NDCG(q) = $\frac{DCG(q)}{IDCG(q)}$ \\ 
				& Here $grel_k$ refers to graded relevance of $k_{th}$ result based on ground truth information. & NDCG(Q)=$\frac{1}{|Q|}\sum_{q\in Q}NDCG(q)$ \\
				\hline
				Spent time & Time spent to accomplish a certain programming task. & \\
				\hline
				Likert & Assessment made by human participants on a numerical scale. & \\
				\hline
				
			\end{tabular}
		\end{threeparttable}
	}
\end{table}

\normalsize

\newpage
\section{Experimental Dataset}\label{appendix:D}
Table \ref{table:study-dataset-selection} shows the datasets used by the primary studies.
\small
\begin{longtable}[!t]{l|c|c|p{3in}}
	\caption{Dataset used by primary studies\vspace{-.35cm}}\label{table:study-dataset-selection}\\
	\hline
	\textbf{ID} & \textbf{Queries} & \textbf{Systems} & \textbf{Origin of queries} \\
	\hline
	\endfirsthead
	\multicolumn{4}{c}%
	{\tablename\ \thetable\ -- \small{Continued from previous page}} \\
	\hline
	\textbf{ID} & \textbf{Queries} & \textbf{Systems} & \textbf{Origin of queries} \\
	\hline
	\endhead
	\hline \multicolumn{4}{r}{\small{Continued on next page}} \\
	\endfoot
	\hline
	\endlastfoot
	S1 & 200 & 04 & Change requests, version control history \\
	\hline
	S2 & 22 & - & Code search logs \\		
	\hline
	S3 & 34 & - & Test cases\\
	\hline
	S4 & 70 & 01 & Existing benchmark \\ 
	\hline
	S5 & 70 & - & Stack Overflow questions \\
	\hline
	S6 & 4,393 & 02 & Bug reports \\ 
	\hline
	S7 & 174 & 01 & Stack Overflow questions, study participants \\ 
	\hline
	S8 & 164 & 05 & Change requests \\
	\hline
	S9 & 217 & 02 & Feature requests, feature description \\
	\hline
	S10 & 282 & 05 & Change requests, study participants\\
	\hline
	S11 & 06 & 01 & Change requests \\
	\hline
	S12 & 28 & - & Study participants \\
	\hline
	S13 & 128 & - & Study participants \\
	\hline
	S14 & 06 & - & Code search logs \\
	\hline
	S15 & 129 & - & Existing benchmark, code search logs, Stack Overflow questions, study participants \\
	\hline
	S16 & 82 & - & Study participants \\
	\hline
	S17 & 4,917 & - & Code search logs \\
	\hline
	S18 & 07 & - & Study participants \\
	\hline
	S19 & 310 & 04 & Existing benchmark, programming tutorial sites \\
	\hline
	S20 & 74 & 24,666 & Existing benchmark, Stack Overflow questions \\
	\hline
	S21 & 4,684 & 03 & Bug reports \\
	\hline
	S22 & 1,675 & 08 & Change requests \\
	\hline
	S23 & 21 & 07 & Bug reports, product descriptions \\
	\hline
	S24 & 04 & - & Existing benchmark \\
	\hline
	S25 & 5,139 & 06 & Bug reports \\
	\hline
	S26 & 300 & 03 & Stack Overflow questions \\
	\hline
	S27 & 08 & 04 & Existing benchmark \\
	\hline
	S28 & 97 & - & Code comment \\
	\hline
	S29 & 424 & 08 & Bug reports \\
	\hline
	S30 & 30 & 03 & Change requests \\
	\hline
	S31 & 08 & - & Code search logs \\
	\hline
	S32 & 1043 & 15 & Bug reports, use cases \\
	\hline
	S33 & 05 & - & Study participants \\
	\hline
	S34 & 20 & 01 & Stack Overflow questions \\
	\hline
	S35 & 54 & - & Existing benchmark \\
	\hline
	S36 & 19 & 01 & Existing benchmark \\
	\hline
	S37 &  14 & - & Study participants \\
	\hline
	S38 & 510 & 02 & Change requests \\
	\hline
	S39 & 175 & 03 & Programming tutorial sites \\
	\hline
	S40 & 1,939 & 08 & Change requests \\
	\hline
	S41 & 30 & 1 & Code search logs \\
	\hline
	S42 & 13 & 100 & Existing benchmark \\
	\hline
	S43 & 09 & 04 & Bug reports \\ 
	\hline
	S44 & 451 & 21 & Bug reports \\
	\hline
	S45 & 1,787 & 04 & Change requests\\
	\hline
	S46 & 45 & - & Hand-picked \\
	\hline
	S47 & 276 & - & Study participants \\
	\hline
	S48 & - & - & Case study \\
	\hline
	S49 & 169 & - & Study participants \\
	\hline
	S50 & 88 & 01 & Study participants, source code\\ 
	\hline
	S51 & 59 & 10 & Regulatory documents \\
	\hline
	S52 & 383 & 01 & Requirement documents \\
	\hline
	S53 & 5 & 01 & Task description \\
	\hline
	S54 & 372 & 04 & Feature description \\
	\hline
	S55 & 12 & - & Study participants\\
	\hline
	S56 & 47 & 04 & Task description \\
	\hline
	S57 & 531 & 05 & Bug reports \\
	\hline
	S58 & 209 & 737 & Existing benchmark \\
	\hline
	S59 & 43 & 23 & Study participants \\
	\hline
	S60 & 20 & 1 & Bug reports \\
	\hline
	S61 & 1,546 & 6 & Existing benchmark \\
	\hline
	S62 & 1,546 & 6 & Existing benchmark \\
	\hline
	S63 & 54 & 625 & Existing benchmark \\
	\hline
	S64 & 43 & 1 & Industry partner \\
	\hline
	S65 & 34 & 2,151 & Existing benchmark \\
	\hline
	S66 & 65,103 & 1 & Stack Overflow search logs \\
	\hline
	S67 & 613 & 13 & Existing benchmark \\
	\hline
	S68 & 620 & 13 & Bug reports \\
	\hline
	S69 & 121 & 23 & Feature descriptions \\
	\hline
	S70 & 121 & 1 & Stack Overflow questions\\
	\hline
\end{longtable}
\normalsize

\newpage
\section{Contributing Authors}\label{appendix:E}
Table \ref{table:paper-authors} shows the authors who contributed to at least two primary studies.
\small
\begin{table}[H]
	\centering
	\caption{Contributing authors}\label{table:paper-authors}
	\vspace{-.35cm}
	%\resizebox{3in}{!}{%
	\begin{threeparttable}
		\begin{tabular}{l|l}
			\hline
			\textbf{Author} & \textbf{\# Studies}\\
			\hline
			A. Marcus&7 \\
			\hline
			M. M. Rahman&6\\
			\hline
			C. Roy&6\\
			\hline
			D. Lo&5\\
			\hline
			S. Haiduc&4\\
			\hline
			S. Wang&4\\
			\hline
			L. Pollock&3\\
			\hline
			G. Bavota&3\\
			\hline
			C. Lopes&3\\
			\hline
			R. Oliveto&3\\
			\hline
			Z. Li&2\\
			\hline
			O. Chaparro&2\\
			\hline
			Y. Zou&2\\
			\hline
			A. Kak&2\\
			\hline
			A. Lucia &2\\
			\hline
			K. Vijay-Shanker&2\\
			\hline
			Y. Zhang&2\\
			\hline
			E. Hill&2\\
			\hline
			F. Pérez&2\\
			\hline
			O. Lemos&2\\
			\hline
			D. Shepherd&2\\
			\hline
			J. Zhao&2\\
			\hline
			L. Jiang&2\\
			\hline
			J. Font&2\\
			\hline
			X. Peng&2\\
			\hline
			W. Zhao&2\\
			\hline
			S. Jain&2\\
			\hline
			T. Fritz&2\\
			\hline
			T. Menzies&2\\
			\hline
			T. Wang&2\\
			\hline
			G. Yin&2\\
			\hline
			B. Sisman&2\\
			\hline
			K. Kevic&2\\
			\hline
			C. Cetina &2\\
			\hline
		\end{tabular}
	\end{threeparttable}
	%	}
\end{table}
\normalsize

\newpage
\renewcommand*{\bibfont}{\normalsize}
\bibliographystyle{ACM-Reference-Format}
\bibliography{MyBibTex}

\end{document}